\documentclass[draft]{agujournal2019}
\usepackage{url} 
\usepackage{lineno}
\usepackage{soul}
\usepackage{siunitx}
\usepackage{ragged2e}
\usepackage{booktabs}
\usepackage[version=4]{mhchem}

\draftfalse

\journalname{Journal of Geophysical Research: Solid Earth}
\newcommand{\hbrg}[4]{Mg$_{#1}$Si$_{#2}$O$_{#3}$H$_{#4}$}

\begin{document}
\justifying

\title{Hydrogen diffusion in the lower mantle revealed by machine learning potentials}

%
%

\authors{Yihang Peng\affil{1}, Jie Deng\affil{1}}

\affiliation{1}{Department of Geosciences, Princeton University, Princeton, NJ 08544, USA.}

\correspondingauthor{Yihang Peng}{yhpeng@princeton.edu}
\correspondingauthor{Jie Deng}{jie.deng@princeton.edu}



\begin{keypoints}
\item A machine learning potential of \textit{ab initio} quality has been developed for hydrous MgSiO$_3$ system
\item Hydrogen diffusion is sluggish in both bridgmanite and post-perovskite, leading to heterogeneous water distribution in the lower mantle
\item {Proton diffusion may be associated with the variation of electrical conductivity in the deep mantle}
\end{keypoints}

%
%


\begin{abstract}

Hydrogen may be incorporated into nominally anhydrous minerals including bridgmanite and post-perovskite as defects, making the Earth's deep mantle a potentially significant water reservoir. The diffusion of hydrogen and its contribution to the electrical conductivity in the lower mantle are rarely explored and remain largely unconstrained. Here we calculate hydrogen diffusivity in hydrous bridgmanite and post-perovskite, using molecular dynamics simulations driven by machine learning potentials of \textit{ab initio} quality. Our findings reveal that hydrogen diffusivity significantly increases with increasing temperature and decreasing pressure, and is considerably sensitive to hydrogen incorporation mechanism. Among the four defect mechanisms examined, (Mg + 2H)\textsubscript{Si} and (Al + H)\textsubscript{Si} show similar patterns and yield the highest hydrogen diffusivity. Hydrogen diffusion is generally faster in post-perovskite than in bridgmanite, and these two phases exhibit distinct diffusion anisotropies. Overall, hydrogen diffusion is slow on geological time scales and may result in heterogeneous water distribution in the lower mantle. Additionally, the proton conductivity of bridgmanite for (Mg + 2H)\textsubscript{Si} and (Al + H)\textsubscript{Si} defects aligns with the same order of magnitude of lower mantle conductivity, suggesting that the water distribution in the lower mantle may be inferred by examining the heterogeneity of electrical conductivity.
 
\end{abstract}

\section*{Plain Language Summary}

Water or hydrogen may be trapped in the Earth's deep mantle, affecting the state and evolution of our planet. However, the mobility of hydrogen in the lower mantle remains poorly understood. By using advanced machine learning-driven simulations, we find that hydrogen in lower-mantle silicates diffuses faster at higher temperatures and lower pressures, and how hydrogen is incorporated into the silicates greatly influences this mobility. Two specific ways of hydrogen incorporation were found to allow hydrogen to move very efficiently. On a geological scale, the transport of hydrogen is slow, suggesting that water may be unevenly distributed in the Earth's lower mantle. The electrical conductivity associated with the movement of hydrogen aligns with previous observations of the Earth's mantle. Variations in electrical conductivity in the lower mantle may inform where water is located in the Earth's deep interior.

%
%

\section{Introduction}

Water is widely present in the Earth's deep interior \cite{Ohtani2021a}, and affects the physical and chemical properties of the Earth's materials, including rheology \cite{Karato1986a}, seismic wave velocities \cite{Karato1995a}, phase transitions \cite{Ohtani2006a}, and electrical conductivity \cite{Karato1990a}. Therefore, investigating the behavior of hydrogen in mantle phases is important for understanding the dynamics and evolution of the Earth. {(Mg,Fe)SiO$_3$} bridgmanite (Brg) is believed to be the dominant phase in the Earth's lower mantle. {This material remains stable across a broad range of depths, from 660 km discontinuity down to the bottom of the mantle, where the pressure and temperature (P/T) conditions may change from $\sim$23 GPa and 2000 K to $\sim$136 GPa and 4000 K {\cite{fiquet_melting_2010}}}. Bridgmanite may transform into post-perovskite (pPv) at the lowermost mantle conditions {{\cite<above 125 GPa and 2500 K, >{Murakami2004a}}}. Water or hydrogen can be incorporated into bridgmanite and post-perovskite as defects \cite{Fu2019a, Townsend2016a}, potentially forming volatile-rich reservoirs in the lower mantle \cite{Hallis2015a}. {The water storage capacity in the lower mantle is still under debate, since studies on the water solubility in bridgmanite have not reached a consensus. As summarized by {\citeA{lu_solubility_2023}}, experimental results show water content in bridgmanite spanning over three orders of magnitude. Even the most recent studies have yielded results ranging from less than 100 ppm {\cite{liu_bridgmanite_2021, ishii_aluminum_2022}} to over 1000 ppm {\cite{Fu2019a, yang_nanosims_2023}}. The solubility of water in bridgmanite is potentially influenced by a combination of factors including chemical composition, pressure, temperature, and oxygen fugacity, and is likely to have a strong depth dependence {\cite{yang_nanosims_2023, lu_solubility_2023, Muir2018}}.}

Owing to its light atomic weight {and small radius}, hydrogen may be highly mobile and characterized with large diffusivity. The possible incorporation of hydrogen into bridgmanite and post-perovskite and its presumably large diffusivity may affect the geophysical and geochemical properties of the deep mantle, including electrical conductivity.

Electrical conductivity of the Earth's lower mantle exhibits significant variations both radially and laterally revealed by geomagnetic observations \cite<e.g.,>{Olsen1999b, Verhoeven2021a, Tarits2010a, Khan2011}. {Using three-dimensional electromagnetic inversion, {\citeA{Tarits2010a}} reported a laterally heterogeneous conductivity structure at the depth range of 900-1400 km. In addition, with resolution limited to the depth range of $\sim$500--1200 km, {\citeA{Khan2011}} found lateral variations in electrical conductivity throughout the mantle, and the magnitude of the variation reaches 1 S\;m$^{-1}$.} Yet, the source of these heterogeneities remains enigmatic. The electrical conductivity of the lower mantle is sensitive to the presence of a trace amount of hydrogen due to its high mobility \cite{Karato1990a}. Consequently, the heterogeneous electrical conductivity may be caused by the heterogeneous distribution of water \cite{Hae2006a,Ohtani2021a, Sun2015a, Zhou2022a}. Assessing this hypothesis requires a thorough understanding of hydrogen diffusivity in bridgmanite and post-perovskite, as both the distribution of water in the lower mantle and hydrogen's contribution to electrical conductivity are governed by the diffusion of hydrogen. However, neither theoretical nor experimental data of hydrogen diffusivity in bridgmanite or post-perovskite are currently available.

In this study, we perform molecular dynamics (MD) simulations driven by machine learning potentials to investigate the hydrogen diffusion in bridgmanite and post-perovskite. Calculating diffusivity using \textit{ab initio} molecular dynamics (AIMD) simulations is straightforward and well-developed, with no {\it ad hoc} diffusion pathway assumed. {In these simulations, atoms move in a periodic box driven by Newtonian dynamics, with interatomic forces computed by the density functional theory (DFT), and diffusivity is directly derived from trajectories of atoms.} Deriving a statistically robust result, however, is very costly, especially when the diffusion of atoms is sluggish \cite<e.g.,>{Caracas2017a}. Moreover, a system composed of thousands of atoms or even tens of thousands of atoms is required to simulate bridgmanite and post-perovskite with realistic water contents of less than thousands of ppm, which exceeds the capacity of \textit{ab initio} methods, {where computational cost scales to the third or fourth power of the number of electrons}.

Machine learning potentials (MLPs) are an emerging method allowing to simulating large systems and long {simulation} trajectories with high efficiency ({the cost scales linearly with the number of atoms}) while maintaining \textit{ab initio} accuracy \cite{Zhang2018a}. MLPs have recently proven successful in accurately capturing various crystal defect properties, including their kinetics \cite{Freitas2022}. Our MLP for water-bearing bridgmanite and post-perovskite is developed by training an artificial neural network model with the \textit{ab initio} data, following the protocol described in \citeA{Deng2023}. {Trained with atomic coordinates, energies, stresses, and atomic forces from AIMD simulations, the model can predict potential energy and force field based on the local environment of atoms, which are further used to perform MD simulations.} The long trajectories and large systems made possible by the MLP facilitate a detailed examination of hydrogen diffusion under wide pressure, temperature, and compositional conditions. The results inform the distribution of water in the lower mantle and the contributions of hydrogen to the electrical conductivity of bridgmanite and post-perovskite, which are further compared against geophysical observations.

\section{Methods}
\subsection{Building a Machine Learning Potential}

We adopt the DeePMD approach \cite{Wang2018a,Zhang2018a} to build an MLP from \textit{ab initio} data. In this approach, the total potential energy is written as the sum of the energy of each atom, $E=\sum_{i=1}^N E_i=\sum_{i=1}^N \epsilon (\chi_i)$, where $N$ is the total number of atoms, $E_i$ is the contribution to the total energy from atom i, and $\chi_i$ is a descriptor representing the local atomic environment and is a function of the coordinates of the atoms in the vicinity of atom $i$.

In this study, the descriptor is parameterized using an embedding network with three hidden layers of 25, 50, and 100 nodes, and the form of the potential energy surface is determined using a three-layer fitting network with 240 nodes in each layer. The atomic local environment is described with a cutoff radius of 6 Å. The loss function is defined as
\begin{linenomath*}
\begin{equation}
    L\left(p_\epsilon, p_f, p_{\xi}\right)=p_\epsilon \Delta \epsilon^2+\frac{p_f}{3 N} \sum_i\left|\Delta
    \mathbf{F}_i\right|+\frac{p_{\xi}}{9}\|\Delta \xi\|^2
\end{equation}
\end{linenomath*}
, where $p_\epsilon$, $p_f$, $p_\xi$ are tunable prefactors for the difference between the MLP prediction and training data. Here $\epsilon$ is the energy per atom, $\mathbf{F}_i$ is the atomic force of atom $i$, $\xi$ is the virial tensor divided by $N$, and $N$ is the number of atoms. We follow \citeA{Deng2023} to increase both $p_\epsilon$ and $p_\xi$ from 0.02 to 1 while decreasing $p_f$ from 1000 to 1 over the course of training.

Our MLP is built upon an earlier version for pure MgSiO$_3$ \cite{Deng2023}. This new MLP targets at a wider compositional space, covering not only pure MgSiO$_3$, but also pure H$_2$O, pure H$_2$ (liquid only), and intermediate compositions with bulk compositions of $a \mathrm{MgSiO_3} + b \mathrm{H_2O}$, $a \mathrm{MgSiO_3} + b\mathrm{H}_2$ and of $a \mathrm{MgSiO_3} + c \mathrm{OH}$, where $a = 27$ or 32, $b = 1$ to 9, $c = 1$ to 18. In addition, we explicitly consider the intermediate composition obtained by charge-coupled cation substitution, i.e., $a \mathrm{MgSiO_3} +b \mathrm{H} - c\mathrm{Mg}$, where $a = 32$, $b = 2c$, corresponding to (2H)\textsubscript{Mg}; $a \mathrm{MgSiO_3} + b \mathrm{H} - c \mathrm{Si}$, where $a = 32$, $b = 4c$, corresponding to (4H)\textsubscript{Si}; $a \mathrm{MgSiO_3} +b \mathrm{H} +c \mathrm{Mg} - d \mathrm{Si}$, where $a = 32$, $b + 2c = 4d$, corresponding to (Mg + 2H)\textsubscript{Si}.

For all compositions, we employ an iterative training scheme proposed by \citeA{Deng2023} to ensure that the potential energy surface is sufficiently sampled. This scheme mainly entails enhanced sampling using LAMMPS \cite{Plimpton1995a} interfaced with PLUMED 2 \cite{Tribello2014a}, principal component analysis on the MD trajectories with the farthest point sampling technique \cite{Cheng2020a,Imbalzano2018a}, recalculation with DFT at higher accuracy for selected frames, and refining the MLP with recalculated new frames. The iterative training scheme ensures that the training set is succinct and balanced. Indeed, the size of the final training set is very small, consisting of only 9095 configurations. {Here, a ``configuration'' refers to a unique spatial arrangement of atoms, labeled with {\textit{ab initio}} energy and atomic forces for training the MLP.} This training set covers a wide range of temperature and pressure (1000--8000 K and 0--220 GPa), compared to typically tens of thousands of frames for mono-atomic species over much narrower ranges of pressure and temperature conditions \cite{Niu2020a,Yang2021a}. {The electronic temperatures are taken into account during both the DFT calculation and the MLP training processes. To mitigate edge effects during machine learning, we deliberately encompass a pressure range significantly exceeding the highest pressure found in Earth's mantle. This approach ensures robustness. For instance, when the input data covers solely the 0--140 GPa pressure range, the MLP's performance might be compromised, particularly at the core-mantle boundary conditions.} We note that our training set includes a significant number of frames that sample various defects and interfaces, thanks to the multithermal-multibaric technique used \cite{Deng2023, Piaggi2019a}. As such, though not the focus of this study,
our MLP can also be useful for studying the phase transition of the (hydrous) $\mathrm{MgSiO_3}$ system.

\subsection{\textit{ab initio} Molecular Dynamics Simulations}

To train the MLP, we perform AIMD simulations on pure and hydrous $\mathrm{MgSiO_3}$, as well as pure H$_2$ and $\mathrm{H_2O}$ based on DFT under periodic boundary conditions. We adopt the PBEsol approximation \cite{Perdew2008a} using VASP \cite{Kresse1996a} with the projector augmented wave (PAW) method \cite{Kresse1999a}. The physical properties of silicates and oxides calculated using the PBEsol approximation are consistent with experimental data \cite{Deng2021a, Scipioni2017a}. The core radii of O (2s$^2$2p$^4$), Si (3s$^2$3p$^2$), Mg (2p$^6$3s$^2$), and H (1s$^1$) are 0.820 Å, 1.312 Å, 1.058 Å, and 0.370 Å, respectively. AIMD simulations are performed in the NVT ensemble with a fixed number of atoms, volume, and temperature, controlled by the Nosé-Hoover thermostat \cite{Hoover1985a}. Our AIMD simulations run for 5--20 ps with a time step of 1 fs. The Mermin functional is used to presume that ions and electrons reach thermal equilibrium \cite{Mermin1965a}. For the initial dataset, we employ a 500-eV energy cutoff for AIMD simulations and sample the Brillouin zone at the $\Gamma$ point. The energy, force, and stress of selected configurations for building the MLP are recalculated at higher precision with the following changes. The energy cutoff is increased from 500 eV to 800 eV, the precision for finding the self-consistent solution to the Kohn-Sham equations is increased from 10$^{-4}$ eV to 10$^{-6}$ eV, and sampling of the Brillouin zone is increased from the $\Gamma$ point only to a $2 \times 2 \times 2$ Monkhorst-Pack mesh. We emphasize that this high-accuracy recalculation is crucial for enhancing the precision and scope of the MLP \cite{Deng2021b}.

\subsection{Molecular Dynamics Simulations Driven by Machine Learning Potentials}

Hydrogen may be incorporated into nominally anhydrous minerals as Si or Mg vacancies through charge-coupled substitution mechanisms. We introduce hydrogen defects by substituting Mg and Si with H while maintaining charge balance. For the MgSiO$_3$ system, we explore three representative hydrogen incorporation mechanisms: (2H)\textsubscript{Mg}, (4H)\textsubscript{Si}, and (Mg + 2H)\textsubscript{Si}, where the subscript indicates the lattice site of the defect, and inside the parentheses are atoms after substitution. These defects are also found in other mantle silicates such as wadsleyite and ringwoodite \cite{Kudoh1999a, Panero2013a}. {{\citeA{Muir2018}} presents evidence that water is preferentially incorporated into bridgmanite as (Al + H)\textsubscript{Si} due to its high configurational entropy. It is worth noting that this defect shares some commonalities with the (Mg + 2H)\textsubscript{Si}: both have a cation carrying a positive charge but less than that of Si, and achieve charge balance by supplementing protons. Given the possibility of (Al + H)\textsubscript{Si} being an energy-favorable defect, we employed \textit{ab initio} calculations combined with the machine learning force field (MLFF) method using VASP to simulate the Al-bearing bridgmanite with (Al + H)\textsubscript{Si} defects. We observed that (Al + H)\textsubscript{Si} and (Mg + 2H)\textsubscript{Si} exhibit remarkably similar hydrogen diffusion patterns, with closely matching hydrogen diffusion coefficients. Our detailed methods for simulating (Al + H)\textsubscript{Si} are presented in the Supporting Information. Owing to the similarity in hydrogen diffusion between (Al + H)\textsubscript{Si} and (Mg + 2H)\textsubscript{Si}, we propose that they can be treated analogously, and the following discussions about (Mg + 2H)\textsubscript{Si} might well apply to (Al + H)\textsubscript{Si} too.}

We consider a range of supercells for $\mathrm{MgSiO_3}$ bridgmanite and post-perovskite, containing 80 ($2 \times 2 \times 1$ for {\it Pbnm} bridgmanite and $4 \times 1 \times 1$ for {\it Cmcm} post-perovskite) to 81,920 ($16 \times 16 \times 16$ for {\it Pbnm} bridgmanite and $32 \times 8 \times 16$ for {\it Cmcm} post-perovskite) atoms. The largest supercell considered enables constructing systems with water content as low as 175 ppm. The distribution of water in the supercell is initially uniform, i.e., all defects are evenly distributed in the supercell, except for the data in Figure~\ref{fig:anisotropy}, where the defect sites are randomly selected. We carefully study the relative stability of different configurations with random defect distributions and their impact on the hydrogen diffusion coefficient, concluding that while defect distribution marginally affects diffusion anisotropy, it has minimal impact on the bulk diffusivity. The computational details are provided in the Supporting Information.

MD simulations are performed using LAMMPS \cite{Plimpton1995a} and DeePMD-kit \cite{Wang2018a} under periodic boundary conditions. For each defect mechanism, we consider three representative temperatures (2000, 3000, and 4000 K) and three pressures (25, 75, and 140 GPa), covering the P/T conditions of the entire lower mantle. Additionally, we conduct simulations along an adiabatic geotherm \cite{Katsura2010a}, ranging from 700 km to 2900 km in depth. Detailed simulation setups can be found in Table S1 and Table S2. Initially, we relax systems under an NPT ensemble for 25 ps at target temperatures and pressures. The resulting configurations are then used for simulations under an NVT ensemble employing the Nosé-Hoover
thermostat \cite{Hoover1985a}, generating 1-ns trajectories from which hydrogen diffusivity is derived. The timestep of all simulations is 0.5 fs.

\subsection{Hydrogen diffusion coefficient and proton conductivity}
Hydrogen diffusivity, $D_\mathrm{H}$, is derived as the slope of mean square displacement (MSD) by,
\begin{linenomath*}
\begin{equation}\label{eq:MSD}
    D_\mathrm{H} = \lim _{t \rightarrow \infty} \frac{\mathrm{MSD}}{6 t}=\lim _{t \rightarrow \infty}
    \frac{\left \langle\left[\vec{r}\left(t+t_0\right)-\vec{r}\left(t_0\right)\right]^2\right \rangle_\mathrm{H}}{6 t}
\end{equation}
\end{linenomath*}
where $\vec{r}(t)$ is the particle trajectories continuous in Cartesian space, and $\langle \cdots \rangle_\mathrm{H}$ represents an average over all hydrogen atoms and over time with different origins \cite{Karki2015a}. Only when MSD is a linear function of time do we intercept the segment of MSD and calculate the diffusion coefficient by linear fitting (Figure~S1). The diffusion coefficients as a function of temperature {obtained from our simulations} are fitted by Arrhenius equation,
\begin{linenomath*}
\begin{equation}\label{eq:Arrhenius}
    D=D_0 \mathrm{e}^{-\frac{\Delta H}{R T}}
\end{equation}
\end{linenomath*}
where $D_0$ is the pre-exponential factor, $R$ is the ideal gas constant, {$T$ is the temperature in Kelvin}, and $\Delta H$ is the activation enthalpy.

The hydrogen diffusion coefficient and the proton conductivity $\sigma_\mathrm{H}$, i.e., the electrical conductivity (unit: S\;m$^-1$) contributed by hydrogen diffusion, are related by the Nernst-Einstein equation \cite{Karato1990a},
\begin{linenomath*}
\begin{equation}\label{eq:NE}
    \sigma_{\mathrm{H}}=\frac{F^2}{R T} c_{\mathrm{H}} z^2 D
\end{equation}
\end{linenomath*}
{where $F$ is the Faraday's constant in C\;mol$^{-1}$; $R$ is the ideal gas constant in J\;mol$^{-1}$\;K$^{-1}$; $T$ is the temperature in Kelvin; $c_\mathrm{H}$ is the molar concentration of hydrogen in mol\;m$^{-3}$; $z$ is the charge of the ion in C; and $D$ is the hydrogen diffusion coefficient in m$^2$\;s$^{-1}$. We use a nominal charge of $z = +1 e$ for proton following previous literature {\cite<e.g.,>{Hae2006a,Sun2015a, Novella2017}}.}

\begin{figure*}[!t]
    \centering
    \noindent\includegraphics[width=\textwidth]{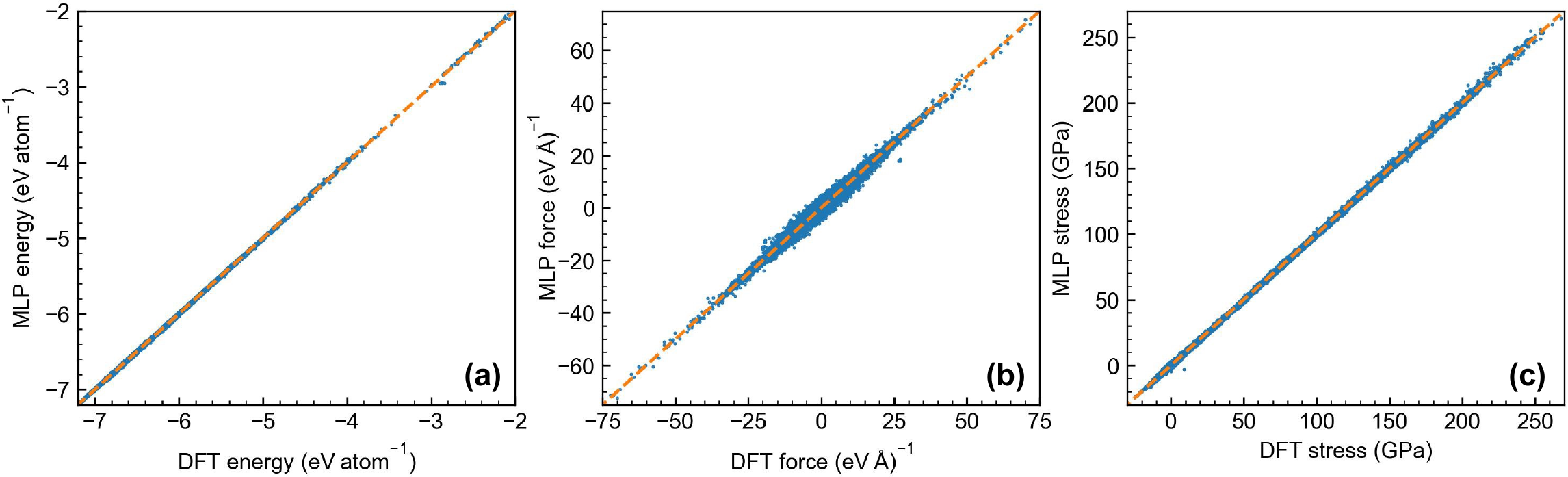}
    \caption{Comparisons of energies (a), atomic forces (b), and stresses (c) between density functional theory (DFT) and the machine learning potential (MLP) for all the test data (50,587 configurations) at temperatures ranging from 1000 to 8000 K and pressures from $\sim$0 GPa to 220 GPa. 50,587 energies, 49,422,324 force components, and 910,566 stress components are included in these comparisons. The {orange} dashed lines are guides for perfect matches.}\label{fig:benchmark}
\end{figure*}

\section{Results}
\subsection{Benchmarks of the machine learning potential}
We compare the energies, atomic forces, and stresses from the MLP to those from DFT simulations for 50,587 configurations that are not included in the training set (Figure~\ref{fig:benchmark}). The root-mean-square errors of the energies, atomic forces, and stresses are 6.39 meV atom$^{-1}$, 0.28 eV Å$^{-1}$, and 0.45 GPa, respectively. These uncertainties are comparable to the typical precision of AIMD simulations \cite{Deng2021a}.

To further validate the reliability of the MLP, we compare the MSD and hydrogen diffusivity obtained from \textit{ab initio} calculations to that from our MLP (Figure~S1). {Figure~S1b shows that the uncertainty in MSD significantly increases over time, since larger values of $t$ in Equation~{\ref{eq:MSD}} mean a smaller dataset for averaging, resulting in larger statistical errors. Therefore, we chose to calculate the diffusion coefficient from the MSD over a short time segment ($t < 30$ ps). For the same initial condition, DFT and MLP methods yield nearly identical MSDs in the selected region. The discrepancy in the DFT and MLP results for $t > 30$ ps can be attributed to the significantly increased statistical error in MSD, and the mean values of MSD calculated from DFT and MLP methods still fall within the error bars of each other.} As shown in Figure~S1c, hydrogen diffusivities derived from the two methods are highly consistent, indicating that our MLP has accuracy comparable to \textit{ab initio} calculations in computing the diffusivity. We emphasize that in this example the simulation using MLP is approximately 5000 times faster than that driven by DFT.

\subsection{{Convergence and uncertainty of hydrogen diffusion coefficient}}

To investigate the impact of system size on diffusivity, we construct a series of supercells of hydrous bridgmanite that contain 81 to 41,472 atoms, incorporating hydrogen via (2H)\textsubscript{Mg} mechanism. The results suggest that hydrogen diffusivities initially increase with system sizes before eventually converging to a constant. The positive correlation between system size and diffusivity is reminiscent of the liquid phase, for which long-range hydrodynamic interactions result in a linear decay with the reciprocal spatial scale of the simulation system characterized by the Oseen tensors \cite{Yeh2004a}. The associated standard deviation (SD) drops significantly as the system expands (Figure~S2). We characterize uncertainties following the approach proposed by \citeA{He2018a},
\begin{linenomath*}
\begin{equation}\label{eq:sd}
    \mathrm{SD}=\bar{D} \times \left( \frac{A}{\sqrt{N_{\mathrm{eff}}}}+B \right)
\end{equation}
\end{linenomath*}
, where $\bar{D}$ and SD are the average value and standard deviation of diffusion coefficients, respectively; $N_{\mathrm{eff}}$ denotes the total number of effective ion hops; $A$ and $B$ are constants. A larger supercell with a given water content contains more hydrogen atoms, leading to a larger number of effective hydrogen hops, and consequently reducing uncertainties of diffusivity as shown in Figure~S2. The diffusion coefficients reported in Table~\ref{tab:1}, Table S1, and Table S2 are based on systems consisting of at least 4000 atoms to guarantee size convergence.

To examine how long of a simulation is sufficient to calculate an adequately accurate diffusion coefficient, we run a long simulation for 10 ns of hydrous bridgmanite ({\hbrg{960}{1024}{3072}{128}}) at 2000 K and 25 GPa.
We take the first 1, 2, 3, 4, 5, and 7 ns segment from this 10 ns trajectory, and then calculate the respective hydrogen diffusion coefficients ({Figure~S3}). The data point at 1 ns is the average result of 10 independent simulations, while the other data points are obtained from single simulations. The results show that as the simulation time increases, the calculated hydrogen diffusion coefficient gradually converges to the results of the 10 ns simulation. However, even with as short as 1-ns simulation, the maximum deviation is within 20\%, and the average value is nearly identical to the 10-ns simulation result. Therefore, we confirm that a 1-ns trajectory is sufficient to obtain an accurate result of the hydrogen diffusion coefficients.

The uncertainties of the calculated diffusivities are evaluated by performing 10 independent simulations with different initial velocity distributions or different defect distributions in the supercell (see Supporting Information). The error bars calculated by the above method are represented by solid lines in the figures. However, due to the limitations of computational resources, it is computationally impractical to perform 10 independent simulations for all data reported in this paper. Therefore, based on our existing results, we estimated the uncertainties of the hydrogen diffusion coefficients for similar systems. The estimated error bars are represented by dashed lines in the figures.

\subsection{Hydrogen diffusion coefficients in bridgmanite and post-perovskite}

Figure~S1a illustrates an example of MSD results for hydrous bridgmanite (\hbrg{15}{16}{48}{2}) at 3000 K and 25 GPa. During the initial few tens of femtoseconds, MSD as a function of time resembles that of ballistic motion, {which means that the motion of particles is predominantly governed by their initial velocities, rather than by interactions with their surroundings or thermal fluctuations.} For the calculation of diffusion coefficients, we exclude this ballistic stage and only consider the portion of MSD which is a linear function of time.

We obtain hydrogen diffusion coefficients for both bridgmanite and post-perovskite with four representative water concentrations ($\sim$175 ppm, $\sim$700 ppm, $\sim$1400 ppm, and $\sim$1.13 wt\%), three hydrogen incorporation mechanisms (i.e., (2H)\textsubscript{Mg}, (4H)\textsubscript{Si}, and (Mg + 2H)\textsubscript{Si}), and various pressures and temperatures (Table~\ref{tab:1}, Table S1, and Table S2).

\begin{figure*}[!t]
    \centering
    \noindent\includegraphics[width=0.9\textwidth]{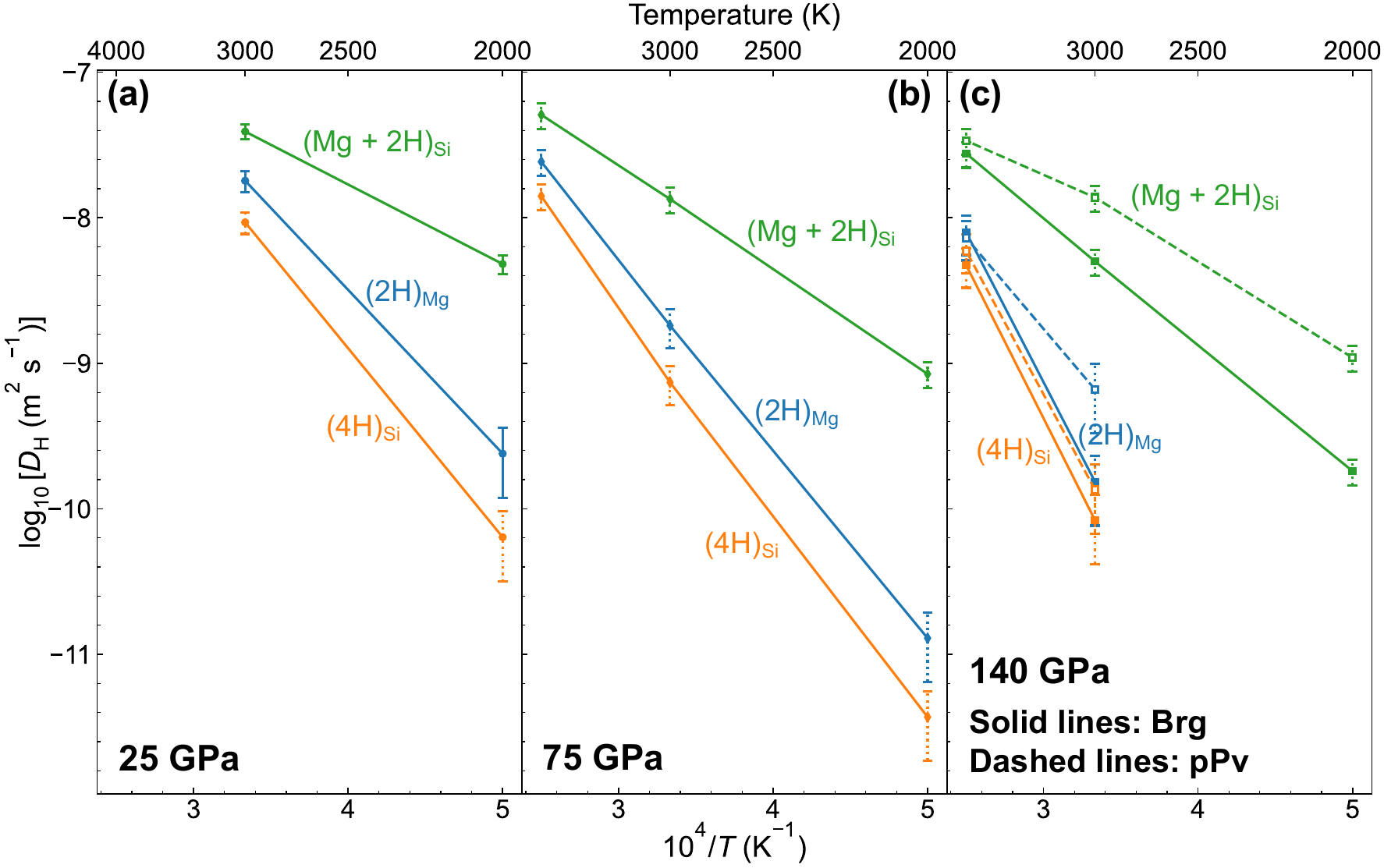}
    \caption{Hydrogen diffusion coefficients in hydrous bridgmanite (Brg, solid lines) and post-perovskite (pPv, dashed lines) as a function of reciprocal temperature (2000 K, 3000 K, and 4000 K) for different hydrogen incorporation mechanisms at 25 GPa (a), 75 GPa (b), and 140 GPa (c). The water content of all systems is 1400 ppm. The (2H)\textsubscript{Mg} defect (\hbrg{8128}{8192}{24576}{128}), the (4H)\textsubscript{Si} defect (\hbrg{8192}{8160}{24576}{128}), and the (Mg + 2H)\textsubscript{Si} defect (\hbrg{8256}{8128}{24576}{128}) are shown in blue, orange, and green, respectively. The solid error bars represent the 2SD of 10 independent simulations. The dashed error bars are estimated based on the error bars of systems with similar diffusion coefficients that have undergone repeated simulations.}\label{fig:summary}
\end{figure*}

As an example, Figure~\ref{fig:summary} summarizes the results for systems that contain 1400 ppm of water. Similar trends are observed for other water concentrations. Calculations for post-perovskite are conducted only at 140 GPa \cite{Murakami2004a}. Bridgmanite melts at 4000 K and 25 GPa, and thus no data is displayed under this P/T condition. At 2000 K and 140 GPa, The slopes of all MSD-time plots are close to zero because no effective ion hop is observed during the total simulation time of 1 ns, and thus corresponding diffusivities cannot be derived.

\begin{table*}[!t]
    \centering
    \caption{Pre-exponential factor ($D_0$) and activation enthalpy ($\Delta H$) of hydrogen diffusion in bridgmanite (Brg) and post-perovskite (pPv) fitted by the Arrhenius equation. The water content ($C_{\rm water}$) is obtained by dividing the mass of available H$_2$O by the total mass of all atoms. There are three runs of temperatures of 2000 K, 3000 K, and 4000 K for each pressure unless the system is molten or no effective ion hop is observed. All diffusion coefficients as well as uncertainties of $\Delta H$ and $D_0$ are provided in Table S1 of the Supporting Information.}\label{tab:1}
    \resizebox{\textwidth}{!}{
    \begin{tabular}{ccccccc}
    \toprule
    Defect      & Pressure (GPa) & Mineral & Chemical composition      & $C_{\rm water}$ (wt\%) & $\Delta H$ (kJ mol$^{-1}$) & $\log_{10} [D_0\; (\si{m^{2}s^{-1}})]$ \\ \midrule
    (2H)\textsubscript{Mg}      & 25             & Brg     & $\rm Mg_{16368}Si_{16384}O_{49152}H_{32}$& 0.0175          & 174.5
& -4.57
\\
    (2H)\textsubscript{Mg}      & 25             & Brg     & $\rm Mg_{8160}Si_{8192}O_{24576}H_{64}$& 0.0702          & 212.5
& -4.02
\\
    (2H)\textsubscript{Mg}      & 25             & Brg     & $\rm Mg_{8128}Si_{8192}O_{24576}H_{128}$& 0.140           & 215.4
& -4.00
\\
    (2H)\textsubscript{Mg}      & 75             & Brg     & $\rm Mg_{8128}Si_{8192}O_{24576}H_{128}$& 0.140           & 250.1
& -4.36
\\
    (2H)\textsubscript{Mg}      & 140            & Brg     & $\rm Mg_{8128}Si_{8192}O_{24576}H_{128}$& 0.140           & 393.5
& -2.96
\\
    (2H)\textsubscript{Mg}      & 25             & Brg     & $\rm Mg_{960}Si_{1024}O_{3072}H_{128}$& 1.14            & 159.8
& -5.19
\\
    (2H)\textsubscript{Mg}      & 75             & Brg     & $\rm Mg_{960}Si_{1024}O_{3072}H_{128}$& 1.14            & 198.8
& -5.25
\\
    (2H)\textsubscript{Mg}      & 140            & Brg     & $\rm Mg_{960}Si_{1024}O_{3072}H_{128}$& 1.14            & 265.4
& -4.65
\\
    (4H)\textsubscript{Si}      & 25             & Brg     & $\rm Mg_{8192}Si_{8160}O_{24576}H_{128}$& 0.140           & 248.5
& -3.70
\\
    (4H)\textsubscript{Si}      & 75             & Brg     & $\rm Mg_{8192}Si_{8160}O_{24576}H_{128}$& 0.140           & 262.2
& -4.48
\\
    (4H)\textsubscript{Si}      & 140            & Brg     & $\rm Mg_{8192}Si_{8160}O_{24576}H_{128}$& 0.140           & 402.4
& -3.07
\\
    (4H)\textsubscript{Si}      & 25             & Brg     & $\rm Mg_{1024}Si_{992}O_{3072}H_{128}$& 1.13            & 163.5
& -5.05
\\
    (4H)\textsubscript{Si}      & 75             & Brg     & $\rm Mg_{1024}Si_{992}O_{3072}H_{128}$& 1.13            & 199.3
& -5.07
\\
    (4H)\textsubscript{Si}      & 140            & Brg     & $\rm Mg_{1024}Si_{992}O_{3072}H_{128}$& 1.13            & 197.5
& -5.40
\\
    (Mg + 2H)\textsubscript{Si}       & 25             & Brg     & $\rm Mg_{8256}Si_{8128}O_{24576}H_{128}$& 0.140           & 104.5
& -5.59
\\
    (Mg + 2H)\textsubscript{Si}       & 75             & Brg     & $\rm Mg_{8256}Si_{8128}O_{24576}H_{128}$& 0.140           & 136.6
& -5.50
\\
    (Mg + 2H)\textsubscript{Si}       & 140            & Brg     & $\rm Mg_{8256}Si_{8128}O_{24576}H_{128}$& 0.140           & 166.8
& -5.39
\\
    (Mg + 2H)\textsubscript{Si}       & 25             & Brg     & $\rm Mg_{1088}Si_{960}O_{3072}H_{128}$& 1.12            & 82.9
& -5.92
\\
    (Mg + 2H)\textsubscript{Si}       & 75             & Brg     & $\rm Mg_{1088}Si_{960}O_{3072}H_{128}$& 1.12            & 104.9
& -5.88
\\
    (Mg + 2H)\textsubscript{Si}       & 140            & Brg     & $\rm Mg_{1088}Si_{960}O_{3072}H_{128}$& 1.12            & 114.8
& -6.01
\\
    (2H)\textsubscript{Mg}      & 140            & pPv     & $\rm Mg_{8128}Si_{8192}O_{24576}H_{128}$& 0.140           & 239.2
& -5.01
\\
    (2H)\textsubscript{Mg}      & 140            & pPv     & $\rm Mg_{960}Si_{1024}O_{3072}H_{128}$& 1.14            & 256.1
& -4.85
\\
    (4H)\textsubscript{Si}      & 140            & pPv     & $\rm Mg_{8192}Si_{8160}O_{24576}H_{128}$& 0.140           & 376.7
& -3.31
\\
    (4H)\textsubscript{Si}      & 140            & pPv     & $\rm Mg_{1024}Si_{992}O_{3072}H_{128}$& 1.13            & 182.9
& -5.61
\\
    (Mg + 2H)\textsubscript{Si}       & 140            & pPv     & $\rm Mg_{8256}Si_{8128}O_{24576}H_{128}$& 0.140           & 115.9
& -5.91
\\
    (Mg + 2H)\textsubscript{Si}       & 140            & pPv     & $\rm Mg_{1088}Si_{960}O_{3072}H_{128}$& 1.12            & 101.6& -6.20\\ \bottomrule
    \end{tabular}
    }
\end{table*}

\section{Discussion}
\subsection{Factors affecting hydrogen diffusivity in bridgmanite and post-perovskite}
\subsubsection{Temperature and Pressure}

Hydrogen diffusivity exhibits a significant increase with temperature, following an Arrhenius-type behavior,  while it decreases with compression (Figure~\ref{fig:summary}). Table~\ref{tab:1} presents pre-exponential factors and activation enthalpies for various defects, pressures, mineral phases, and water contents, calculated based on our hydrogen diffusivity results by fitting into Eq. \ref{eq:Arrhenius}. The opposing effects of pressure and temperature along the geotherm \cite{Katsura2010a} leads to a near-constant hydrogen diffusivity throughout most of the lower mantle (see Table S2).

\begin{figure*}[!t]
    \centering
    \noindent\includegraphics[width=0.95\textwidth]{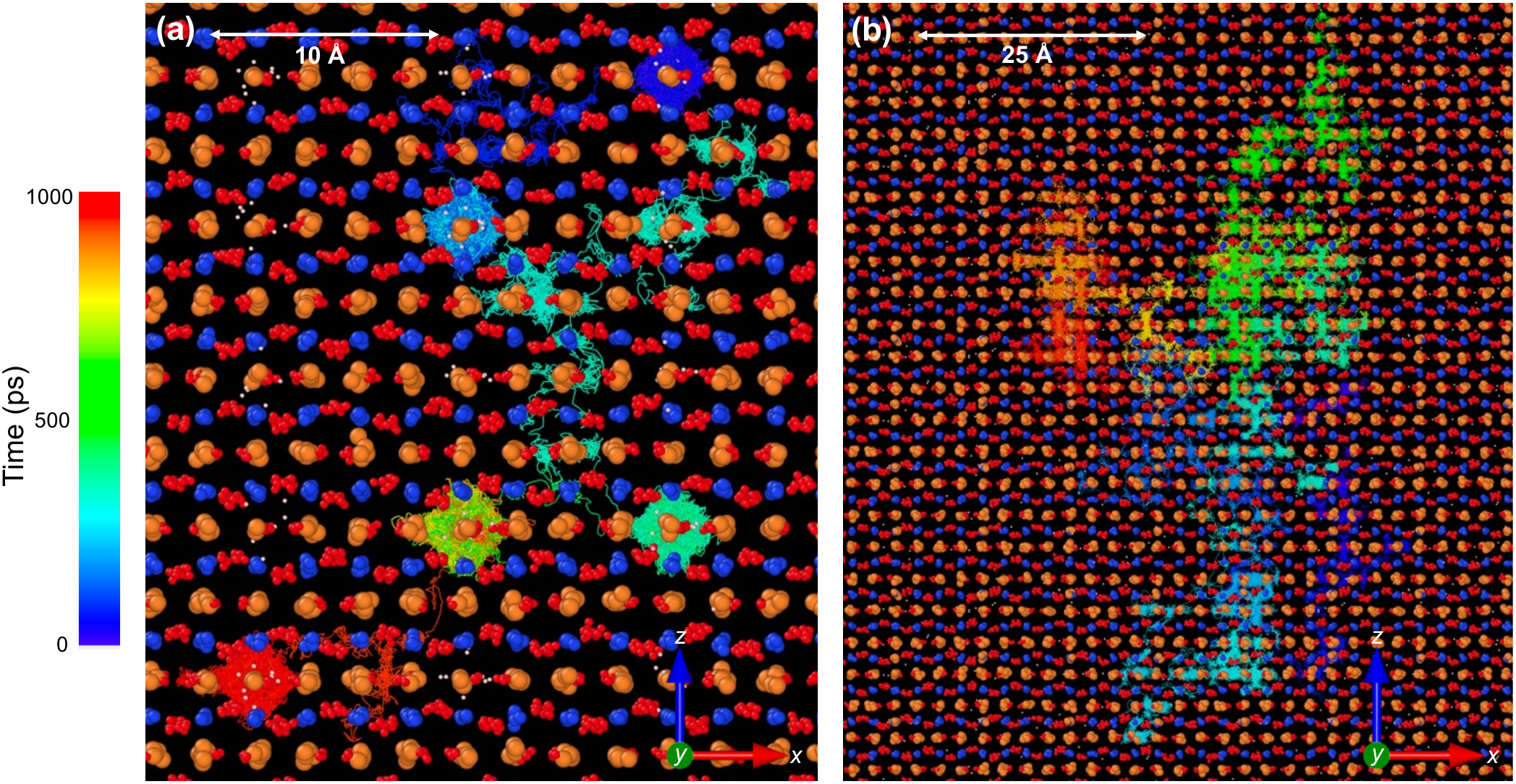}
    \caption{Color mapping of the trajectories of two hydrogen atoms over time in hydrous bridgmanite for (a) the (2H)\textsubscript{Mg} defect ({\hbrg{960}{1024}{3072}{128}}) and (b) the (Mg + 2H)\textsubscript{Si} defect ({\hbrg{1088}{960}{3072}{128}}) at 2000 K and 25 GPa. As the simulation time increases from 0 ps to 1000 ps, the color of the trajectory changes along the visible spectrum. Orange, blue, red, and white spheres represent magnesium, silicon, oxygen, and hydrogen, respectively. The atomic positions along the three crystallographic axes as a function of time are shown in {Figure~S5}.}\label{fig:mapping}
\end{figure*}

\subsubsection{Hydrogen incorporation mechanisms}\label{sec:micro_mech}

Figure~\ref{fig:summary} reveals that the diffusivity of hydrogen is the highest when incorporated via the (Mg + 2H)\textsubscript{Si} defect. {Additionally, the (Al + H)\textsubscript{Si} defect exhibits a hydrogen diffusivity close to (Mg + 2H)\textsubscript{Si}, far surpassing that of the other two defect mechanisms (Figure~S8). To explore the underlying reason, we examine the color mapping trajectories of hydrogen diffusion in hydrous bridgmanite systems with (2H)\textsubscript{Mg}, (Mg + 2H)\textsubscript{Si}, and (Al + H)\textsubscript{Si} defects (Figure~\ref{fig:mapping} and Figure~S7).} We also plot the atomic positions (Figure~S5) along the three crystallographic axes as a function of time corresponding to the trajectories shown in Figure~\ref{fig:mapping}. For the (2H)\textsubscript{Mg} defect ({Figure~\ref{fig:mapping}a}, {Figure~S5a}), the hydrogen atom spends the majority of the time (about 95\%) bound within Mg vacancy sites, while the remaining time is spent jumping and migrating through the interstitial sites. Specifically, the hydrogen atom stays in a Mg vacancy site for 72 ps until it starts a diffusion migration towards another Mg vacancy site. After staying for 228 ps, this hydrogen atom begins its next migration and arrives at another vacancy site at 341 ps. This migration crosses multiple interstitial sites, undergoing several consecutive jumps from one interstitial site to another. After staying for 75 ps, the hydrogen atom moves to the next vacancy site through a rapid diffusion jump that takes only 2 ps, and then jumps to the final vacancy 511 ps later. For the (Mg + 2H)\textsubscript{Si} defect (Figure~\ref{fig:mapping}b, Figure~S5b, Figure~S7b) {and the (Al + H)\textsubscript{Si} defect (Figure~S7a)}, however, the hydrogen atoms quickly jump through the interstitial sites and do not exhibit prolonged stays at defect sites, showing significantly higher mobility. In this case, the hydrogen diffusivity in bridgmanite exhibits a profound similarity to that in the one-dimensional water channels of stishovite-water superstructures \cite{Li2023}. Furthermore, the elevated hydrogen diffusivity may lead to an increase in configurational entropy, a reduction in defect formation free energy, and thus an enhanced stability of (Mg + 2H)\textsubscript{Si} and (Al + H)\textsubscript{Si} defects in water-bearing bridgmanite \cite{Li2023, Muir2018}.

Generally, hydrogen atoms predominantly occupy vacancy or interstitial sites between diffusion jumps. Diffusion occurs through rapid jumps between vacancy or interstitial sites (i.e., effective ion hops), with jump lengths reaching up to several angstroms. The microscopic processes of hydrogen diffusion in bridgmanite and post-perovskite resemble those in wadsleyite and ringwoodite {\cite{Caracas2017a}}.

In fact, this microscopic mechanism is closely related to the activation enthalpy in the Arrhenius equation Eq.~\ref{eq:Arrhenius}. The activation enthalpy is the energy difference between the system when hydrogen occupies the most energetically favorable position and when hydrogen passes the saddle point connecting the two energy minima \cite{Ingrin2006a}. Activation enthalpies of (Mg + 2H)\textsubscript{Si} are notably lower than those of (2H)\textsubscript{Mg} and (4H)\textsubscript{Si}, for both bridgmanite and post-perovskite (Table~\ref{tab:1} and Figure~S4). {Given that the ionic radius of Mg$^{2+}$ ($\sim$86 pm) and Al$^{3+}$ ($\sim$67.5 pm) is larger than that of Si$^{4+}$ ($\sim$54 pm) \cite{Slater1964a}, Mg and Al can effectively take up most of the space within a Si vacancy, making it difficult for hydrogen atoms to fit in. As such, (Mg + 2H)\textsubscript{Si} and (Al + H)\textsubscript{Si} have relatively higher potential energies, and lower activation enthalpies, leading to the fastest hydrogen diffusion across all considered temperatures, pressures, and mineral phases.}

\begin{figure*}[!t]
    \centering
    \noindent\includegraphics[width=\textwidth]{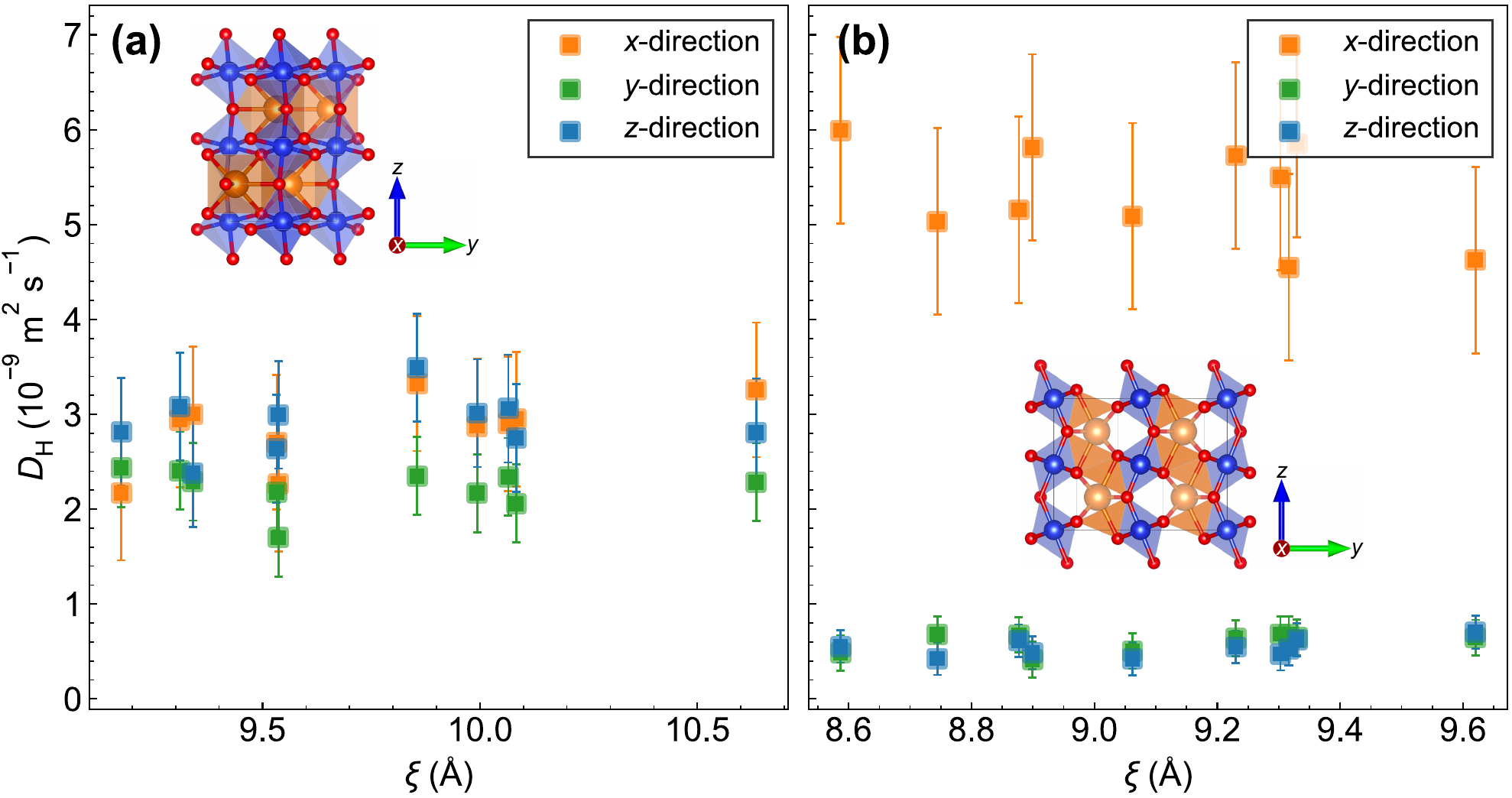}
    \caption{The anisotropy of hydrogen diffusion in hydrous bridgmanite ({\hbrg{8128}{8192}{24576}{128}}) and post-perovskite ({\hbrg{8128}{8192}{24576}{128}}) shown by plotting hydrogen diffusion coefficients along three different crystallographic axes, $x$ (orange), $y$ (green), and $z$ (blue) of 10 random defect distributions as a function of average inter-defect distance of nearest neighbors ($\xi$), defined in the Supporting Information. Simulations are performed at 4000 K and 140 GPa. Orange, blue, and red spheres represent magnesium, silicon, and oxygen, respectively. The error bars represent 2SD of data along each direction.}\label{fig:anisotropy}
\end{figure*}

\subsubsection{Bridgmanite vs. post-perovskite}\label{sec:anisotropy}
Bridgmanite and post-perovskite exhibit significantly different characteristics of hydrogen diffusion. Hydrogen diffuses overall faster in post-perovskite than in bridgmanite, particularly at low temperatures (Figure~\ref{fig:summary}c). Furthermore, these two minerals show completely different hydrogen diffusivities along their crystallographic axes ({Figure~\ref{fig:anisotropy}}). In bridgmanite, for initial configurations with various defect distributions, the error bars of diffusion coefficients along the three directions overlap with each other, exhibiting nearly isotropic hydrogen diffusion. Changes in the defect distribution will affect the order of diffusivity magnitudes along different directions. In post-perovskite, hydrogen atoms prefer diffusing along {[100]} ($x$-direction), showing a significant anisotropy. {The diffusion coefficient along the [100] direction is nearly an order of magnitude faster than along the [010] and [001] directions.} This pattern of anisotropy remains regardless of defect incorporation mechanisms, defect distributions in the supercell, and defect densities considered. 

Similar to the anisotropic diffusion of protons in rutile \cite{Farver2010a} and stishovite-water superstructures \cite{Li2023}, the large open channels (see {Figure~\ref{fig:anisotropy}b}) along the {[100] direction in the post-perovskite structure likely provide ample space for hydrogen diffusion pathways, resulting in more efficient hydrogen diffusion along this direction. Different from {\citeA{Li2023}}, the open channels in post-perovskite are an inherent feature of its crystallography, occurring between the edge-sharing SiO$_6$ octahedra, rather than being formed due to the incorporation of hydrogen defects. The structural features of post-perovskite result in anisotropic hydrogen diffusion, facilitated by large open channels that enhance diffusivity. Consequently, post-perovskite yields a higher hydrogen diffusivity than bridgmanite, despite its higher density.}

\subsubsection{Water concentration}\label{sec:cwater}

\begin{figure}[!ht]
    \centering
    \noindent\includegraphics[width=0.6\textwidth]{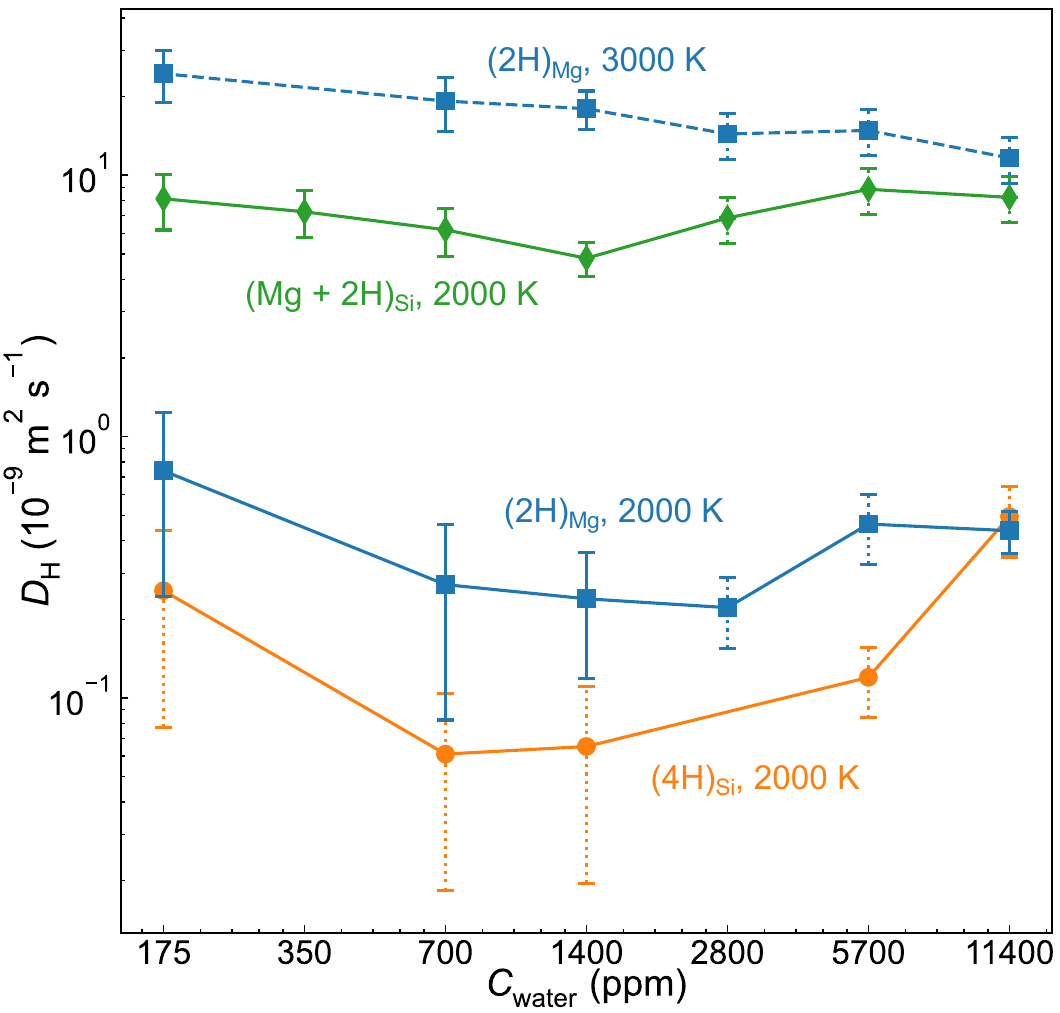}
    \caption{Hydrogen diffusion coefficients in hydrous bridgmanite (chemical compositions are shown in Table~\ref{tab:1}) as a function of water concentration at 2000 K (solid lines) and 3000 K (dashed line). The pressure is 25 GPa. (2H)\textsubscript{Mg}, (4H)\textsubscript{Si}, and (Mg + 2H)\textsubscript{Si} defects are shown in blue, orange, and green, respectively. {The solid error bars represent the 2SD of 10 independent simulations. The dashed error bars are estimated based on the error bars of systems with similar diffusion coefficients that have undergone repeated simulations.}}\label{fig:water}
\end{figure}

Water concentration also affects the hydrogen diffusivity in the host mineral {\cite<e.g.,>{Caracas2017a}}. We find that the effects of water content are moderate and nonlinear ({Figure~\ref{fig:water}}). At 3000 K, the hydrogen diffusivity decreases with increasing water content; while at 2000 K, simulations of all three defect mechanisms show a trend where the diffusivity first increases and then decreases with increasing water content.

To understand the underlying reason for this phenomenon, we compare the microscopic mechanisms of hydrogen diffusion in hydrous bridgmanite systems with the same (2H)\textsubscript{Mg} defect but different water contents.
{Figure~\ref{fig:mapping}a} and Figure~S6a present the hydrogen trajectories of {\hbrg{960}{1024}{3072}{128}} ($C_{\rm water} = 1.14$ wt\%) and {\hbrg{16368}{16384}{49152}{32}} ($C_{\rm water} = 0.0175$ wt\%), respectively. We find that in the simulation with $1.14$ wt\% water, most ($\sim$80\%) of hydrogen atoms hop at least twice, passing through multiple vacancy sites, and spend the majority of the simulation time ($\sim$95\%) residing within these vacancy sites. In contrast, for simulations with $0.0175$ wt\% water, only 3 out of the 32 hydrogen atoms successfully migrate to neighboring sites. The other 29 atoms either remain at the initial sites throughout the 1-ns simulation or attempt to migrate but return to their initial positions quickly. Additionally, hydrogen atoms with successful diffusion migration often reside in interstitial sites for a significant amount of time (up to $\sim$500 ps) (Figure~S6a).

{We believe that water content primarily affects the diffusion coefficient by changing the distance between neighboring vacancy sites. Let us consider two systems with different water contents while the attempt frequency of hydrogen diffusion migration is the same. In the water-rich system, due to the short distance between vacancies, these attempts are more likely to succeed in transporting the hydrogen atom to a new site, achieving net diffusion. This is the reason why we observe more hydrogen atom migrations in Figure~\ref{fig:mapping}a than Figure~S6a. In contrast, in the water-poor system, the hydrogen atom struggles to find a stable vacancy site to settle in and frequently returns to the initial defect site under the attraction of a local negative charge. Since the Coulombic force diminishes rapidly with distance ($\propto r^{-2}$), once hydrogen reaches a distance far enough to escape the attraction, it lingers in the interstitial site for a long time and diffuses rapidly (the interstitial site has high potential energy) until it reaches the next low-energy vacancy site (Figure~S6a).}

In high-temperature scenarios, the situation is straightforward: since the hydrogen atom is very light, it has high velocity, easily escapes the Coulombic attraction, and diffuses quickly in interstitial sites. An increase in water content (or a decrease in the average distance between vacancies) results in hydrogen being more constrained within the vacancy sites, leading to more pauses during diffusion ({Figure~\ref{fig:mapping}a}). Consequently, the diffusion coefficient decreases monotonically with increasing water content.

In the case of low temperatures, the situation becomes more intricate. The velocity of hydrogen is lower, making it harder to overcome Coulombic attraction and achieve a successful migration attempt. In high water content scenarios ($>$1400 ppm), nearby vacancy sites do not act as obstacles to diffusion. Instead, a nearby attractive vacancy can transform a failed migration attempt into a successful one. This is consistent with previous studies on wadsleyite and ringwoodite with high water content, where increasing the defect density may shorten the hopping distance of hydrogen and consequently enhance its diffusion {\cite{Caracas2017a}}. However, when the water content decreases to a certain threshold ($\sim$1400 ppm), the nearest defects are too distant to have any significant influence on hydrogen during a migration attempt. The facilitative effect of vacancies disappears, and the only remaining influence is their inhibitory effect on the rapid diffusion in interstitial sites, degenerating to the situation at high temperatures. As a result, the diffusion coefficient of hydrogen first increases and then decreases with increasing water content, reaching a minimum of around 1400 ppm.

In addition, higher water concentration may cause additional hydrogen to occupy interstitial sites, obstructing some diffusion pathways, and decelerating hydrogen diffusion, similar to the saturation effect in metal-hydrogen systems {\cite<e.g.,>{Shelyapina2022a}}. With respect to the (Mg + 2H)\textsubscript{Si} defect, since hydrogen atoms mainly reside in interstitial sites, this can also be a potential factor affecting hydrogen diffusivity.

In summary, hydrogen diffusion as a function of water concentration in mantle minerals is controlled by many competing factors and thus exhibits a complicated dependence on water contents. Nevertheless, for all conditions considered in this study, the variance of diffusion coefficients due to varying water concentrations does not exceed one order of magnitude (Figure~\ref{fig:water}). Overall, the water content does not have a significant impact on the hydrogen diffusion coefficient. Based on this finding, we conclude that our hydrogen diffusivity results are applicable to the lower mantle given the similar plausible range of water content \cite{Ohtani2021a}.

\begin{figure}[!ht]
    \centering
    \noindent\includegraphics[width=0.6\textwidth]{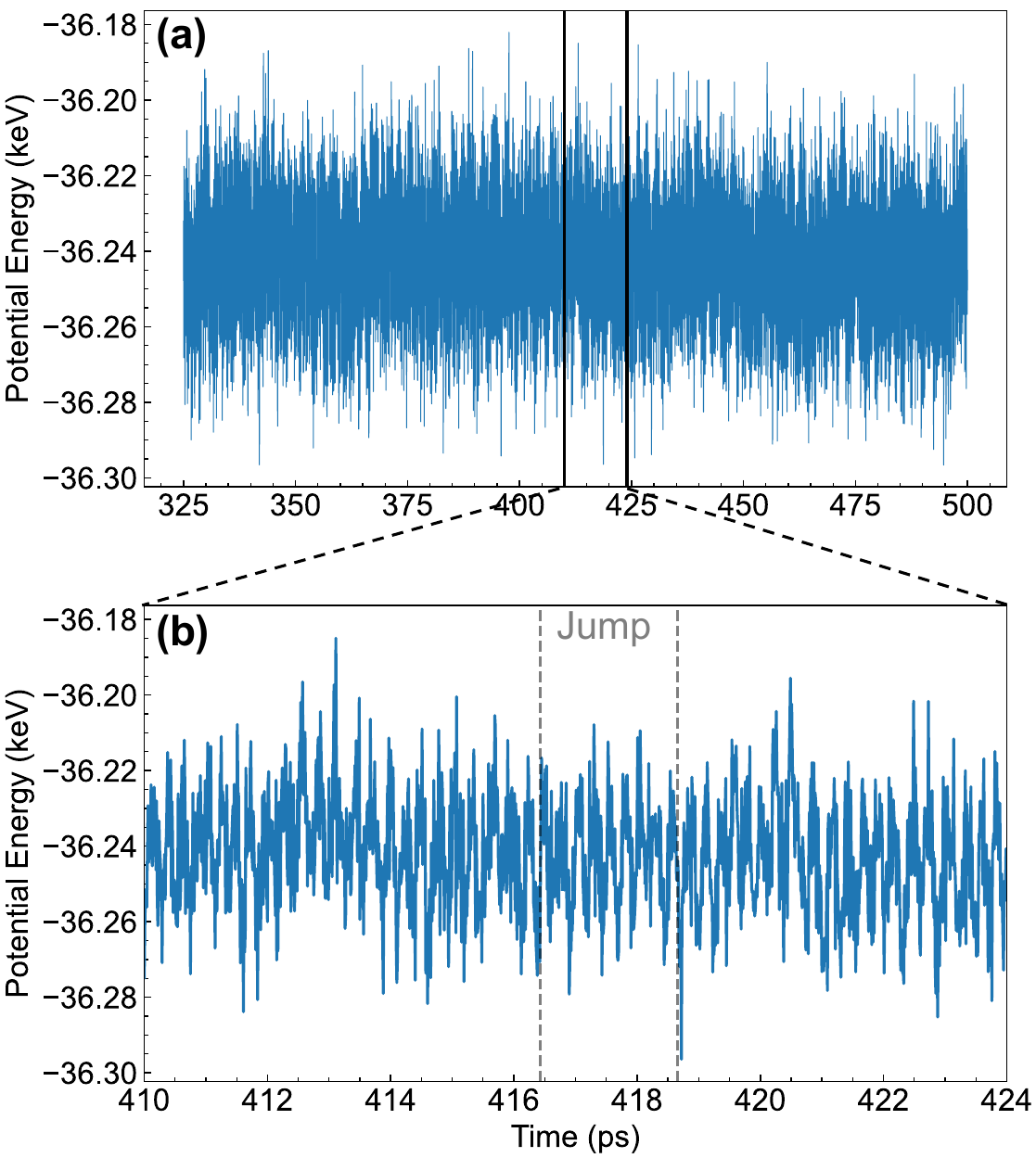}
    \caption{Potential energy fluctuations of hydrous bridgmanite (\hbrg{960}{1024}{3072}{128}) at 2500 K and 25 GPa during the MD simulation, as described in Section~\ref{sec:micro_mech}, Figure~\ref{fig:mapping}, and Figure~S5a.}\label{fig:fluctuation}
\end{figure}

\subsection{Energy fluctuation}

We also monitor potential energy fluctuations upon hydrogen diffusion. The total potential energy of the system does not exhibit a clear local energy maximum or a saddle point during the diffusion jumps of hydrogen (Figure~\ref{fig:fluctuation}) described in Section~\ref{sec:micro_mech}, Figure~\ref{fig:mapping}, and Figure~S5, in contrast to the classical transition state theory \cite{Henkelman2000a}. The absence of an energy saddle point may be explained by the low concentration of hydrogen (thousands of ppm) considered in this study, which dilutes the effect of diffusion on the total potential energy of the entire system. Another possibility is that other atoms within the system also adjust their positions when diffusion occurs. This behavior highlights the intricate interplay between the positions of all atoms in the lattice and the total energy of the system, rather than just the positions of diffusing atoms. This result is consistent with the energy fluctuation observed during hydrogen diffusion in wadsleyite and ringwoodite \cite{Caracas2017a}.

\subsection{{Self- and chemical diffusion of hydrogen in mantle silicates}}\label{sec:chemical}

{Self-diffusion is the process where chemically identical atoms move within a homogeneous material, while chemical diffusion involves the movement of atoms from areas of high concentration to low concentration. The values of self-diffusion and chemical diffusion coefficients are generally not the same.} The above discussions focus on the self-diffusion of hydrogen since the simulated system does not exhibit a concentration gradient of both hydrogen and defects. As the transport of water in the Earth's interior often occurs with a concentration gradient, the chemical diffusion of hydrogen in mantle silicates becomes crucial.

Different mechanisms have been proposed for the chemical diffusion of hydrogen in mantle silicates. First, protons may diffuse in interstitial sites and the proton flux is coupled with a counter flux of polarons, facilitated by the mobile excess charge on ferric iron (\ce{Fe^2+ <=> Fe^3+ + e^-}) \cite<e.g.,>{Mackwell1990, Kohlstedt1998, Hae2006a}. {Here, polarons are quasiparticles formed by an excess charge carrier (electron or hole) localized within a potential well, self-generated by displacing the surrounding ions {\cite{franchini_polarons_2021}}.} Since the diffusivity of polarons significantly surpasses that of protons \cite{Schmalzried1974}, the chemical diffusivity $\widetilde{D}_{\mathrm{H}}$ can be written as \cite{Kohlstedt1998, Hae2006a}.
\begin{linenomath*}
\begin{equation}\label{eq:chemical}
    \widetilde{D}_{\mathrm{H}}=\frac{2 D_{\mathrm{P}} D_{\mathrm{H}}}{D_{\mathrm{P}}+D_{\mathrm{H}}} \approx 2 D_{\mathrm{H}}
\end{equation}
\end{linenomath*}
where $D_{\mathrm{H}}$ and $D_{\mathrm{P}}$ are the self-diffusion coefficients of hydrogen and polaron, respectively. Thus, the chemical diffusivity of hydrogen can be directly estimated using the self-diffusivity obtained in this study. Second, hydrogen diffusion can be coupled with the diffusion of intrinsic point defects such as Mg and Si vacancies \cite<e.g.,>{Kohlstedt1998, Demouchy2003}. In this case, the chemical diffusion of hydrogen is much slower (two orders of magnitude slower than the first mechanism \cite{Kohlstedt1998}), and the self-diffusivity of hydrogen may not effectively constrain its chemical diffusivity. Third, activated hydrogen atoms may dissociate from a thermal vacancy and migrate as an interstitial particle until it is trapped again at another thermal vacancy, known as the dissociative mechanism \cite{Novella2017, Frank1956}. This closely resembles the self-diffusion of hydrogen in our simulations (see Section~\ref{sec:micro_mech} and Section~\ref{sec:cwater}). This dissociative mechanism can well explain why the self-/chemical diffusivity of hydrogen \cite<e.g.,>{Novella2017, Hae2006a} is close to the diffusivity of small polarons \cite{Novella2017}, but much higher than those of Mg and Si in mantle silicates \cite<e.g.,>{Xu2011}.

Furthermore, we note that Al might be extensively present in the lower mantle \cite<e.g.,>{Irifune1994} and thus obtain the diffusion coefficient of hydrogen in Al-bearing bridgmanite at 2000 K and 25 GPa (see Supporting Information). Intriguingly, this diffusion is independent of any vacancy movement, as the incorporation of hydrogen does not rely on the presence of such vacancies. Given that Al\textsuperscript{3+} bears one less positive charge than Si\textsuperscript{4+}, and the bridgmanite crystal remains electrically neutral, the non-protonated Al\textsuperscript{3+} ions might be associated with Fe\textsuperscript{3+} \cite{Mccammon1997}. This suggests that the chemical diffusion of hydrogen in Al-bearing bridgmanite could also potentially involve the flux of excess charge on ferric iron, showing a similar pattern with the self-diffusion.

\subsection{The water distribution in the lower mantle}\label{sec:distribution}

\begin{figure}[!ht]
    \centering
    \noindent\includegraphics[width=0.6\textwidth]{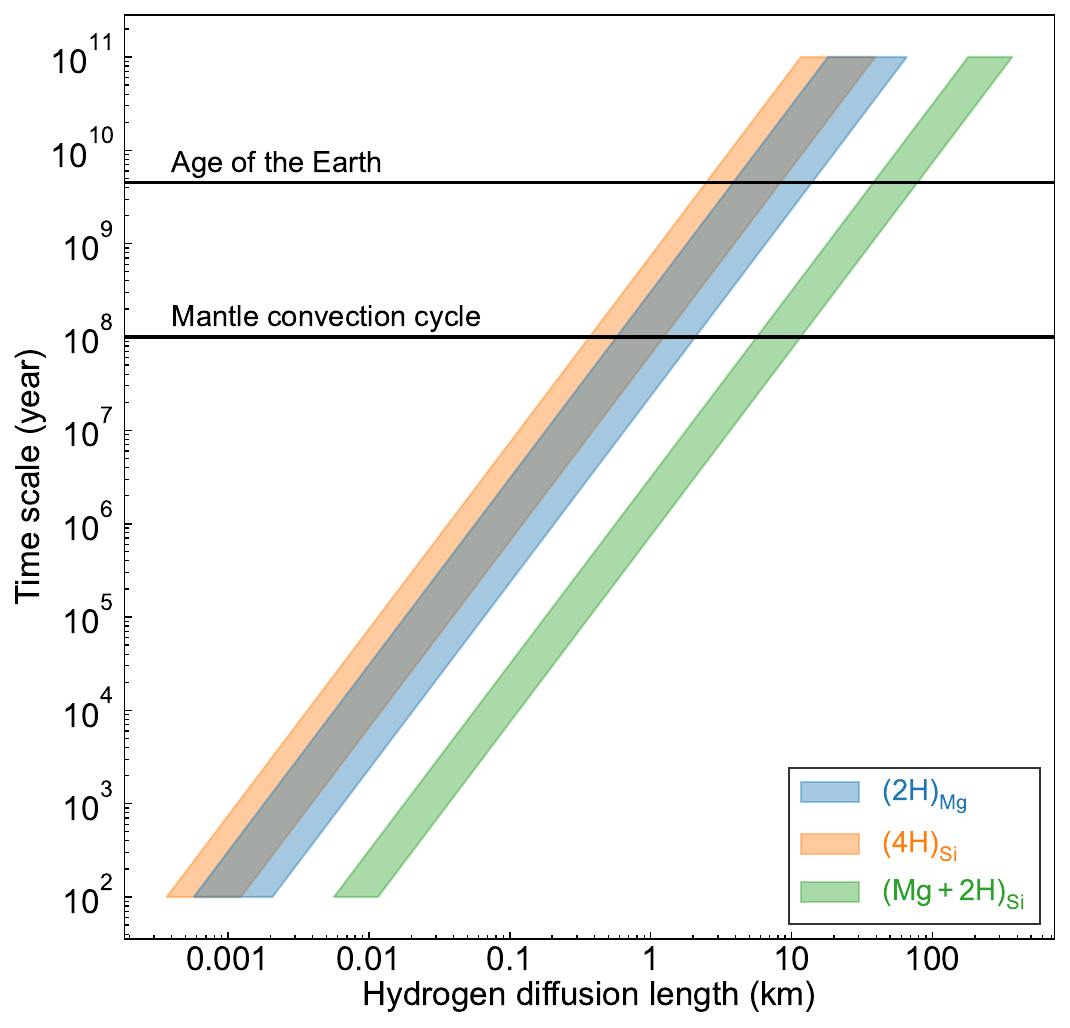}
    \caption{Hydrogen diffusion time scales ($t$) as a function of hydrogen diffusion length scales ($L_d=\sqrt{2 \widetilde{D}_{\mathrm{H}} t}$). The left and right boundaries of the shaded area correspond to the minimum and maximum values obtained from simulations along the mantle geotherm \cite{Katsura2010a} considering various hydrogen incorporation mechanisms, and water contents (Table S2). The estimated mantle convection cycle \cite<100 Ma,>{Holzapfel2005a} and the age of the Earth (4.5 Ga) are also shown. (2H)\textsubscript{Mg}, (4H)\textsubscript{Si}, and (Mg + 2H)\textsubscript{Si} defects are shown in blue, orange, and green, respectively.}\label{fig:length}
\end{figure}

Hydrous minerals in the subducted oceanic crust may transport a substantial amount of water into the Earth's deep interior, reaching as deep down as the core-mantle boundary \cite<e.g.,>{Ohtani2015a}. {In addition, the low D/H ratio in Baffin Island and Icelandic lavas suggest the existence of primordial hydrogen reservoirs in the lower mantle that inherited the D/H ratio directly from the protosolar nebula {\cite{Hallis2015a}}.} Furthermore, previous experimental and observational studies suggest a heterogeneous water distribution in the mantle transition zone \cite{Hae2006a, Sun2015a, Zhou2022a}.

Bridgmanite is the most abundant mineral phase of the lower mantle. {{\citeA{hernandez_incorporation_2013}} found that the partition coefficient of water between ferropericlase and bridgmanite is around 90 at 24 GPa using {\textit{ab initio}} methods. This suggests that the water content in bridgmanite is about two orders of magnitude higher than in coexisting ferropericlase, the second most abundant lower mantle mineral, aligning with average estimates of water solubility in them. {\cite{lu_solubility_2023}}. Furthermore, {\citeA{Muir2018}} calculated water distribution among various lower mantle mineral phases, revealing that for a few hundred ppm of water in the lower mantle, most of the water is partitioned into Al-bearing bridgmanite.} Therefore, the main medium for water transport in the lower mantle may be bridgmanite. To evaluate the effect of hydrogen diffusivity on the distribution of water in the lower mantle, we calculate hydrogen diffusion length scales for bridgmanite and post-perovskite, based on hydrogen diffusion coefficients derived along the mantle geotherm \cite{Katsura2010a} (Figure~\ref{fig:length}). {The diffusion length scale is given by $L_d=\sqrt{2 \widetilde{D}_{\mathrm{H}} t}$, where $\widetilde{D}_{\mathrm{H}}$ is the estimated chemical diffusivity of hydrogen ({Eq.~\ref{eq:chemical}}) and $t$ represents the time scale, considering the first chemical diffusion mechanism as discussed in {Section~\ref{sec:chemical}}.} With an assumed mantle convection velocity of 5 cm per year, the estimated time scale for one mantle convection cycle is $\sim$100 Ma \cite{Holzapfel2005a}. The corresponding hydrogen diffusion length scales in bridgmanite and post-perovskite are less than 12 km, regardless of depths, water content, and hydrogen incorporation mechanisms {(assuming  (Mg + 2H){\textsubscript{Si}} and (Al + H){\textsubscript{Si}} are analogs regarding hydrogen diffusion). Considering that a portion of water may undergo much slower chemical diffusion coupled with cation vacancies ({Section~\ref{sec:chemical}}), this diffusion length would be even shorter.} Therefore, water brought into the lower mantle by the subducting slabs would be locally confined, leading to a heterogeneous distribution of water throughout the Earth's lower mantle. However, other important factors that might influence hydrogen diffusion behavior in bridgmanite, such as grain boundary diffusion and compositional effects (e.g., iron content), need further study to corroborate this conclusion.

\subsection{The electrical conductivity in the lower mantle}

The electrical conductivity of mantle silicates is mainly contributed by four conduction mechanisms: ionic conduction, hopping (small polaron) conduction, large polaron conduction, and proton conduction \cite{Yoshino2010a}. Due to its low bonding energy and small ionic radius, the energy barrier of the transiting proton is relatively low, causing proton conductivity to potentially dominate the bulk conductivity under wet conditions \cite{Sun2015a,Yoshino2010a}.

\begin{figure*}[!t]
    \centering
    \noindent\includegraphics[width=0.95\textwidth]{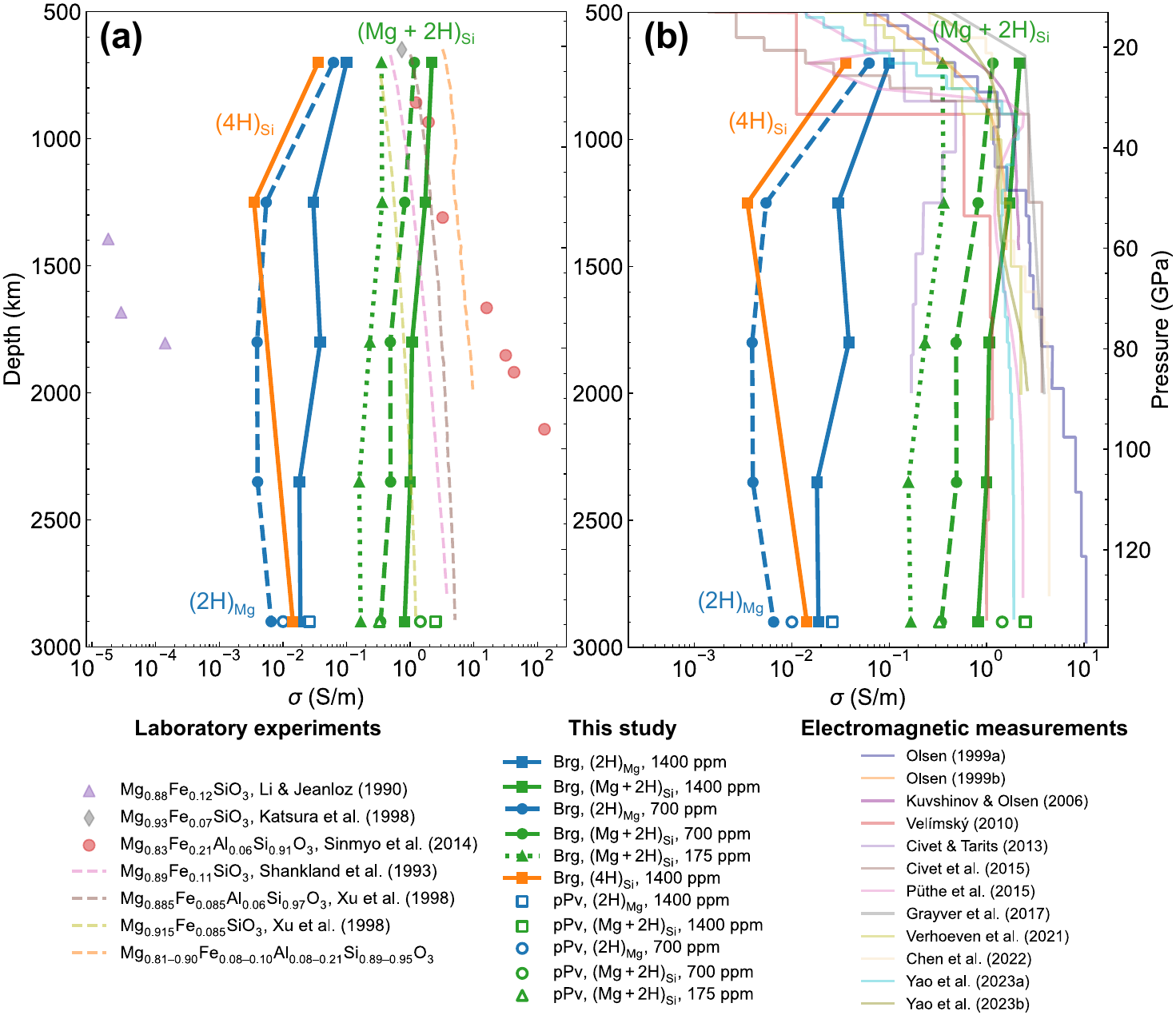}
    \caption{Proton conductivities of bridgmanite (solid symbols) and post-perovskite (empty symbols) compared to experimentally measured electrical conductivities of bridgmanite (a) and bulk conductivities of the lower mantle measured with electromagnetic induction (b). All simulations are performed along an adiabatic geotherm \cite{Katsura2010a}, considering various water contents and hydrogen incorporation mechanisms (1,400 ppm: solid thick lines; 700 ppm: dashed thick lines; 175 ppm: dotted thick lines; (2H)\textsubscript{Mg}: blue; (4H)\textsubscript{Si}: orange; (Mg + 2H)\textsubscript{Si}: green). Some data points for (4H)\textsubscript{Si} defects are not shown since no effective ion hop is observed during the entire simulation time of up to $\sim$ns. Semi-transparent data points denote the experimentally measured conductivity of dry bridgmanite. Semi-transparent dashed lines represent the lower mantle conductivity models based on experimental results. Semi-transparent solid lines show the 1D electrical conductivity profile determined from electromagnetic (EM) induction studies.}\label{fig:EC}
\end{figure*}

In this study, we employ the Nernst-Einstein equation (Eq.~\ref{eq:NE}) to calculate the proton conductivity in bridgmanite and post-perovskite, based on hydrogen diffusivities obtained from MD simulations along the geotherm \cite{Katsura2010a} (Table S2). Proton conductivity results for various systems are plotted as a function of depth, along with experimental results (Figure~\ref{fig:EC}a) and observations (Figure~\ref{fig:EC}b) of the bulk conductivity of the lower mantle. The proton conductivity of bridgmanite is sensitive to the water content and hydrogen incorporation mechanisms. If water content remains constant and the hydrogen incorporation mechanism is consistent throughout the lower mantle, the proton conductivity would show relatively low depth sensitivity due to the near-constant hydrogen diffusivity. The diffusivity of hydrogen is the largest when incorporated as (Mg + 2H)\textsubscript{Si} defects {(or the diffusionally analogous (Al + H)\textsubscript{Si} defects)}, resulting in the highest proton conductivity of up to 1 S m$^{-1}$ (Table S2 and Figure~\ref{fig:EC}). Under the lowermost mantle conditions, the proton conductivity in post-perovskite is higher than that in bridgmanite, with the difference within one order of magnitude.

Figure~\ref{fig:EC}a presents experimentally measured conductivity results of dry bridgmanite with various Fe and Al contents \cite{Katsura1998a,Li1990a,Shankland1993a,Sinmyo2014a,Xu1998a}, except for one study that used bridgmanite samples containing 0--92 ppm water \cite{Yoshino2016a}. These experimental results, although highly scattered, suggest that the conductivity of bridgmanite is sensitive to chemical compositions, e.g., Al and Fe contents. The relatively high conductivity results measured by \citeA{Yoshino2016a}, compared to other studies, may be caused by the presence of water in their samples. We find that the contribution of proton for Earth-relevant water contents is comparable to the sum of all other three conduction mechanisms in bridgmanite (i.e., the conductivity of dry bridgmanite) {when hydrogen is incorporated as (Mg + 2H)\textsubscript{Si} or (Al + H)\textsubscript{Si} defects}. The 1D conductivity-depth profiles in the lower mantle, as determined from electromagnetic induction studies, are shown in Figure~\ref{fig:EC}b \cite{Civet2013a,Civet2015a,Grayver2017a,Kuvshinov2006a,Olsen1999a,Olsen1999b,Puthe2015a,Velimsky2010a,Verhoeven2021a}. Most of these conductivity data fall within two orders of magnitude (0.1--10 S m$^{-1}$) and generally remain nearly constant or slightly increase with increasing depth. {Considering (Mg + 2H)\textsubscript{Si} or (Al + H)\textsubscript{Si} defects, our results of proton conductivity are within the same magnitude as the measured conductivity, supporting the potential role of water in affecting the electrical conductivity of the lower mantle.} 

Previous studies reported strong lateral heterogeneities of the electrical conductivity in the mid-lower mantle regions \cite<e.g.,>{Khan2011,Tarits2010a}. Since the proton conductivity in bridgmanite is sensitive to both temperature and water content, the possible thermal anomalies \cite{Hsieh2020a} as well as the potential heterogeneous water distribution discussed in Section~\ref{sec:distribution} may lead to these heterogeneities of the lower mantle conductivity. {If we consider bridgmanite with relatively high but still reasonable water contents, for example, around 700 ppm (assuming it primarily exists in the form of (Mg + 2H)\textsubscript{Si} or (Al + H)\textsubscript{Si}), the proton conductivity is $\sim$1 S\;m\textsuperscript{-1} at 900 km depth, which aligns perfectly with the heterogeneity in the observed mantle conductivity at a depth of 900 km \cite<1.4--2.4\;S\;m$^{-1}$,>{Khan2011}. Therefore, water may act as a significant charge carrier in the hydrated regions of the lower mantle, and contribute to the heterogeneity of electrical conductivity. Further, by analyzing electrical conductivity variations, we can potentially deduce water distribution in the lower mantle.} It is worth noting that the strong anisotropic hydrogen diffusion in post-perovskite discussed in Section~\ref{sec:anisotropy} may result in anisotropic electrical conductivity, similar to the case with olivine in the upper mantle \cite{Bahr2002a}.

We acknowledge that real-world scenarios involve a multitude of factors influencing hydrogen diffusivity and electrical conductivity. However, considering this is the first study on the diffusion of hydrogen in lower mantle minerals, we position our results as an initial step and believe our current results provide a foundational step towards creating more comprehensive models that can capture the multifaceted nature of diffusion processes in the Earth's mantle.

\section{Conclusions}

We have developed a unified machine learning potential of \textit{ab initio} quality, which, for the first time, enables a comprehensive examination of hydrogen diffusion in bridgmanite and post-perovskite under the Earth's lower mantle conditions. Our systematic analysis of factors influencing hydrogen diffusivity demonstrates that hydrogen diffusion is highly sensitive to defect mechanism, temperature, and pressure, while relatively insensitive to water content. The trade-off between pressure and temperature leads to a nearly constant hydrogen diffusivity in the lower mantle along the geotherm. Among the {four} defect mechanisms examined, (Mg + 2H)\textsubscript{Si} and (Al + H)\textsubscript{Si} show similar patterns and yield the highest hydrogen diffusivity, which may result from the low activation enthalpy. Hydrogen diffuses overall faster in post-perovskite than in bridgmanite, especially at relatively low temperatures. In addition, these two minerals exhibit distinct anisotropies of hydrogen diffusion. Based on our diffusivity results, we conclude that hydrogen diffusion is sluggish on the geological time scale, suggesting that the water distribution in the lower mantle may be localized and highly heterogeneous. We also calculate the electrical conductivities contributed by hydrogen diffusion in bridgmanite and post-perovskite. When hydrogen incorporates via (Mg + 2H)\textsubscript{Si} and (Al + H)\textsubscript{Si} defects, the proton conductivity of bridgmanite falls within the same order of magnitude as the reported lower mantle conductivity profile and can well explain its lateral heterogeneity, indicating that one can potentially deduce water distribution by analyzing electrical conductivity variations in the lower mantle.

\section*{Open Research}

All the data and source codes used in this study are stored at the Open Science Framework (\url{https://osf.io/brwcv/}) via DOI \url{10.17605/OSF.IO/BRWCV} \cite{Peng_data_2023}. Part of our training set and test set for the MLP can be found in \citeA{Deng_data_2023}. For the software packages used in this study, VASP is a commercial code available at \url{www.vasp.at}; DeePMD-kit is developed openly at \url{https://github.com/deepmodeling/deepmd-kit}; LAMMPS is developed openly at \url{https://github.com/lammps/lammps} {and available at {\citeA{plimpton_lammps_2021}}}; PLUMED 2 is developed openly at \url{https://github.com/plumed/plumed2}; ASAP is developed openly at \url{https://github.com/BingqingCheng/ASAP}.

\acknowledgments
J.D. acknowledges National Science Foundation EAR-2242946. {We are grateful to the anonymous reviewers, Dr. John Brodholt, and Dr. Zhixue Du whose comments helped to improve the manuscript.} The simulations presented in this article were performed on computational resources managed and supported by Princeton Research Computing, a consortium of groups including the Princeton Institute for Computational Science and Engineering (PICSciE) and the Office of Information Technology's High-Performance Computing Center and Visualization Laboratory at Princeton University.


\bibliography{hydrogen_diffusion}

\begin{thebibliography}{}

\bibitem [\protect \citeauthoryear {%
Bahr%
\ \BBA {} Simpson%
}{%
Bahr%
\ \BBA {} Simpson%
}{%
{\protect \APACyear {2002}}%
}]{%
Bahr2002a}
\APACinsertmetastar {%
Bahr2002a}%
\begin{APACrefauthors}%
Bahr, K.%
\BCBT {}\ \BBA {} Simpson, F.%
\end{APACrefauthors}%
\unskip\
\newblock
\APACrefYearMonthDay{2002}{}{}.
\newblock
{\BBOQ}\APACrefatitle {Electrical Anisotropy below Slow- and Fast-Moving
  Plates: {{Paleoflow}} in the Upper Mantle?} {Electrical anisotropy below
  slow- and fast-moving plates: {{Paleoflow}} in the upper mantle?}{\BBCQ}
\newblock
\APACjournalVolNumPages{Science (New York, N.Y.)}{295}{5558}{1270--1272}.
\newblock
\begin{APACrefDOI} \doi{10.1126/science.1066161} \end{APACrefDOI}
\PrintBackRefs{\CurrentBib}

\bibitem [\protect \citeauthoryear {%
Caracas%
\ \BBA {} Panero%
}{%
Caracas%
\ \BBA {} Panero%
}{%
{\protect \APACyear {2017}}%
}]{%
Caracas2017a}
\APACinsertmetastar {%
Caracas2017a}%
\begin{APACrefauthors}%
Caracas, R.%
\BCBT {}\ \BBA {} Panero, W\BPBI R.%
\end{APACrefauthors}%
\unskip\
\newblock
\APACrefYearMonthDay{2017}{}{}.
\newblock
{\BBOQ}\APACrefatitle {Hydrogen Mobility in Transition Zone Silicates}
  {Hydrogen mobility in transition zone silicates}.{\BBCQ}
\newblock
\APACjournalVolNumPages{Progress in Earth and Planetary Science}{4}{}{11}.
\newblock
\begin{APACrefDOI} \doi{10.1186/s40645-017-0119-8} \end{APACrefDOI}
\PrintBackRefs{\CurrentBib}

\bibitem [\protect \citeauthoryear {%
Cheng%
\ \protect \BOthers {.}}{%
Cheng%
\ \protect \BOthers {.}}{%
{\protect \APACyear {2020}}%
}]{%
Cheng2020a}
\APACinsertmetastar {%
Cheng2020a}%
\begin{APACrefauthors}%
Cheng, B\BPBI Q.%
, Griffiths, R\BPBI R.%
, Wengert, S.%
, Kunkel, C.%
, Stenczel, T.%
, Zhu, B\BPBI N.%
\BDBL {}Csanyi, G.%
\end{APACrefauthors}%
\unskip\
\newblock
\APACrefYearMonthDay{2020}{}{}.
\newblock
{\BBOQ}\APACrefatitle {Mapping Materials and Molecules} {Mapping materials and
  molecules}.{\BBCQ}
\newblock
\APACjournalVolNumPages{Accounts of Chemical Research}{53}{9}{1981--1991}.
\newblock
\begin{APACrefDOI} \doi{10.1021/acs.accounts.0c00403} \end{APACrefDOI}
\PrintBackRefs{\CurrentBib}

\bibitem [\protect \citeauthoryear {%
Civet%
\ \BBA {} Tarits%
}{%
Civet%
\ \BBA {} Tarits%
}{%
{\protect \APACyear {2013}}%
}]{%
Civet2013a}
\APACinsertmetastar {%
Civet2013a}%
\begin{APACrefauthors}%
Civet, F.%
\BCBT {}\ \BBA {} Tarits, P.%
\end{APACrefauthors}%
\unskip\
\newblock
\APACrefYearMonthDay{2013}{}{}.
\newblock
{\BBOQ}\APACrefatitle {Analysis of Magnetic Satellite Data to Infer the Mantle
  Electrical Conductivity of Telluric Planets in the Solar System} {Analysis of
  magnetic satellite data to infer the mantle electrical conductivity of
  telluric planets in the solar system}.{\BBCQ}
\newblock
\APACjournalVolNumPages{Planetary and Space Science}{84}{}{102--111}.
\newblock
\begin{APACrefDOI} \doi{10.1016/j.pss.2013.05.004} \end{APACrefDOI}
\PrintBackRefs{\CurrentBib}

\bibitem [\protect \citeauthoryear {%
Civet%
, Thebault%
, Verhoeven%
, Langlais%
\BCBL {}\ \BBA {} Saturnino%
}{%
Civet%
\ \protect \BOthers {.}}{%
{\protect \APACyear {2015}}%
}]{%
Civet2015a}
\APACinsertmetastar {%
Civet2015a}%
\begin{APACrefauthors}%
Civet, F.%
, Thebault, E.%
, Verhoeven, O.%
, Langlais, B.%
\BCBL {}\ \BBA {} Saturnino, D.%
\end{APACrefauthors}%
\unskip\
\newblock
\APACrefYearMonthDay{2015}{}{}.
\newblock
{\BBOQ}\APACrefatitle {Electrical Conductivity of the {{Earth}}'s Mantle from
  the First {{Swarm}} Magnetic Field Measurements} {Electrical conductivity of
  the {{Earth}}'s mantle from the first {{Swarm}} magnetic field
  measurements}.{\BBCQ}
\newblock
\APACjournalVolNumPages{Geophysical Research Letters}{42}{9}{3338--3346}.
\newblock
\begin{APACrefDOI} \doi{10.1002/2015gl063397} \end{APACrefDOI}
\PrintBackRefs{\CurrentBib}

\bibitem [\protect \citeauthoryear {%
Demouchy%
\ \BBA {} Mackwell%
}{%
Demouchy%
\ \BBA {} Mackwell%
}{%
{\protect \APACyear {2003}}%
}]{%
Demouchy2003}
\APACinsertmetastar {%
Demouchy2003}%
\begin{APACrefauthors}%
Demouchy, S.%
\BCBT {}\ \BBA {} Mackwell, S.%
\end{APACrefauthors}%
\unskip\
\newblock
\APACrefYearMonthDay{2003}{{\APACmonth{09}}}{}.
\newblock
{\BBOQ}\APACrefatitle {Water diffusion in synthetic iron-free forsterite}
  {Water diffusion in synthetic iron-free forsterite}.{\BBCQ}
\newblock
\APACjournalVolNumPages{Physics and Chemistry of Minerals}{30}{8}{486--494}.
\newblock
\begin{APACrefDOI} \doi{10.1007/s00269-003-0342-2} \end{APACrefDOI}
\PrintBackRefs{\CurrentBib}

\bibitem [\protect \citeauthoryear {%
Deng%
, Niu%
, Hu%
, Chen%
\BCBL {}\ \BBA {} Stixrude%
}{%
Deng%
\ \protect \BOthers {.}}{%
{\protect \APACyear {2023}}%
{\protect \APACexlab {{\protect \BCnt {1}}}}}]{%
Deng2023}
\APACinsertmetastar {%
Deng2023}%
\begin{APACrefauthors}%
Deng, J.%
, Niu, H.%
, Hu, J.%
, Chen, M.%
\BCBL {}\ \BBA {} Stixrude, L.%
\end{APACrefauthors}%
\unskip\
\newblock
\APACrefYearMonthDay{2023{\protect \BCnt {1}}}{{\APACmonth{02}}}{}.
\newblock
{\BBOQ}\APACrefatitle {Melting of {{MgSiO}}{\textsubscript{3}} Determined by
  Machine Learning Potentials} {Melting of {{MgSiO}}{\textsubscript{3}}
  determined by machine learning potentials}.{\BBCQ}
\newblock
\APACjournalVolNumPages{Physical Review B}{107}{6}{064103}.
\newblock
\begin{APACrefDOI} \doi{10.1103/PhysRevB.107.064103} \end{APACrefDOI}
\PrintBackRefs{\CurrentBib}

\bibitem [\protect \citeauthoryear {%
Deng%
, Niu%
, Hu%
, Chen%
\BCBL {}\ \BBA {} Stixrude%
}{%
Deng%
\ \protect \BOthers {.}}{%
{\protect \APACyear {2023}}%
{\protect \APACexlab {{\protect \BCnt {2}}}}}]{%
Deng_data_2023}
\APACinsertmetastar {%
Deng_data_2023}%
\begin{APACrefauthors}%
Deng, J.%
, Niu, H.%
, Hu, J.%
, Chen, M.%
\BCBL {}\ \BBA {} Stixrude, L.%
\end{APACrefauthors}%
\unskip\
\newblock
\APACrefYearMonthDay{2023{\protect \BCnt {2}}}{}{}.
\newblock
\APACrefbtitle {Train and test data} {Train and test data}\ [Dataset].
\newblock
\APACaddressPublisher{}{{OSF}}.
\newblock
\begin{APACrefURL} \url{https://osf.io/dt4xs/} \end{APACrefURL}
\newblock
\begin{APACrefDOI} \doi{10.17605/OSF.IO/DT4XS} \end{APACrefDOI}
\PrintBackRefs{\CurrentBib}

\bibitem [\protect \citeauthoryear {%
Deng%
\ \BBA {} Stixrude%
}{%
Deng%
\ \BBA {} Stixrude%
}{%
{\protect \APACyear {2021}}%
{\protect \APACexlab {{\protect \BCnt {1}}}}}]{%
Deng2021a}
\APACinsertmetastar {%
Deng2021a}%
\begin{APACrefauthors}%
Deng, J.%
\BCBT {}\ \BBA {} Stixrude, L.%
\end{APACrefauthors}%
\unskip\
\newblock
\APACrefYearMonthDay{2021{\protect \BCnt {1}}}{}{}.
\newblock
{\BBOQ}\APACrefatitle {Deep Fractionation of {{Hf}} in a Solidifying Magma
  Ocean and Its Implications for Tungsten Isotopic Heterogeneities in the
  Mantle} {Deep fractionation of {{Hf}} in a solidifying magma ocean and its
  implications for tungsten isotopic heterogeneities in the mantle}.{\BBCQ}
\newblock
\APACjournalVolNumPages{Earth and Planetary Science Letters}{562}{}{8}.
\newblock
\begin{APACrefDOI} \doi{10.1016/j.epsl.2021.116873} \end{APACrefDOI}
\PrintBackRefs{\CurrentBib}

\bibitem [\protect \citeauthoryear {%
Deng%
\ \BBA {} Stixrude%
}{%
Deng%
\ \BBA {} Stixrude%
}{%
{\protect \APACyear {2021}}%
{\protect \APACexlab {{\protect \BCnt {2}}}}}]{%
Deng2021b}
\APACinsertmetastar {%
Deng2021b}%
\begin{APACrefauthors}%
Deng, J.%
\BCBT {}\ \BBA {} Stixrude, L.%
\end{APACrefauthors}%
\unskip\
\newblock
\APACrefYearMonthDay{2021{\protect \BCnt {2}}}{}{}.
\newblock
{\BBOQ}\APACrefatitle {Thermal Conductivity of Silicate Liquid Determined by
  Machine Learning Potentials} {Thermal conductivity of silicate liquid
  determined by machine learning potentials}.{\BBCQ}
\newblock
\APACjournalVolNumPages{Geophysical Research Letters}{48}{17}{10}.
\newblock
\begin{APACrefDOI} \doi{10.1029/2021gl093806} \end{APACrefDOI}
\PrintBackRefs{\CurrentBib}

\bibitem [\protect \citeauthoryear {%
Farver%
}{%
Farver%
}{%
{\protect \APACyear {2010}}%
}]{%
Farver2010a}
\APACinsertmetastar {%
Farver2010a}%
\begin{APACrefauthors}%
Farver, J\BPBI R.%
\end{APACrefauthors}%
\unskip\
\newblock
\APACrefYearMonthDay{2010}{{\APACmonth{01}}}{}.
\newblock
{\BBOQ}\APACrefatitle {Oxygen and Hydrogen Diffusion in Minerals} {Oxygen and
  hydrogen diffusion in minerals}.{\BBCQ}
\newblock
\APACjournalVolNumPages{Reviews in Mineralogy and
  Geochemistry}{72}{1}{447--507}.
\newblock
\begin{APACrefDOI} \doi{10.2138/rmg.2010.72.10} \end{APACrefDOI}
\PrintBackRefs{\CurrentBib}

\bibitem [\protect \citeauthoryear {%
Fiquet%
\ \protect \BOthers {.}}{%
Fiquet%
\ \protect \BOthers {.}}{%
{\protect \APACyear {2010}}%
}]{%
fiquet_melting_2010}
\APACinsertmetastar {%
fiquet_melting_2010}%
\begin{APACrefauthors}%
Fiquet, G.%
, Auzende, A\BPBI L.%
, Siebert, J.%
, Corgne, A.%
, Bureau, H.%
, Ozawa, H.%
\BCBL {}\ \BBA {} Garbarino, G.%
\end{APACrefauthors}%
\unskip\
\newblock
\APACrefYearMonthDay{2010}{{\APACmonth{09}}}{}.
\newblock
{\BBOQ}\APACrefatitle {Melting of {Peridotite} to 140 {Gigapascals}} {Melting
  of {Peridotite} to 140 {Gigapascals}}.{\BBCQ}
\newblock
\APACjournalVolNumPages{Science}{329}{5998}{1516--1518}.
\newblock
\APACrefnote{Publisher: American Association for the Advancement of Science}
\newblock
\begin{APACrefDOI} \doi{10.1126/science.1192448} \end{APACrefDOI}
\PrintBackRefs{\CurrentBib}

\bibitem [\protect \citeauthoryear {%
Franchini%
, Reticcioli%
, Setvin%
\BCBL {}\ \BBA {} Diebold%
}{%
Franchini%
\ \protect \BOthers {.}}{%
{\protect \APACyear {2021}}%
}]{%
franchini_polarons_2021}
\APACinsertmetastar {%
franchini_polarons_2021}%
\begin{APACrefauthors}%
Franchini, C.%
, Reticcioli, M.%
, Setvin, M.%
\BCBL {}\ \BBA {} Diebold, U.%
\end{APACrefauthors}%
\unskip\
\newblock
\APACrefYearMonthDay{2021}{{\APACmonth{07}}}{}.
\newblock
{\BBOQ}\APACrefatitle {Polarons in materials} {Polarons in materials}.{\BBCQ}
\newblock
\APACjournalVolNumPages{Nature Reviews Materials}{6}{7}{560--586}.
\newblock
\begin{APACrefDOI} \doi{10.1038/s41578-021-00289-w} \end{APACrefDOI}
\PrintBackRefs{\CurrentBib}

\bibitem [\protect \citeauthoryear {%
Frank%
\ \BBA {} Turnbull%
}{%
Frank%
\ \BBA {} Turnbull%
}{%
{\protect \APACyear {1956}}%
}]{%
Frank1956}
\APACinsertmetastar {%
Frank1956}%
\begin{APACrefauthors}%
Frank, F\BPBI C.%
\BCBT {}\ \BBA {} Turnbull, D.%
\end{APACrefauthors}%
\unskip\
\newblock
\APACrefYearMonthDay{1956}{{\APACmonth{11}}}{}.
\newblock
{\BBOQ}\APACrefatitle {Mechanism of {Diffusion} of {Copper} in {Germanium}}
  {Mechanism of {Diffusion} of {Copper} in {Germanium}}.{\BBCQ}
\newblock
\APACjournalVolNumPages{Physical Review}{104}{3}{617--618}.
\newblock
\begin{APACrefDOI} \doi{10.1103/PhysRev.104.617} \end{APACrefDOI}
\PrintBackRefs{\CurrentBib}

\bibitem [\protect \citeauthoryear {%
Freitas%
\ \BBA {} Cao%
}{%
Freitas%
\ \BBA {} Cao%
}{%
{\protect \APACyear {2022}}%
}]{%
Freitas2022}
\APACinsertmetastar {%
Freitas2022}%
\begin{APACrefauthors}%
Freitas, R.%
\BCBT {}\ \BBA {} Cao, Y.%
\end{APACrefauthors}%
\unskip\
\newblock
\APACrefYearMonthDay{2022}{{\APACmonth{10}}}{}.
\newblock
{\BBOQ}\APACrefatitle {Machine-Learning Potentials for Crystal Defects}
  {Machine-learning potentials for crystal defects}.{\BBCQ}
\newblock
\APACjournalVolNumPages{MRS Communications}{12}{5}{510--520}.
\newblock
\begin{APACrefDOI} \doi{10.1557/s43579-022-00221-5} \end{APACrefDOI}
\PrintBackRefs{\CurrentBib}

\bibitem [\protect \citeauthoryear {%
Fu%
\ \protect \BOthers {.}}{%
Fu%
\ \protect \BOthers {.}}{%
{\protect \APACyear {2019}}%
}]{%
Fu2019a}
\APACinsertmetastar {%
Fu2019a}%
\begin{APACrefauthors}%
Fu, S\BPBI Y.%
, Yang, J.%
, Karato, S.%
, Vasiliev, A.%
, Presniakov, M\BPBI Y.%
, Gavriliuk, A\BPBI G.%
\BDBL {}Lin, J\BPBI F.%
\end{APACrefauthors}%
\unskip\
\newblock
\APACrefYearMonthDay{2019}{}{}.
\newblock
{\BBOQ}\APACrefatitle {Water Concentration in Single-Crystal
  ({{Al}},{{Fe}})-{{Bearing}} Bridgmanite Grown from the Hydrous Melt:
  {{Implications}} for Dehydration Melting at the Topmost Lower Mantle} {Water
  concentration in single-crystal ({{Al}},{{Fe}})-{{Bearing}} bridgmanite grown
  from the hydrous melt: {{Implications}} for dehydration melting at the
  topmost lower mantle}.{\BBCQ}
\newblock
\APACjournalVolNumPages{Geophysical Research Letters}{46}{17-18}{10346--10357}.
\newblock
\begin{APACrefDOI} \doi{10.1029/2019gl084630} \end{APACrefDOI}
\PrintBackRefs{\CurrentBib}

\bibitem [\protect \citeauthoryear {%
Grayver%
\ \protect \BOthers {.}}{%
Grayver%
\ \protect \BOthers {.}}{%
{\protect \APACyear {2017}}%
}]{%
Grayver2017a}
\APACinsertmetastar {%
Grayver2017a}%
\begin{APACrefauthors}%
Grayver, A\BPBI V.%
, Munch, F\BPBI D.%
, Kuvshinov, A\BPBI V.%
, Khan, A.%
, Sabaka, T\BPBI J.%
\BCBL {}\ \BBA {} {Toffner-Clausen}, L.%
\end{APACrefauthors}%
\unskip\
\newblock
\APACrefYearMonthDay{2017}{}{}.
\newblock
{\BBOQ}\APACrefatitle {Joint Inversion of Satellite-Detected Tidal and
  Magnetospheric Signals Constrains Electrical Conductivity and Water Content
  of the Upper Mantle and Transition Zone} {Joint inversion of
  satellite-detected tidal and magnetospheric signals constrains electrical
  conductivity and water content of the upper mantle and transition
  zone}.{\BBCQ}
\newblock
\APACjournalVolNumPages{Geophysical Research Letters}{44}{12}{6074--6081}.
\newblock
\begin{APACrefDOI} \doi{10.1002/2017gl073446} \end{APACrefDOI}
\PrintBackRefs{\CurrentBib}

\bibitem [\protect \citeauthoryear {%
Hae%
, Ohtani%
, Kubo%
, Koyama%
\BCBL {}\ \BBA {} Utada%
}{%
Hae%
\ \protect \BOthers {.}}{%
{\protect \APACyear {2006}}%
}]{%
Hae2006a}
\APACinsertmetastar {%
Hae2006a}%
\begin{APACrefauthors}%
Hae, R.%
, Ohtani, E.%
, Kubo, T.%
, Koyama, T.%
\BCBL {}\ \BBA {} Utada, H.%
\end{APACrefauthors}%
\unskip\
\newblock
\APACrefYearMonthDay{2006}{}{}.
\newblock
{\BBOQ}\APACrefatitle {Hydrogen Diffusivity in Wadsleyite and Water
  Distribution in the Mantle Transition Zone} {Hydrogen diffusivity in
  wadsleyite and water distribution in the mantle transition zone}.{\BBCQ}
\newblock
\APACjournalVolNumPages{Earth and Planetary Science
  Letters}{243}{1-2}{141--148}.
\newblock
\begin{APACrefDOI} \doi{10.1016/j.epsl.2005.12.035} \end{APACrefDOI}
\PrintBackRefs{\CurrentBib}

\bibitem [\protect \citeauthoryear {%
Hallis%
\ \protect \BOthers {.}}{%
Hallis%
\ \protect \BOthers {.}}{%
{\protect \APACyear {2015}}%
}]{%
Hallis2015a}
\APACinsertmetastar {%
Hallis2015a}%
\begin{APACrefauthors}%
Hallis, L\BPBI J.%
, Huss, G\BPBI R.%
, Nagashima, K.%
, Taylor, G\BPBI J.%
, Halldorsson, S\BPBI A.%
, Hilion, D\BPBI R.%
\BDBL {}Meech, K\BPBI J.%
\end{APACrefauthors}%
\unskip\
\newblock
\APACrefYearMonthDay{2015}{}{}.
\newblock
{\BBOQ}\APACrefatitle {Evidence for Primordial Water in {{Earth}}'s Deep
  Mantle} {Evidence for primordial water in {{Earth}}'s deep mantle}.{\BBCQ}
\newblock
\APACjournalVolNumPages{Science (New York, N.Y.)}{350}{6262}{795--797}.
\newblock
\begin{APACrefDOI} \doi{10.1126/science.aac4834} \end{APACrefDOI}
\PrintBackRefs{\CurrentBib}

\bibitem [\protect \citeauthoryear {%
He%
, Zhu%
, Epstein%
\BCBL {}\ \BBA {} Mo%
}{%
He%
\ \protect \BOthers {.}}{%
{\protect \APACyear {2018}}%
}]{%
He2018a}
\APACinsertmetastar {%
He2018a}%
\begin{APACrefauthors}%
He, X\BPBI F.%
, Zhu, Y\BPBI Z.%
, Epstein, A.%
\BCBL {}\ \BBA {} Mo, Y\BPBI F.%
\end{APACrefauthors}%
\unskip\
\newblock
\APACrefYearMonthDay{2018}{}{}.
\newblock
{\BBOQ}\APACrefatitle {Statistical Variances of Diffusional Properties from Ab
  Initio Molecular Dynamics Simulations} {Statistical variances of diffusional
  properties from ab initio molecular dynamics simulations}.{\BBCQ}
\newblock
\APACjournalVolNumPages{Npj Computational Materials}{4}{}{9}.
\newblock
\begin{APACrefDOI} \doi{10.1038/s41524-018-0074-y} \end{APACrefDOI}
\PrintBackRefs{\CurrentBib}

\bibitem [\protect \citeauthoryear {%
Henkelman%
, Uberuaga%
\BCBL {}\ \BBA {} Jonsson%
}{%
Henkelman%
\ \protect \BOthers {.}}{%
{\protect \APACyear {2000}}%
}]{%
Henkelman2000a}
\APACinsertmetastar {%
Henkelman2000a}%
\begin{APACrefauthors}%
Henkelman, G.%
, Uberuaga, B\BPBI P.%
\BCBL {}\ \BBA {} Jonsson, H.%
\end{APACrefauthors}%
\unskip\
\newblock
\APACrefYearMonthDay{2000}{}{}.
\newblock
{\BBOQ}\APACrefatitle {A Climbing Image Nudged Elastic Band Method for Finding
  Saddle Points and Minimum Energy Paths} {A climbing image nudged elastic band
  method for finding saddle points and minimum energy paths}.{\BBCQ}
\newblock
\APACjournalVolNumPages{Journal of Chemical Physics}{113}{22}{9901--9904}.
\newblock
\begin{APACrefDOI} \doi{10.1063/1.1329672} \end{APACrefDOI}
\PrintBackRefs{\CurrentBib}

\bibitem [\protect \citeauthoryear {%
Hernández%
, Alfè%
\BCBL {}\ \BBA {} Brodholt%
}{%
Hernández%
\ \protect \BOthers {.}}{%
{\protect \APACyear {2013}}%
}]{%
hernandez_incorporation_2013}
\APACinsertmetastar {%
hernandez_incorporation_2013}%
\begin{APACrefauthors}%
Hernández, E\BPBI R.%
, Alfè, D.%
\BCBL {}\ \BBA {} Brodholt, J.%
\end{APACrefauthors}%
\unskip\
\newblock
\APACrefYearMonthDay{2013}{{\APACmonth{02}}}{}.
\newblock
{\BBOQ}\APACrefatitle {The incorporation of water into lower-mantle
  perovskites: {A} first-principles study} {The incorporation of water into
  lower-mantle perovskites: {A} first-principles study}.{\BBCQ}
\newblock
\APACjournalVolNumPages{Earth and Planetary Science Letters}{364}{}{37--43}.
\newblock
\begin{APACrefDOI} \doi{10.1016/j.epsl.2013.01.005} \end{APACrefDOI}
\PrintBackRefs{\CurrentBib}

\bibitem [\protect \citeauthoryear {%
Holzapfel%
, Rubie%
, Frost%
\BCBL {}\ \BBA {} Langenhorst%
}{%
Holzapfel%
\ \protect \BOthers {.}}{%
{\protect \APACyear {2005}}%
}]{%
Holzapfel2005a}
\APACinsertmetastar {%
Holzapfel2005a}%
\begin{APACrefauthors}%
Holzapfel, C.%
, Rubie, D\BPBI C.%
, Frost, D\BPBI J.%
\BCBL {}\ \BBA {} Langenhorst, F.%
\end{APACrefauthors}%
\unskip\
\newblock
\APACrefYearMonthDay{2005}{}{}.
\newblock
{\BBOQ}\APACrefatitle {Fe-{{Mg}} Interdiffusion in
  ({{Mg}},{{Fe}}){{SiO}}{\textsubscript{3}} Perovskite and Lower Mantle
  Reequilibration} {Fe-{{Mg}} interdiffusion in
  ({{Mg}},{{Fe}}){{SiO}}{\textsubscript{3}} perovskite and lower mantle
  reequilibration}.{\BBCQ}
\newblock
\APACjournalVolNumPages{Science (New York, N.Y.)}{309}{5741}{1707--1710}.
\newblock
\begin{APACrefDOI} \doi{10.1126/science.1111895} \end{APACrefDOI}
\PrintBackRefs{\CurrentBib}

\bibitem [\protect \citeauthoryear {%
Hoover%
}{%
Hoover%
}{%
{\protect \APACyear {1985}}%
}]{%
Hoover1985a}
\APACinsertmetastar {%
Hoover1985a}%
\begin{APACrefauthors}%
Hoover, W\BPBI G.%
\end{APACrefauthors}%
\unskip\
\newblock
\APACrefYearMonthDay{1985}{}{}.
\newblock
{\BBOQ}\APACrefatitle {Canonical Dynamics - Equilibrium Phase-Space
  Distributions} {Canonical dynamics - equilibrium phase-space
  distributions}.{\BBCQ}
\newblock
\APACjournalVolNumPages{Physical Review A}{31}{3}{1695--1697}.
\newblock
\begin{APACrefDOI} \doi{10.1103/PhysRevA.31.1695} \end{APACrefDOI}
\PrintBackRefs{\CurrentBib}

\bibitem [\protect \citeauthoryear {%
Hsieh%
\ \protect \BOthers {.}}{%
Hsieh%
\ \protect \BOthers {.}}{%
{\protect \APACyear {2020}}%
}]{%
Hsieh2020a}
\APACinsertmetastar {%
Hsieh2020a}%
\begin{APACrefauthors}%
Hsieh, W\BPBI P.%
, Ishii, T.%
, Chao, K\BPBI H.%
, Tsuchiya, J.%
, Deschamps, F.%
\BCBL {}\ \BBA {} Ohtani, E.%
\end{APACrefauthors}%
\unskip\
\newblock
\APACrefYearMonthDay{2020}{}{}.
\newblock
{\BBOQ}\APACrefatitle {Spin Transition of Iron in Delta-({{Al}},{{Fe}}){{OOH}}
  Induces Thermal Anomalies in Earth's Lower Mantle} {Spin transition of iron
  in delta-({{Al}},{{Fe}}){{OOH}} induces thermal anomalies in earth's lower
  mantle}.{\BBCQ}
\newblock
\APACjournalVolNumPages{Geophysical Research Letters}{47}{4}{10}.
\newblock
\begin{APACrefDOI} \doi{10.1029/2020gl087036} \end{APACrefDOI}
\PrintBackRefs{\CurrentBib}

\bibitem [\protect \citeauthoryear {%
Imbalzano%
\ \protect \BOthers {.}}{%
Imbalzano%
\ \protect \BOthers {.}}{%
{\protect \APACyear {2018}}%
}]{%
Imbalzano2018a}
\APACinsertmetastar {%
Imbalzano2018a}%
\begin{APACrefauthors}%
Imbalzano, G.%
, Anelli, A.%
, Giofre, D.%
, Klees, S.%
, Behler, J.%
\BCBL {}\ \BBA {} Ceriotti, M.%
\end{APACrefauthors}%
\unskip\
\newblock
\APACrefYearMonthDay{2018}{}{}.
\newblock
{\BBOQ}\APACrefatitle {Automatic Selection of Atomic Fingerprints and Reference
  Configurations for Machine-Learning Potentials} {Automatic selection of
  atomic fingerprints and reference configurations for machine-learning
  potentials}.{\BBCQ}
\newblock
\APACjournalVolNumPages{Journal of Chemical Physics}{148}{24}{9}.
\newblock
\begin{APACrefDOI} \doi{10.1063/1.5024611} \end{APACrefDOI}
\PrintBackRefs{\CurrentBib}

\bibitem [\protect \citeauthoryear {%
Ingrin%
\ \BBA {} Blanchard%
}{%
Ingrin%
\ \BBA {} Blanchard%
}{%
{\protect \APACyear {2006}}%
}]{%
Ingrin2006a}
\APACinsertmetastar {%
Ingrin2006a}%
\begin{APACrefauthors}%
Ingrin, J.%
\BCBT {}\ \BBA {} Blanchard, M.%
\end{APACrefauthors}%
\unskip\
\newblock
\APACrefYearMonthDay{2006}{{\APACmonth{01}}}{}.
\newblock
{\BBOQ}\APACrefatitle {Diffusion of {{Hydrogen}} in {{Minerals}}} {Diffusion of
  {{Hydrogen}} in {{Minerals}}}.{\BBCQ}
\newblock
\APACjournalVolNumPages{Reviews in Mineralogy and
  Geochemistry}{62}{1}{291--320}.
\newblock
\begin{APACrefDOI} \doi{10.2138/rmg.2006.62.13} \end{APACrefDOI}
\PrintBackRefs{\CurrentBib}

\bibitem [\protect \citeauthoryear {%
Irifune%
}{%
Irifune%
}{%
{\protect \APACyear {1994}}%
}]{%
Irifune1994}
\APACinsertmetastar {%
Irifune1994}%
\begin{APACrefauthors}%
Irifune, T.%
\end{APACrefauthors}%
\unskip\
\newblock
\APACrefYearMonthDay{1994}{{\APACmonth{07}}}{}.
\newblock
{\BBOQ}\APACrefatitle {Absence of an Aluminous Phase in the Upper Part of the
  {{Earth}}'s Lower Mantle} {Absence of an aluminous phase in the upper part of
  the {{Earth}}'s lower mantle}.{\BBCQ}
\newblock
\APACjournalVolNumPages{Nature}{370}{6485}{131--133}.
\newblock
\begin{APACrefDOI} \doi{10.1038/370131a0} \end{APACrefDOI}
\PrintBackRefs{\CurrentBib}

\bibitem [\protect \citeauthoryear {%
Ishii%
, Ohtani%
\BCBL {}\ \BBA {} Shatskiy%
}{%
Ishii%
\ \protect \BOthers {.}}{%
{\protect \APACyear {2022}}%
}]{%
ishii_aluminum_2022}
\APACinsertmetastar {%
ishii_aluminum_2022}%
\begin{APACrefauthors}%
Ishii, T.%
, Ohtani, E.%
\BCBL {}\ \BBA {} Shatskiy, A.%
\end{APACrefauthors}%
\unskip\
\newblock
\APACrefYearMonthDay{2022}{{\APACmonth{04}}}{}.
\newblock
{\BBOQ}\APACrefatitle {Aluminum and hydrogen partitioning between bridgmanite
  and high-pressure hydrous phases: {Implications} for water storage in the
  lower mantle} {Aluminum and hydrogen partitioning between bridgmanite and
  high-pressure hydrous phases: {Implications} for water storage in the lower
  mantle}.{\BBCQ}
\newblock
\APACjournalVolNumPages{Earth and Planetary Science Letters}{583}{}{117441}.
\newblock
\begin{APACrefDOI} \doi{10.1016/j.epsl.2022.117441} \end{APACrefDOI}
\PrintBackRefs{\CurrentBib}

\bibitem [\protect \citeauthoryear {%
Karato%
}{%
Karato%
}{%
{\protect \APACyear {1990}}%
}]{%
Karato1990a}
\APACinsertmetastar {%
Karato1990a}%
\begin{APACrefauthors}%
Karato, S.%
\end{APACrefauthors}%
\unskip\
\newblock
\APACrefYearMonthDay{1990}{}{}.
\newblock
{\BBOQ}\APACrefatitle {The Role of Hydrogen in the Electrical-Conductivity of
  the Upper Mantle} {The role of hydrogen in the electrical-conductivity of the
  upper mantle}.{\BBCQ}
\newblock
\APACjournalVolNumPages{Nature}{347}{6290}{272--273}.
\newblock
\begin{APACrefDOI} \doi{10.1038/347272a0} \end{APACrefDOI}
\PrintBackRefs{\CurrentBib}

\bibitem [\protect \citeauthoryear {%
Karato%
}{%
Karato%
}{%
{\protect \APACyear {1995}}%
}]{%
Karato1995a}
\APACinsertmetastar {%
Karato1995a}%
\begin{APACrefauthors}%
Karato, S.%
\end{APACrefauthors}%
\unskip\
\newblock
\APACrefYearMonthDay{1995}{}{}.
\newblock
{\BBOQ}\APACrefatitle {Effects of Water on Seismic-Wave Velocities in the
  Upper-Mantle} {Effects of water on seismic-wave velocities in the
  upper-mantle}.{\BBCQ}
\newblock
\APACjournalVolNumPages{Proceedings of the Japan Academy Series B-Physical and
  Biological Sciences}{71}{2}{61--66}.
\newblock
\begin{APACrefDOI} \doi{10.2183/pjab.71.61} \end{APACrefDOI}
\PrintBackRefs{\CurrentBib}

\bibitem [\protect \citeauthoryear {%
Karato%
, Paterson%
\BCBL {}\ \BBA {} Fitz~Gerald%
}{%
Karato%
\ \protect \BOthers {.}}{%
{\protect \APACyear {1986}}%
}]{%
Karato1986a}
\APACinsertmetastar {%
Karato1986a}%
\begin{APACrefauthors}%
Karato, S.%
, Paterson, M\BPBI S.%
\BCBL {}\ \BBA {} Fitz~Gerald, J\BPBI D.%
\end{APACrefauthors}%
\unskip\
\newblock
\APACrefYearMonthDay{1986}{}{}.
\newblock
{\BBOQ}\APACrefatitle {Rheology of Synthetic Olivine Aggregates - Influence of
  Grain-Size and Water} {Rheology of synthetic olivine aggregates - influence
  of grain-size and water}.{\BBCQ}
\newblock
\APACjournalVolNumPages{Journal of Geophysical Research-Solid Earth and
  Planets}{91}{B8}{8151--8176}.
\newblock
\begin{APACrefDOI} \doi{10.1029/JB091iB08p08151} \end{APACrefDOI}
\PrintBackRefs{\CurrentBib}

\bibitem [\protect \citeauthoryear {%
Karki%
}{%
Karki%
}{%
{\protect \APACyear {2015}}%
}]{%
Karki2015a}
\APACinsertmetastar {%
Karki2015a}%
\begin{APACrefauthors}%
Karki, B\BPBI B.%
\end{APACrefauthors}%
\unskip\
\newblock
\APACrefYearMonthDay{2015}{}{}.
\newblock
{\BBOQ}\APACrefatitle {First-Principles Computation of Mantle Materials in
  Crystalline and Amorphous Phases} {First-principles computation of mantle
  materials in crystalline and amorphous phases}.{\BBCQ}
\newblock
\APACjournalVolNumPages{Physics of the Earth and Planetary
  Interiors}{240}{}{43--69}.
\newblock
\begin{APACrefDOI} \doi{10.1016/j.pepi.2014.11.004} \end{APACrefDOI}
\PrintBackRefs{\CurrentBib}

\bibitem [\protect \citeauthoryear {%
Katsura%
, Sato%
\BCBL {}\ \BBA {} Ito%
}{%
Katsura%
\ \protect \BOthers {.}}{%
{\protect \APACyear {1998}}%
}]{%
Katsura1998a}
\APACinsertmetastar {%
Katsura1998a}%
\begin{APACrefauthors}%
Katsura, T.%
, Sato, K.%
\BCBL {}\ \BBA {} Ito, E.%
\end{APACrefauthors}%
\unskip\
\newblock
\APACrefYearMonthDay{1998}{}{}.
\newblock
{\BBOQ}\APACrefatitle {Electrical Conductivity of Silicate Perovskite at
  Lower-Mantle Conditions} {Electrical conductivity of silicate perovskite at
  lower-mantle conditions}.{\BBCQ}
\newblock
\APACjournalVolNumPages{Nature}{395}{6701}{493--495}.
\newblock
\begin{APACrefDOI} \doi{10.1038/26736} \end{APACrefDOI}
\PrintBackRefs{\CurrentBib}

\bibitem [\protect \citeauthoryear {%
Katsura%
, Yoneda%
, Yamazaki%
, Yoshino%
\BCBL {}\ \BBA {} Ito%
}{%
Katsura%
\ \protect \BOthers {.}}{%
{\protect \APACyear {2010}}%
}]{%
Katsura2010a}
\APACinsertmetastar {%
Katsura2010a}%
\begin{APACrefauthors}%
Katsura, T.%
, Yoneda, A.%
, Yamazaki, D.%
, Yoshino, T.%
\BCBL {}\ \BBA {} Ito, E.%
\end{APACrefauthors}%
\unskip\
\newblock
\APACrefYearMonthDay{2010}{}{}.
\newblock
{\BBOQ}\APACrefatitle {Adiabatic Temperature Profile in the Mantle} {Adiabatic
  temperature profile in the mantle}.{\BBCQ}
\newblock
\APACjournalVolNumPages{Physics of the Earth and Planetary
  Interiors}{183}{1-2}{212--218}.
\newblock
\begin{APACrefDOI} \doi{10.1016/j.pepi.2010.07.001} \end{APACrefDOI}
\PrintBackRefs{\CurrentBib}

\bibitem [\protect \citeauthoryear {%
Khan%
, Kuvshinov%
\BCBL {}\ \BBA {} Semenov%
}{%
Khan%
\ \protect \BOthers {.}}{%
{\protect \APACyear {2011}}%
}]{%
Khan2011}
\APACinsertmetastar {%
Khan2011}%
\begin{APACrefauthors}%
Khan, A.%
, Kuvshinov, A.%
\BCBL {}\ \BBA {} Semenov, A.%
\end{APACrefauthors}%
\unskip\
\newblock
\APACrefYearMonthDay{2011}{}{}.
\newblock
{\BBOQ}\APACrefatitle {On the heterogeneous electrical conductivity structure
  of the {Earth}’s mantle with implications for transition zone water
  content} {On the heterogeneous electrical conductivity structure of the
  {Earth}’s mantle with implications for transition zone water
  content}.{\BBCQ}
\newblock
\APACjournalVolNumPages{Journal of Geophysical Research: Solid
  Earth}{116}{B1}{}.
\newblock
\begin{APACrefDOI} \doi{10.1029/2010JB007458} \end{APACrefDOI}
\PrintBackRefs{\CurrentBib}

\bibitem [\protect \citeauthoryear {%
Kohlstedt%
\ \BBA {} Mackwell%
}{%
Kohlstedt%
\ \BBA {} Mackwell%
}{%
{\protect \APACyear {1998}}%
}]{%
Kohlstedt1998}
\APACinsertmetastar {%
Kohlstedt1998}%
\begin{APACrefauthors}%
Kohlstedt, D\BPBI L.%
\BCBT {}\ \BBA {} Mackwell, S\BPBI J.%
\end{APACrefauthors}%
\unskip\
\newblock
\APACrefYearMonthDay{1998}{}{}.
\newblock
{\BBOQ}\APACrefatitle {Diffusion of hydrogen and intrinsic point defects in
  olivine} {Diffusion of hydrogen and intrinsic point defects in
  olivine}.{\BBCQ}
\newblock
\APACjournalVolNumPages{Zeitschrift fur Physikalische
  Chemie}{207}{1-2}{147--162}.
\newblock
\begin{APACrefDOI} \doi{10.1524/zpch.1998.207.part_1_2.147} \end{APACrefDOI}
\PrintBackRefs{\CurrentBib}

\bibitem [\protect \citeauthoryear {%
Kresse%
\ \BBA {} Furthmuller%
}{%
Kresse%
\ \BBA {} Furthmuller%
}{%
{\protect \APACyear {1996}}%
}]{%
Kresse1996a}
\APACinsertmetastar {%
Kresse1996a}%
\begin{APACrefauthors}%
Kresse, G.%
\BCBT {}\ \BBA {} Furthmuller, J.%
\end{APACrefauthors}%
\unskip\
\newblock
\APACrefYearMonthDay{1996}{}{}.
\newblock
{\BBOQ}\APACrefatitle {Efficient Iterative Schemes for Ab Initio Total-Energy
  Calculations Using a Plane-Wave Basis Set} {Efficient iterative schemes for
  ab initio total-energy calculations using a plane-wave basis set}.{\BBCQ}
\newblock
\APACjournalVolNumPages{Physical Review B}{54}{16}{11169--11186}.
\newblock
\begin{APACrefDOI} \doi{10.1103/PhysRevB.54.11169} \end{APACrefDOI}
\PrintBackRefs{\CurrentBib}

\bibitem [\protect \citeauthoryear {%
Kresse%
\ \BBA {} Joubert%
}{%
Kresse%
\ \BBA {} Joubert%
}{%
{\protect \APACyear {1999}}%
}]{%
Kresse1999a}
\APACinsertmetastar {%
Kresse1999a}%
\begin{APACrefauthors}%
Kresse, G.%
\BCBT {}\ \BBA {} Joubert, D.%
\end{APACrefauthors}%
\unskip\
\newblock
\APACrefYearMonthDay{1999}{}{}.
\newblock
{\BBOQ}\APACrefatitle {From Ultrasoft Pseudopotentials to the Projector
  Augmented-Wave Method} {From ultrasoft pseudopotentials to the projector
  augmented-wave method}.{\BBCQ}
\newblock
\APACjournalVolNumPages{Physical Review B}{59}{3}{1758--1775}.
\newblock
\begin{APACrefDOI} \doi{10.1103/PhysRevB.59.1758} \end{APACrefDOI}
\PrintBackRefs{\CurrentBib}

\bibitem [\protect \citeauthoryear {%
Kudoh%
\ \BBA {} Inoue%
}{%
Kudoh%
\ \BBA {} Inoue%
}{%
{\protect \APACyear {1999}}%
}]{%
Kudoh1999a}
\APACinsertmetastar {%
Kudoh1999a}%
\begin{APACrefauthors}%
Kudoh, Y.%
\BCBT {}\ \BBA {} Inoue, T.%
\end{APACrefauthors}%
\unskip\
\newblock
\APACrefYearMonthDay{1999}{}{}.
\newblock
{\BBOQ}\APACrefatitle {Mg-Vacant Structural Modules and Dilution of the
  Symmetry of Hydrous Wadsleyite,
  {$\beta$}-{{Mg}}{\textsubscript{2-x}}{{SiH}}{\textsubscript{2x}}{{O}}{\textsubscript{4}}
  with 0.00{$\leq$}x{$\leq$}0.25} {Mg-vacant structural modules and dilution of
  the symmetry of hydrous wadsleyite,
  {$\beta$}-{{Mg}}{\textsubscript{2-x}}{{SiH}}{\textsubscript{2x}}{{O}}{\textsubscript{4}}
  with 0.00{$\leq$}x{$\leq$}0.25}.{\BBCQ}
\newblock
\APACjournalVolNumPages{Physics and Chemistry of Minerals}{26}{5}{382--388}.
\newblock
\begin{APACrefDOI} \doi{10.1007/s002690050198} \end{APACrefDOI}
\PrintBackRefs{\CurrentBib}

\bibitem [\protect \citeauthoryear {%
Kuvshinov%
\ \BBA {} Olsen%
}{%
Kuvshinov%
\ \BBA {} Olsen%
}{%
{\protect \APACyear {2006}}%
}]{%
Kuvshinov2006a}
\APACinsertmetastar {%
Kuvshinov2006a}%
\begin{APACrefauthors}%
Kuvshinov, A.%
\BCBT {}\ \BBA {} Olsen, N.%
\end{APACrefauthors}%
\unskip\
\newblock
\APACrefYearMonthDay{2006}{}{}.
\newblock
{\BBOQ}\APACrefatitle {A Global Model of Mantle Conductivity Derived from 5
  Years of {{CHAMP}}, {{Orsted}}, and {{SAC-C}} Magnetic Data} {A global model
  of mantle conductivity derived from 5 years of {{CHAMP}}, {{Orsted}}, and
  {{SAC-C}} magnetic data}.{\BBCQ}
\newblock
\APACjournalVolNumPages{Geophysical Research Letters}{33}{18}{5}.
\newblock
\begin{APACrefDOI} \doi{10.1029/2006gl027083} \end{APACrefDOI}
\PrintBackRefs{\CurrentBib}

\bibitem [\protect \citeauthoryear {%
J.~Li%
\ \protect \BOthers {.}}{%
J.~Li%
\ \protect \BOthers {.}}{%
{\protect \APACyear {2023}}%
}]{%
Li2023}
\APACinsertmetastar {%
Li2023}%
\begin{APACrefauthors}%
Li, J.%
, Lin, Y.%
, Meier, T.%
, Liu, Z.%
, Yang, W.%
, Mao, H\BHBI k.%
\BDBL {}Hu, Q.%
\end{APACrefauthors}%
\unskip\
\newblock
\APACrefYearMonthDay{2023}{{\APACmonth{09}}}{}.
\newblock
{\BBOQ}\APACrefatitle {Silica-water superstructure and one-dimensional
  superionic conduit in {Earth}’s mantle} {Silica-water superstructure and
  one-dimensional superionic conduit in {Earth}’s mantle}.{\BBCQ}
\newblock
\APACjournalVolNumPages{Science Advances}{9}{35}{eadh3784}.
\newblock
\begin{APACrefDOI} \doi{10.1126/sciadv.adh3784} \end{APACrefDOI}
\PrintBackRefs{\CurrentBib}

\bibitem [\protect \citeauthoryear {%
X\BPBI Y.~Li%
\ \BBA {} Jeanloz%
}{%
X\BPBI Y.~Li%
\ \BBA {} Jeanloz%
}{%
{\protect \APACyear {1990}}%
}]{%
Li1990a}
\APACinsertmetastar {%
Li1990a}%
\begin{APACrefauthors}%
Li, X\BPBI Y.%
\BCBT {}\ \BBA {} Jeanloz, R.%
\end{APACrefauthors}%
\unskip\
\newblock
\APACrefYearMonthDay{1990}{}{}.
\newblock
{\BBOQ}\APACrefatitle {Laboratory Studies of the Electrical-Conductivity of
  Silicate Perovskites at High-Pressures and Temperatures} {Laboratory studies
  of the electrical-conductivity of silicate perovskites at high-pressures and
  temperatures}.{\BBCQ}
\newblock
\APACjournalVolNumPages{Journal of Geophysical Research-Solid Earth and
  Planets}{95}{B4}{5067--5078}.
\newblock
\begin{APACrefDOI} \doi{10.1029/JB095iB04p05067} \end{APACrefDOI}
\PrintBackRefs{\CurrentBib}

\bibitem [\protect \citeauthoryear {%
Liu%
\ \protect \BOthers {.}}{%
Liu%
\ \protect \BOthers {.}}{%
{\protect \APACyear {2021}}%
}]{%
liu_bridgmanite_2021}
\APACinsertmetastar {%
liu_bridgmanite_2021}%
\begin{APACrefauthors}%
Liu, Z.%
, Fei, H.%
, Chen, L.%
, McCammon, C.%
, Wang, L.%
, Liu, R.%
\BDBL {}Katsura, T.%
\end{APACrefauthors}%
\unskip\
\newblock
\APACrefYearMonthDay{2021}{{\APACmonth{09}}}{}.
\newblock
{\BBOQ}\APACrefatitle {Bridgmanite is nearly dry at the top of the lower
  mantle} {Bridgmanite is nearly dry at the top of the lower mantle}.{\BBCQ}
\newblock
\APACjournalVolNumPages{Earth and Planetary Science Letters}{570}{}{117088}.
\newblock
\begin{APACrefDOI} \doi{10.1016/j.epsl.2021.117088} \end{APACrefDOI}
\PrintBackRefs{\CurrentBib}

\bibitem [\protect \citeauthoryear {%
Lu%
\ \BBA {} Li%
}{%
Lu%
\ \BBA {} Li%
}{%
{\protect \APACyear {2023}}%
}]{%
lu_solubility_2023}
\APACinsertmetastar {%
lu_solubility_2023}%
\begin{APACrefauthors}%
Lu, W.%
\BCBT {}\ \BBA {} Li, Y.%
\end{APACrefauthors}%
\unskip\
\newblock
\APACrefYearMonthDay{2023}{{\APACmonth{12}}}{}.
\newblock
{\BBOQ}\APACrefatitle {Solubility of water in bridgmanite} {Solubility of water
  in bridgmanite}.{\BBCQ}
\newblock
\APACjournalVolNumPages{Acta Geochimica}{42}{6}{998--1006}.
\newblock
\begin{APACrefDOI} \doi{10.1007/s11631-023-00642-6} \end{APACrefDOI}
\PrintBackRefs{\CurrentBib}

\bibitem [\protect \citeauthoryear {%
Mackwell%
\ \BBA {} Kohlstedt%
}{%
Mackwell%
\ \BBA {} Kohlstedt%
}{%
{\protect \APACyear {1990}}%
}]{%
Mackwell1990}
\APACinsertmetastar {%
Mackwell1990}%
\begin{APACrefauthors}%
Mackwell, S\BPBI J.%
\BCBT {}\ \BBA {} Kohlstedt, D\BPBI L.%
\end{APACrefauthors}%
\unskip\
\newblock
\APACrefYearMonthDay{1990}{}{}.
\newblock
{\BBOQ}\APACrefatitle {Diffusion of hydrogen in olivine: {Implications} for
  water in the mantle} {Diffusion of hydrogen in olivine: {Implications} for
  water in the mantle}.{\BBCQ}
\newblock
\APACjournalVolNumPages{Journal of Geophysical Research: Solid
  Earth}{95}{B4}{5079--5088}.
\newblock
\begin{APACrefDOI} \doi{10.1029/JB095iB04p05079} \end{APACrefDOI}
\PrintBackRefs{\CurrentBib}

\bibitem [\protect \citeauthoryear {%
McCammon%
}{%
McCammon%
}{%
{\protect \APACyear {1997}}%
}]{%
Mccammon1997}
\APACinsertmetastar {%
Mccammon1997}%
\begin{APACrefauthors}%
McCammon, C.%
\end{APACrefauthors}%
\unskip\
\newblock
\APACrefYearMonthDay{1997}{{\APACmonth{06}}}{}.
\newblock
{\BBOQ}\APACrefatitle {Perovskite as a possible sink for ferric iron in the
  lower mantle} {Perovskite as a possible sink for ferric iron in the lower
  mantle}.{\BBCQ}
\newblock
\APACjournalVolNumPages{Nature}{387}{6634}{694--696}.
\newblock
\begin{APACrefDOI} \doi{10.1038/42685} \end{APACrefDOI}
\PrintBackRefs{\CurrentBib}

\bibitem [\protect \citeauthoryear {%
Mermin%
}{%
Mermin%
}{%
{\protect \APACyear {1965}}%
}]{%
Mermin1965a}
\APACinsertmetastar {%
Mermin1965a}%
\begin{APACrefauthors}%
Mermin, N\BPBI D.%
\end{APACrefauthors}%
\unskip\
\newblock
\APACrefYearMonthDay{1965}{}{}.
\newblock
{\BBOQ}\APACrefatitle {Thermal Properties of the Inhomogeneous Electron Gas}
  {Thermal properties of the inhomogeneous electron gas}.{\BBCQ}
\newblock
\APACjournalVolNumPages{Physical Review}{137}{5A}{1441-+}.
\newblock
\begin{APACrefDOI} \doi{10.1103/PhysRev.137.A1441} \end{APACrefDOI}
\PrintBackRefs{\CurrentBib}

\bibitem [\protect \citeauthoryear {%
Muir%
\ \BBA {} Brodholt%
}{%
Muir%
\ \BBA {} Brodholt%
}{%
{\protect \APACyear {2018}}%
}]{%
Muir2018}
\APACinsertmetastar {%
Muir2018}%
\begin{APACrefauthors}%
Muir, J\BPBI M\BPBI R.%
\BCBT {}\ \BBA {} Brodholt, J\BPBI P.%
\end{APACrefauthors}%
\unskip\
\newblock
\APACrefYearMonthDay{2018}{{\APACmonth{02}}}{}.
\newblock
{\BBOQ}\APACrefatitle {Water Distribution in the Lower Mantle: {{Implications}}
  for Hydrolytic Weakening} {Water distribution in the lower mantle:
  {{Implications}} for hydrolytic weakening}.{\BBCQ}
\newblock
\APACjournalVolNumPages{Earth and Planetary Science Letters}{484}{}{363--369}.
\newblock
\begin{APACrefDOI} \doi{10.1016/j.epsl.2017.11.051} \end{APACrefDOI}
\PrintBackRefs{\CurrentBib}

\bibitem [\protect \citeauthoryear {%
Murakami%
, Hirose%
, Kawamura%
, Sata%
\BCBL {}\ \BBA {} Ohishi%
}{%
Murakami%
\ \protect \BOthers {.}}{%
{\protect \APACyear {2004}}%
}]{%
Murakami2004a}
\APACinsertmetastar {%
Murakami2004a}%
\begin{APACrefauthors}%
Murakami, M.%
, Hirose, K.%
, Kawamura, K.%
, Sata, N.%
\BCBL {}\ \BBA {} Ohishi, Y.%
\end{APACrefauthors}%
\unskip\
\newblock
\APACrefYearMonthDay{2004}{}{}.
\newblock
{\BBOQ}\APACrefatitle {Post-Perovskite Phase Transition in
  {{MgSiO}}{\textsubscript{3}}} {Post-perovskite phase transition in
  {{MgSiO}}{\textsubscript{3}}}.{\BBCQ}
\newblock
\APACjournalVolNumPages{Science (New York, N.Y.)}{304}{5672}{855--858}.
\newblock
\begin{APACrefDOI} \doi{10.1126/science.1095932} \end{APACrefDOI}
\PrintBackRefs{\CurrentBib}

\bibitem [\protect \citeauthoryear {%
Niu%
, Bonati%
, Piaggi%
\BCBL {}\ \BBA {} Parrinello%
}{%
Niu%
\ \protect \BOthers {.}}{%
{\protect \APACyear {2020}}%
}]{%
Niu2020a}
\APACinsertmetastar {%
Niu2020a}%
\begin{APACrefauthors}%
Niu, H.%
, Bonati, L.%
, Piaggi, P\BPBI M.%
\BCBL {}\ \BBA {} Parrinello, M.%
\end{APACrefauthors}%
\unskip\
\newblock
\APACrefYearMonthDay{2020}{{\APACmonth{05}}}{}.
\newblock
{\BBOQ}\APACrefatitle {Ab Initio Phase Diagram and Nucleation of Gallium} {Ab
  initio phase diagram and nucleation of gallium}.{\BBCQ}
\newblock
\APACjournalVolNumPages{Nature Communications}{11}{2654}{}.
\newblock
\begin{APACrefDOI} \doi{10.1038/s41467-020-16372-9} \end{APACrefDOI}
\PrintBackRefs{\CurrentBib}

\bibitem [\protect \citeauthoryear {%
Novella%
\ \protect \BOthers {.}}{%
Novella%
\ \protect \BOthers {.}}{%
{\protect \APACyear {2017}}%
}]{%
Novella2017}
\APACinsertmetastar {%
Novella2017}%
\begin{APACrefauthors}%
Novella, D.%
, Jacobsen, B.%
, Weber, P\BPBI K.%
, Tyburczy, J\BPBI A.%
, Ryerson, F\BPBI J.%
\BCBL {}\ \BBA {} Du~Frane, W\BPBI L.%
\end{APACrefauthors}%
\unskip\
\newblock
\APACrefYearMonthDay{2017}{{\APACmonth{07}}}{}.
\newblock
{\BBOQ}\APACrefatitle {Hydrogen self-diffusion in single crystal olivine and
  electrical conductivity of the {Earth}’s mantle} {Hydrogen self-diffusion
  in single crystal olivine and electrical conductivity of the {Earth}’s
  mantle}.{\BBCQ}
\newblock
\APACjournalVolNumPages{Scientific Reports}{7}{1}{5344}.
\newblock
\begin{APACrefDOI} \doi{10.1038/s41598-017-05113-6} \end{APACrefDOI}
\PrintBackRefs{\CurrentBib}

\bibitem [\protect \citeauthoryear {%
Ohtani%
}{%
Ohtani%
}{%
{\protect \APACyear {2015}}%
}]{%
Ohtani2015a}
\APACinsertmetastar {%
Ohtani2015a}%
\begin{APACrefauthors}%
Ohtani, E.%
\end{APACrefauthors}%
\unskip\
\newblock
\APACrefYearMonthDay{2015}{}{}.
\newblock
{\BBOQ}\APACrefatitle {Hydrous Minerals and the Storage of Water in the Deep
  Mantle} {Hydrous minerals and the storage of water in the deep
  mantle}.{\BBCQ}
\newblock
\APACjournalVolNumPages{Chemical Geology}{418}{}{6--15}.
\newblock
\begin{APACrefDOI} \doi{10.1016/j.chemgeo.2015.05.005} \end{APACrefDOI}
\PrintBackRefs{\CurrentBib}

\bibitem [\protect \citeauthoryear {%
Ohtani%
}{%
Ohtani%
}{%
{\protect \APACyear {2021}}%
}]{%
Ohtani2021a}
\APACinsertmetastar {%
Ohtani2021a}%
\begin{APACrefauthors}%
Ohtani, E.%
\end{APACrefauthors}%
\unskip\
\newblock
\APACrefYearMonthDay{2021}{}{}.
\newblock
{\BBOQ}\APACrefatitle {Hydration and Dehydration in Earth's Interior}
  {Hydration and dehydration in earth's interior}.{\BBCQ}
\newblock
\BIn{} R.~Jeanloz\ \BBA {} K\BPBI H.~Freeman\ (\BEDS), \APACrefbtitle {Annual
  Review of Earth and Planetary Sciences, Vol 49, 2021} {Annual review of earth
  and planetary sciences, vol 49, 2021}\ (\BVOL~49, \BPGS\ 253--278).
\newblock
\APACaddressPublisher{{Palo Alto}}{{Annual Reviews}}.
\newblock
\begin{APACrefDOI} \doi{10.1146/annurev-earth-080320-062509} \end{APACrefDOI}
\PrintBackRefs{\CurrentBib}

\bibitem [\protect \citeauthoryear {%
Ohtani%
\ \BBA {} Litasov%
}{%
Ohtani%
\ \BBA {} Litasov%
}{%
{\protect \APACyear {2006}}%
}]{%
Ohtani2006a}
\APACinsertmetastar {%
Ohtani2006a}%
\begin{APACrefauthors}%
Ohtani, E.%
\BCBT {}\ \BBA {} Litasov, K\BPBI D.%
\end{APACrefauthors}%
\unskip\
\newblock
\APACrefYearMonthDay{2006}{}{}.
\newblock
{\BBOQ}\APACrefatitle {The Effect of Water on Mantle Phase Transitions} {The
  effect of water on mantle phase transitions}.{\BBCQ}
\newblock
\APACjournalVolNumPages{Water in Nominally Anhydrous Minerals}{62}{}{397--419}.
\newblock
\begin{APACrefDOI} \doi{10.2138/rmg.2006.62.17} \end{APACrefDOI}
\PrintBackRefs{\CurrentBib}

\bibitem [\protect \citeauthoryear {%
Olsen%
}{%
Olsen%
}{%
{\protect \APACyear {1999}}%
{\protect \APACexlab {{\protect \BCnt {1}}}}}]{%
Olsen1999a}
\APACinsertmetastar {%
Olsen1999a}%
\begin{APACrefauthors}%
Olsen, N.%
\end{APACrefauthors}%
\unskip\
\newblock
\APACrefYearMonthDay{1999{\protect \BCnt {1}}}{}{}.
\newblock
{\BBOQ}\APACrefatitle {Induction Studies with Satellite Data} {Induction
  studies with satellite data}.{\BBCQ}
\newblock
\APACjournalVolNumPages{Surveys in Geophysics}{20}{3-4}{309--340}.
\newblock
\begin{APACrefDOI} \doi{10.1023/a:1006611303582} \end{APACrefDOI}
\PrintBackRefs{\CurrentBib}

\bibitem [\protect \citeauthoryear {%
Olsen%
}{%
Olsen%
}{%
{\protect \APACyear {1999}}%
{\protect \APACexlab {{\protect \BCnt {2}}}}}]{%
Olsen1999b}
\APACinsertmetastar {%
Olsen1999b}%
\begin{APACrefauthors}%
Olsen, N.%
\end{APACrefauthors}%
\unskip\
\newblock
\APACrefYearMonthDay{1999{\protect \BCnt {2}}}{}{}.
\newblock
{\BBOQ}\APACrefatitle {Long-Period (30 Days-1 Year) Electromagnetic Sounding
  and the Electrical Conductivity of the Lower Mantle beneath {{Europe}}}
  {Long-period (30 days-1 year) electromagnetic sounding and the electrical
  conductivity of the lower mantle beneath {{Europe}}}.{\BBCQ}
\newblock
\APACjournalVolNumPages{Geophysical Journal International}{138}{1}{179--187}.
\newblock
\begin{APACrefDOI} \doi{10.1046/j.1365-246x.1999.00854.x} \end{APACrefDOI}
\PrintBackRefs{\CurrentBib}

\bibitem [\protect \citeauthoryear {%
Panero%
, Smyth%
, Pigott%
, Liu%
\BCBL {}\ \BBA {} Frost%
}{%
Panero%
\ \protect \BOthers {.}}{%
{\protect \APACyear {2013}}%
}]{%
Panero2013a}
\APACinsertmetastar {%
Panero2013a}%
\begin{APACrefauthors}%
Panero, W\BPBI R.%
, Smyth, J\BPBI R.%
, Pigott, J\BPBI S.%
, Liu, Z\BPBI X.%
\BCBL {}\ \BBA {} Frost, D\BPBI J.%
\end{APACrefauthors}%
\unskip\
\newblock
\APACrefYearMonthDay{2013}{}{}.
\newblock
{\BBOQ}\APACrefatitle {Hydrous Ringwoodite to 5 {{K}} and 35 {{GPa}}:
  {{Multiple}} Hydrogen Bonding Sites Resolved with {{FTIR}} Spectroscopy}
  {Hydrous ringwoodite to 5 {{K}} and 35 {{GPa}}: {{Multiple}} hydrogen bonding
  sites resolved with {{FTIR}} spectroscopy}.{\BBCQ}
\newblock
\APACjournalVolNumPages{American Mineralogist}{98}{4}{637--642}.
\newblock
\begin{APACrefDOI} \doi{10.2138/am.2013.3978} \end{APACrefDOI}
\PrintBackRefs{\CurrentBib}

\bibitem [\protect \citeauthoryear {%
Peng%
\ \BBA {} Deng%
}{%
Peng%
\ \BBA {} Deng%
}{%
{\protect \APACyear {2023}}%
}]{%
Peng_data_2023}
\APACinsertmetastar {%
Peng_data_2023}%
\begin{APACrefauthors}%
Peng, Y.%
\BCBT {}\ \BBA {} Deng, J.%
\end{APACrefauthors}%
\unskip\
\newblock
\APACrefYearMonthDay{2023}{}{}.
\newblock
\APACrefbtitle {Hydrogen Diffusion in the Lower Mantle Revealed by Machine
  Learning Potentials} {Hydrogen diffusion in the lower mantle revealed by
  machine learning potentials}\ [Dataset].
\newblock
\APACaddressPublisher{}{{OSF}}.
\newblock
\begin{APACrefURL} \url{https://osf.io/brwcv/} \end{APACrefURL}
\newblock
\begin{APACrefDOI} \doi{10.17605/OSF.IO/BRWCV} \end{APACrefDOI}
\PrintBackRefs{\CurrentBib}

\bibitem [\protect \citeauthoryear {%
Perdew%
\ \protect \BOthers {.}}{%
Perdew%
\ \protect \BOthers {.}}{%
{\protect \APACyear {2008}}%
}]{%
Perdew2008a}
\APACinsertmetastar {%
Perdew2008a}%
\begin{APACrefauthors}%
Perdew, J\BPBI P.%
, Ruzsinszky, A.%
, Csonka, G\BPBI I.%
, Vydrov, O\BPBI A.%
, Scuseria, G\BPBI E.%
, Constantin, L\BPBI A.%
\BDBL {}Burke, K.%
\end{APACrefauthors}%
\unskip\
\newblock
\APACrefYearMonthDay{2008}{}{}.
\newblock
{\BBOQ}\APACrefatitle {Restoring the Density-Gradient Expansion for Exchange in
  Solids and Surfaces} {Restoring the density-gradient expansion for exchange
  in solids and surfaces}.{\BBCQ}
\newblock
\APACjournalVolNumPages{Physical Review Letters}{100}{13}{4}.
\newblock
\begin{APACrefDOI} \doi{10.1103/PhysRevLett.100.136406} \end{APACrefDOI}
\PrintBackRefs{\CurrentBib}

\bibitem [\protect \citeauthoryear {%
Piaggi%
\ \BBA {} Parrinello%
}{%
Piaggi%
\ \BBA {} Parrinello%
}{%
{\protect \APACyear {2019}}%
}]{%
Piaggi2019a}
\APACinsertmetastar {%
Piaggi2019a}%
\begin{APACrefauthors}%
Piaggi, P\BPBI M.%
\BCBT {}\ \BBA {} Parrinello, M.%
\end{APACrefauthors}%
\unskip\
\newblock
\APACrefYearMonthDay{2019}{}{}.
\newblock
{\BBOQ}\APACrefatitle {Multithermal-Multibaric Molecular Simulations from a
  Variational Principle} {Multithermal-multibaric molecular simulations from a
  variational principle}.{\BBCQ}
\newblock
\APACjournalVolNumPages{Physical Review Letters}{122}{5}{6}.
\newblock
\begin{APACrefDOI} \doi{10.1103/PhysRevLett.122.050601} \end{APACrefDOI}
\PrintBackRefs{\CurrentBib}

\bibitem [\protect \citeauthoryear {%
Plimpton%
}{%
Plimpton%
}{%
{\protect \APACyear {1995}}%
}]{%
Plimpton1995a}
\APACinsertmetastar {%
Plimpton1995a}%
\begin{APACrefauthors}%
Plimpton, S.%
\end{APACrefauthors}%
\unskip\
\newblock
\APACrefYearMonthDay{1995}{}{}.
\newblock
{\BBOQ}\APACrefatitle {Fast Parallel Algorithms for Short-Range
  Molecular-Dynamics} {Fast parallel algorithms for short-range
  molecular-dynamics}.{\BBCQ}
\newblock
\APACjournalVolNumPages{Journal of Computational Physics}{117}{1}{1--19}.
\newblock
\begin{APACrefDOI} \doi{10.1006/jcph.1995.1039} \end{APACrefDOI}
\PrintBackRefs{\CurrentBib}

\bibitem [\protect \citeauthoryear {%
Plimpton%
, Kohlmeyer%
, Thompson%
, Moore%
\BCBL {}\ \BBA {} Berger%
}{%
Plimpton%
\ \protect \BOthers {.}}{%
{\protect \APACyear {2021}}%
}]{%
plimpton_lammps_2021}
\APACinsertmetastar {%
plimpton_lammps_2021}%
\begin{APACrefauthors}%
Plimpton, S.%
, Kohlmeyer, A.%
, Thompson, A.%
, Moore, S.%
\BCBL {}\ \BBA {} Berger, R.%
\end{APACrefauthors}%
\unskip\
\newblock
\APACrefYearMonthDay{2021}{{\APACmonth{09}}}{}.
\newblock
\APACrefbtitle {{LAMMPS} {Stable} release 29 {September} 2021} {{LAMMPS}
  {Stable} release 29 {September} 2021}\ [Software].
\newblock
\APACaddressPublisher{}{Zenodo}.
\newblock
\begin{APACrefURL} \url{https://zenodo.org/records/6386596} \end{APACrefURL}
\newblock
\begin{APACrefDOI} \doi{10.5281/zenodo.6386596} \end{APACrefDOI}
\PrintBackRefs{\CurrentBib}

\bibitem [\protect \citeauthoryear {%
Puthe%
, Kuvshinov%
, Khan%
\BCBL {}\ \BBA {} Olsen%
}{%
Puthe%
\ \protect \BOthers {.}}{%
{\protect \APACyear {2015}}%
}]{%
Puthe2015a}
\APACinsertmetastar {%
Puthe2015a}%
\begin{APACrefauthors}%
Puthe, C.%
, Kuvshinov, A.%
, Khan, A.%
\BCBL {}\ \BBA {} Olsen, N.%
\end{APACrefauthors}%
\unskip\
\newblock
\APACrefYearMonthDay{2015}{}{}.
\newblock
{\BBOQ}\APACrefatitle {A New Model of {{Earth}}'s Radial Conductivity Structure
  Derived from over 10 Yr of Satellite and Observatory Magnetic Data} {A new
  model of {{Earth}}'s radial conductivity structure derived from over 10 yr of
  satellite and observatory magnetic data}.{\BBCQ}
\newblock
\APACjournalVolNumPages{Geophysical Journal International}{203}{3}{1864--1872}.
\newblock
\begin{APACrefDOI} \doi{10.1093/gji/ggv407} \end{APACrefDOI}
\PrintBackRefs{\CurrentBib}

\bibitem [\protect \citeauthoryear {%
Schmalzried%
\ \BBA {} Schmalzried%
}{%
Schmalzried%
\ \BBA {} Schmalzried%
}{%
{\protect \APACyear {1974}}%
}]{%
Schmalzried1974}
\APACinsertmetastar {%
Schmalzried1974}%
\begin{APACrefauthors}%
Schmalzried, H.%
\BCBT {}\ \BBA {} Schmalzried, H.%
\end{APACrefauthors}%
\unskip\
\newblock
\APACrefYear{1974}.
\newblock
\APACrefbtitle {Solid state reactions} {Solid state reactions}.
\newblock
\APACaddressPublisher{Weinheim, Bergstr}{Verl. Chemie}.
\PrintBackRefs{\CurrentBib}

\bibitem [\protect \citeauthoryear {%
Scipioni%
, Stixrude%
\BCBL {}\ \BBA {} Desjarlais%
}{%
Scipioni%
\ \protect \BOthers {.}}{%
{\protect \APACyear {2017}}%
}]{%
Scipioni2017a}
\APACinsertmetastar {%
Scipioni2017a}%
\begin{APACrefauthors}%
Scipioni, R.%
, Stixrude, L.%
\BCBL {}\ \BBA {} Desjarlais, M\BPBI P.%
\end{APACrefauthors}%
\unskip\
\newblock
\APACrefYearMonthDay{2017}{}{}.
\newblock
{\BBOQ}\APACrefatitle {Electrical Conductivity of {{SiO}}{\textsubscript{2}} at
  Extreme Conditions and Planetary Dynamos} {Electrical conductivity of
  {{SiO}}{\textsubscript{2}} at extreme conditions and planetary
  dynamos}.{\BBCQ}
\newblock
\APACjournalVolNumPages{Proceedings of the National Academy of Sciences of the
  United States of America}{114}{34}{9009--9013}.
\newblock
\begin{APACrefDOI} \doi{10.1073/pnas.1704762114} \end{APACrefDOI}
\PrintBackRefs{\CurrentBib}

\bibitem [\protect \citeauthoryear {%
Shankland%
, Peyronneau%
\BCBL {}\ \BBA {} Poirier%
}{%
Shankland%
\ \protect \BOthers {.}}{%
{\protect \APACyear {1993}}%
}]{%
Shankland1993a}
\APACinsertmetastar {%
Shankland1993a}%
\begin{APACrefauthors}%
Shankland, T\BPBI J.%
, Peyronneau, J.%
\BCBL {}\ \BBA {} Poirier, J\BPBI P.%
\end{APACrefauthors}%
\unskip\
\newblock
\APACrefYearMonthDay{1993}{}{}.
\newblock
{\BBOQ}\APACrefatitle {Electrical-Conductivity of the Earths Lower Mantle}
  {Electrical-conductivity of the earths lower mantle}.{\BBCQ}
\newblock
\APACjournalVolNumPages{Nature}{366}{6454}{453--455}.
\newblock
\begin{APACrefDOI} \doi{10.1038/366453a0} \end{APACrefDOI}
\PrintBackRefs{\CurrentBib}

\bibitem [\protect \citeauthoryear {%
Shelyapina%
}{%
Shelyapina%
}{%
{\protect \APACyear {2022}}%
}]{%
Shelyapina2022a}
\APACinsertmetastar {%
Shelyapina2022a}%
\begin{APACrefauthors}%
Shelyapina, M\BPBI G.%
\end{APACrefauthors}%
\unskip\
\newblock
\APACrefYearMonthDay{2022}{}{}.
\newblock
{\BBOQ}\APACrefatitle {Hydrogen Diffusion on, into and in Magnesium Probed by
  {{DFT}}: {{A}} Review} {Hydrogen diffusion on, into and in magnesium probed
  by {{DFT}}: {{A}} review}.{\BBCQ}
\newblock
\APACjournalVolNumPages{Hydrogen}{3}{3}{285--302}.
\newblock
\begin{APACrefDOI} \doi{10.3390/hydrogen3030017} \end{APACrefDOI}
\PrintBackRefs{\CurrentBib}

\bibitem [\protect \citeauthoryear {%
Sinmyo%
, Pesce%
, Greenberg%
, McCammon%
\BCBL {}\ \BBA {} Dubrovinsky%
}{%
Sinmyo%
\ \protect \BOthers {.}}{%
{\protect \APACyear {2014}}%
}]{%
Sinmyo2014a}
\APACinsertmetastar {%
Sinmyo2014a}%
\begin{APACrefauthors}%
Sinmyo, R.%
, Pesce, G.%
, Greenberg, E.%
, McCammon, C.%
\BCBL {}\ \BBA {} Dubrovinsky, L.%
\end{APACrefauthors}%
\unskip\
\newblock
\APACrefYearMonthDay{2014}{}{}.
\newblock
{\BBOQ}\APACrefatitle {Lower Mantle Electrical Conductivity Based on
  Measurements of {{Al}}, {{Fe-bearing}} Perovskite under Lower Mantle
  Conditions} {Lower mantle electrical conductivity based on measurements of
  {{Al}}, {{Fe-bearing}} perovskite under lower mantle conditions}.{\BBCQ}
\newblock
\APACjournalVolNumPages{Earth and Planetary Science Letters}{393}{}{165--172}.
\newblock
\begin{APACrefDOI} \doi{10.1016/j.epsl.2014.02.049} \end{APACrefDOI}
\PrintBackRefs{\CurrentBib}

\bibitem [\protect \citeauthoryear {%
Slater%
}{%
Slater%
}{%
{\protect \APACyear {1964}}%
}]{%
Slater1964a}
\APACinsertmetastar {%
Slater1964a}%
\begin{APACrefauthors}%
Slater, J\BPBI C.%
\end{APACrefauthors}%
\unskip\
\newblock
\APACrefYearMonthDay{1964}{}{}.
\newblock
{\BBOQ}\APACrefatitle {Atomic Radii in Crystals} {Atomic radii in
  crystals}.{\BBCQ}
\newblock
\APACjournalVolNumPages{Journal of Chemical Physics}{41}{10}{3199}.
\newblock
\begin{APACrefDOI} \doi{10.1063/1.1725697} \end{APACrefDOI}
\PrintBackRefs{\CurrentBib}

\bibitem [\protect \citeauthoryear {%
Sun%
, Yoshino%
, Sakamoto%
\BCBL {}\ \BBA {} Yurimoto%
}{%
Sun%
\ \protect \BOthers {.}}{%
{\protect \APACyear {2015}}%
}]{%
Sun2015a}
\APACinsertmetastar {%
Sun2015a}%
\begin{APACrefauthors}%
Sun, W.%
, Yoshino, T.%
, Sakamoto, N.%
\BCBL {}\ \BBA {} Yurimoto, H.%
\end{APACrefauthors}%
\unskip\
\newblock
\APACrefYearMonthDay{2015}{}{}.
\newblock
{\BBOQ}\APACrefatitle {Hydrogen Self-Diffusivity in Single Crystal Ringwoodite:
  {{Implications}} for Water Content and Distribution in the Mantle Transition
  Zone} {Hydrogen self-diffusivity in single crystal ringwoodite:
  {{Implications}} for water content and distribution in the mantle transition
  zone}.{\BBCQ}
\newblock
\APACjournalVolNumPages{Geophysical Research Letters}{42}{16}{6582--6589}.
\newblock
\begin{APACrefDOI} \doi{10.1002/2015gl064486} \end{APACrefDOI}
\PrintBackRefs{\CurrentBib}

\bibitem [\protect \citeauthoryear {%
Tarits%
\ \BBA {} Mandea%
}{%
Tarits%
\ \BBA {} Mandea%
}{%
{\protect \APACyear {2010}}%
}]{%
Tarits2010a}
\APACinsertmetastar {%
Tarits2010a}%
\begin{APACrefauthors}%
Tarits, P.%
\BCBT {}\ \BBA {} Mandea, M.%
\end{APACrefauthors}%
\unskip\
\newblock
\APACrefYearMonthDay{2010}{}{}.
\newblock
{\BBOQ}\APACrefatitle {The Heterogeneous Electrical Conductivity Structure of
  the Lower Mantle} {The heterogeneous electrical conductivity structure of the
  lower mantle}.{\BBCQ}
\newblock
\APACjournalVolNumPages{Physics of the Earth and Planetary
  Interiors}{183}{1-2}{115--125}.
\newblock
\begin{APACrefDOI} \doi{10.1016/j.pepi.2010.08.002} \end{APACrefDOI}
\PrintBackRefs{\CurrentBib}

\bibitem [\protect \citeauthoryear {%
Townsend%
, Tsuchiya%
, Bina%
\BCBL {}\ \BBA {} Jacobsen%
}{%
Townsend%
\ \protect \BOthers {.}}{%
{\protect \APACyear {2016}}%
}]{%
Townsend2016a}
\APACinsertmetastar {%
Townsend2016a}%
\begin{APACrefauthors}%
Townsend, J\BPBI P.%
, Tsuchiya, J.%
, Bina, C\BPBI R.%
\BCBL {}\ \BBA {} Jacobsen, S\BPBI D.%
\end{APACrefauthors}%
\unskip\
\newblock
\APACrefYearMonthDay{2016}{}{}.
\newblock
{\BBOQ}\APACrefatitle {Water Partitioning between Bridgmanite and
  Postperovskite in the Lowermost Mantle} {Water partitioning between
  bridgmanite and postperovskite in the lowermost mantle}.{\BBCQ}
\newblock
\APACjournalVolNumPages{Earth and Planetary Science Letters}{454}{}{20--27}.
\newblock
\begin{APACrefDOI} \doi{10.1016/j.epsl.2016.08.009} \end{APACrefDOI}
\PrintBackRefs{\CurrentBib}

\bibitem [\protect \citeauthoryear {%
Tribello%
, Bonomi%
, Branduardi%
, Camilloni%
\BCBL {}\ \BBA {} Bussi%
}{%
Tribello%
\ \protect \BOthers {.}}{%
{\protect \APACyear {2014}}%
}]{%
Tribello2014a}
\APACinsertmetastar {%
Tribello2014a}%
\begin{APACrefauthors}%
Tribello, G\BPBI A.%
, Bonomi, M.%
, Branduardi, D.%
, Camilloni, C.%
\BCBL {}\ \BBA {} Bussi, G.%
\end{APACrefauthors}%
\unskip\
\newblock
\APACrefYearMonthDay{2014}{}{}.
\newblock
{\BBOQ}\APACrefatitle {{{PLUMED}} 2: {{New}} Feathers for an Old Bird}
  {{{PLUMED}} 2: {{New}} feathers for an old bird}.{\BBCQ}
\newblock
\APACjournalVolNumPages{Computer Physics Communications}{185}{2}{604--613}.
\newblock
\begin{APACrefDOI} \doi{10.1016/j.cpc.2013.09.018} \end{APACrefDOI}
\PrintBackRefs{\CurrentBib}

\bibitem [\protect \citeauthoryear {%
Velimsky%
}{%
Velimsky%
}{%
{\protect \APACyear {2010}}%
}]{%
Velimsky2010a}
\APACinsertmetastar {%
Velimsky2010a}%
\begin{APACrefauthors}%
Velimsky, J.%
\end{APACrefauthors}%
\unskip\
\newblock
\APACrefYearMonthDay{2010}{}{}.
\newblock
{\BBOQ}\APACrefatitle {Electrical Conductivity in the Lower Mantle:
  {{Constraints}} from {{CHAMP}} Satellite Data by Time-Domain {{EM}} Induction
  Modelling} {Electrical conductivity in the lower mantle: {{Constraints}} from
  {{CHAMP}} satellite data by time-domain {{EM}} induction modelling}.{\BBCQ}
\newblock
\APACjournalVolNumPages{Physics of the Earth and Planetary
  Interiors}{180}{3-4}{111--117}.
\newblock
\begin{APACrefDOI} \doi{10.1016/j.pepi.2010.02.007} \end{APACrefDOI}
\PrintBackRefs{\CurrentBib}

\bibitem [\protect \citeauthoryear {%
Verhoeven%
, Thebault%
, Saturnino%
, Houliez%
\BCBL {}\ \BBA {} Langlais%
}{%
Verhoeven%
\ \protect \BOthers {.}}{%
{\protect \APACyear {2021}}%
}]{%
Verhoeven2021a}
\APACinsertmetastar {%
Verhoeven2021a}%
\begin{APACrefauthors}%
Verhoeven, O.%
, Thebault, E.%
, Saturnino, D.%
, Houliez, A.%
\BCBL {}\ \BBA {} Langlais, B.%
\end{APACrefauthors}%
\unskip\
\newblock
\APACrefYearMonthDay{2021}{}{}.
\newblock
{\BBOQ}\APACrefatitle {Electrical Conductivity and Temperature of the
  {{Earth}}'s Mantle Inferred from {{Bayesian}} Inversion of {{Swarm}} Vector
  Magnetic Data} {Electrical conductivity and temperature of the {{Earth}}'s
  mantle inferred from {{Bayesian}} inversion of {{Swarm}} vector magnetic
  data}.{\BBCQ}
\newblock
\APACjournalVolNumPages{Physics of the Earth and Planetary
  Interiors}{314}{}{11}.
\newblock
\begin{APACrefDOI} \doi{10.1016/j.pepi.2021.106702} \end{APACrefDOI}
\PrintBackRefs{\CurrentBib}

\bibitem [\protect \citeauthoryear {%
Wang%
, Zhang%
, Han%
\BCBL {}\ \BBA {} E%
}{%
Wang%
\ \protect \BOthers {.}}{%
{\protect \APACyear {2018}}%
}]{%
Wang2018a}
\APACinsertmetastar {%
Wang2018a}%
\begin{APACrefauthors}%
Wang, H.%
, Zhang, L\BPBI F.%
, Han, J\BPBI Q.%
\BCBL {}\ \BBA {} E, W\BPBI N.%
\end{APACrefauthors}%
\unskip\
\newblock
\APACrefYearMonthDay{2018}{}{}.
\newblock
{\BBOQ}\APACrefatitle {{{DeePMD-kit}}: {{A}} Deep Learning Package for
  Many-Body Potential Energy Representation and Molecular Dynamics}
  {{{DeePMD-kit}}: {{A}} deep learning package for many-body potential energy
  representation and molecular dynamics}.{\BBCQ}
\newblock
\APACjournalVolNumPages{Computer Physics Communications}{228}{}{178--184}.
\newblock
\begin{APACrefDOI} \doi{10.1016/j.cpc.2018.03.016} \end{APACrefDOI}
\PrintBackRefs{\CurrentBib}

\bibitem [\protect \citeauthoryear {%
J.~Xu%
\ \protect \BOthers {.}}{%
J.~Xu%
\ \protect \BOthers {.}}{%
{\protect \APACyear {2011}}%
}]{%
Xu2011}
\APACinsertmetastar {%
Xu2011}%
\begin{APACrefauthors}%
Xu, J.%
, Yamazaki, D.%
, Katsura, T.%
, Wu, X.%
, Remmert, P.%
, Yurimoto, H.%
\BCBL {}\ \BBA {} Chakraborty, S.%
\end{APACrefauthors}%
\unskip\
\newblock
\APACrefYearMonthDay{2011}{}{}.
\newblock
{\BBOQ}\APACrefatitle {Silicon and magnesium diffusion in a single crystal of
  {MgSiO3} perovskite} {Silicon and magnesium diffusion in a single crystal of
  {MgSiO3} perovskite}.{\BBCQ}
\newblock
\APACjournalVolNumPages{Journal of Geophysical Research: Solid
  Earth}{116}{B12}{}.
\newblock
\begin{APACrefDOI} \doi{10.1029/2011JB008444} \end{APACrefDOI}
\PrintBackRefs{\CurrentBib}

\bibitem [\protect \citeauthoryear {%
Y\BPBI S.~Xu%
, McCammon%
\BCBL {}\ \BBA {} Poe%
}{%
Y\BPBI S.~Xu%
\ \protect \BOthers {.}}{%
{\protect \APACyear {1998}}%
}]{%
Xu1998a}
\APACinsertmetastar {%
Xu1998a}%
\begin{APACrefauthors}%
Xu, Y\BPBI S.%
, McCammon, C.%
\BCBL {}\ \BBA {} Poe, B\BPBI T.%
\end{APACrefauthors}%
\unskip\
\newblock
\APACrefYearMonthDay{1998}{}{}.
\newblock
{\BBOQ}\APACrefatitle {The Effect of Alumina on the Electrical Conductivity of
  Silicate Perovskite} {The effect of alumina on the electrical conductivity of
  silicate perovskite}.{\BBCQ}
\newblock
\APACjournalVolNumPages{Science (New York, N.Y.)}{282}{5390}{922--924}.
\newblock
\begin{APACrefDOI} \doi{10.1126/science.282.5390.922} \end{APACrefDOI}
\PrintBackRefs{\CurrentBib}

\bibitem [\protect \citeauthoryear {%
M.~Yang%
, Karmakar%
\BCBL {}\ \BBA {} Parrinello%
}{%
M.~Yang%
\ \protect \BOthers {.}}{%
{\protect \APACyear {2021}}%
}]{%
Yang2021a}
\APACinsertmetastar {%
Yang2021a}%
\begin{APACrefauthors}%
Yang, M.%
, Karmakar, T.%
\BCBL {}\ \BBA {} Parrinello, M.%
\end{APACrefauthors}%
\unskip\
\newblock
\APACrefYearMonthDay{2021}{{\APACmonth{08}}}{}.
\newblock
{\BBOQ}\APACrefatitle {Liquid-Liquid Critical Point in Phosphorus}
  {Liquid-liquid critical point in phosphorus}.{\BBCQ}
\newblock
\APACjournalVolNumPages{Physical Review Letters}{127}{080603}{}.
\newblock
\begin{APACrefDOI} \doi{10.1103/PhysRevLett.127.080603} \end{APACrefDOI}
\PrintBackRefs{\CurrentBib}

\bibitem [\protect \citeauthoryear {%
Y\BHBI N.~Yang%
\ \protect \BOthers {.}}{%
Y\BHBI N.~Yang%
\ \protect \BOthers {.}}{%
{\protect \APACyear {2023}}%
}]{%
yang_nanosims_2023}
\APACinsertmetastar {%
yang_nanosims_2023}%
\begin{APACrefauthors}%
Yang, Y\BHBI N.%
, Du, Z.%
, Lu, W.%
, Qi, Y.%
, Zhang, Y\BHBI Q.%
, Zhang, W\BHBI F.%
\BCBL {}\ \BBA {} Zhang, P\BHBI F.%
\end{APACrefauthors}%
\unskip\
\newblock
\APACrefYearMonthDay{2023}{}{}.
\newblock
{\BBOQ}\APACrefatitle {{NanoSIMS} analysis of water content in bridgmanite at
  the micron scale: {An} experimental approach to probe water in {Earth}’s
  deep mantle} {{NanoSIMS} analysis of water content in bridgmanite at the
  micron scale: {An} experimental approach to probe water in {Earth}’s deep
  mantle}.{\BBCQ}
\newblock
\APACjournalVolNumPages{Frontiers in Chemistry}{11}{}{}.
\newblock
\begin{APACrefDOI} \doi{10.3389/fchem.2023.1166593} \end{APACrefDOI}
\PrintBackRefs{\CurrentBib}

\bibitem [\protect \citeauthoryear {%
Yeh%
\ \BBA {} Hummer%
}{%
Yeh%
\ \BBA {} Hummer%
}{%
{\protect \APACyear {2004}}%
}]{%
Yeh2004a}
\APACinsertmetastar {%
Yeh2004a}%
\begin{APACrefauthors}%
Yeh, I\BPBI C.%
\BCBT {}\ \BBA {} Hummer, G.%
\end{APACrefauthors}%
\unskip\
\newblock
\APACrefYearMonthDay{2004}{}{}.
\newblock
{\BBOQ}\APACrefatitle {System-Size Dependence of Diffusion Coefficients and
  Viscosities from Molecular Dynamics Simulations with Periodic Boundary
  Conditions} {System-size dependence of diffusion coefficients and viscosities
  from molecular dynamics simulations with periodic boundary
  conditions}.{\BBCQ}
\newblock
\APACjournalVolNumPages{Journal of Physical Chemistry
  B}{108}{40}{15873--15879}.
\newblock
\begin{APACrefDOI} \doi{10.1021/jp0477147} \end{APACrefDOI}
\PrintBackRefs{\CurrentBib}

\bibitem [\protect \citeauthoryear {%
Yoshino%
}{%
Yoshino%
}{%
{\protect \APACyear {2010}}%
}]{%
Yoshino2010a}
\APACinsertmetastar {%
Yoshino2010a}%
\begin{APACrefauthors}%
Yoshino, T.%
\end{APACrefauthors}%
\unskip\
\newblock
\APACrefYearMonthDay{2010}{}{}.
\newblock
{\BBOQ}\APACrefatitle {Laboratory Electrical Conductivity Measurement of Mantle
  Minerals} {Laboratory electrical conductivity measurement of mantle
  minerals}.{\BBCQ}
\newblock
\APACjournalVolNumPages{Surveys in Geophysics}{31}{2}{163--206}.
\newblock
\begin{APACrefDOI} \doi{10.1007/s10712-009-9084-0} \end{APACrefDOI}
\PrintBackRefs{\CurrentBib}

\bibitem [\protect \citeauthoryear {%
Yoshino%
, Kamada%
, Zhao%
, Ohtani%
\BCBL {}\ \BBA {} Hirao%
}{%
Yoshino%
\ \protect \BOthers {.}}{%
{\protect \APACyear {2016}}%
}]{%
Yoshino2016a}
\APACinsertmetastar {%
Yoshino2016a}%
\begin{APACrefauthors}%
Yoshino, T.%
, Kamada, S.%
, Zhao, C\BPBI C.%
, Ohtani, E.%
\BCBL {}\ \BBA {} Hirao, N.%
\end{APACrefauthors}%
\unskip\
\newblock
\APACrefYearMonthDay{2016}{}{}.
\newblock
{\BBOQ}\APACrefatitle {Electrical Conductivity Model of {{Al-bearing}}
  Bridgmanite with Implications for the Electrical Structure of the {{Earth}}'s
  Lower Mantle} {Electrical conductivity model of {{Al-bearing}} bridgmanite
  with implications for the electrical structure of the {{Earth}}'s lower
  mantle}.{\BBCQ}
\newblock
\APACjournalVolNumPages{Earth and Planetary Science Letters}{434}{}{208--219}.
\newblock
\begin{APACrefDOI} \doi{10.1016/j.epsl.2015.11.032} \end{APACrefDOI}
\PrintBackRefs{\CurrentBib}

\bibitem [\protect \citeauthoryear {%
Zhang%
, Han%
, Wang%
, Car%
\BCBL {}\ \BBA {} Weinan%
}{%
Zhang%
\ \protect \BOthers {.}}{%
{\protect \APACyear {2018}}%
}]{%
Zhang2018a}
\APACinsertmetastar {%
Zhang2018a}%
\begin{APACrefauthors}%
Zhang, L\BPBI F.%
, Han, J\BPBI Q.%
, Wang, H.%
, Car, R.%
\BCBL {}\ \BBA {} Weinan, E.%
\end{APACrefauthors}%
\unskip\
\newblock
\APACrefYearMonthDay{2018}{}{}.
\newblock
{\BBOQ}\APACrefatitle {Deep Potential Molecular Dynamics: {{A}} Scalable Model
  with the Accuracy of Quantum Mechanics} {Deep potential molecular dynamics:
  {{A}} scalable model with the accuracy of quantum mechanics}.{\BBCQ}
\newblock
\APACjournalVolNumPages{Physical Review Letters}{120}{14}{6}.
\newblock
\begin{APACrefDOI} \doi{10.1103/PhysRevLett.120.143001} \end{APACrefDOI}
\PrintBackRefs{\CurrentBib}

\bibitem [\protect \citeauthoryear {%
Zhou%
\ \protect \BOthers {.}}{%
Zhou%
\ \protect \BOthers {.}}{%
{\protect \APACyear {2022}}%
}]{%
Zhou2022a}
\APACinsertmetastar {%
Zhou2022a}%
\begin{APACrefauthors}%
Zhou, W\BPBI Y.%
, Hao, M.%
, Zhang, J\BPBI S.%
, Chen, B.%
, Wang, R\BPBI J.%
\BCBL {}\ \BBA {} Schmandt, B.%
\end{APACrefauthors}%
\unskip\
\newblock
\APACrefYearMonthDay{2022}{}{}.
\newblock
{\BBOQ}\APACrefatitle {Constraining Composition and Temperature Variations in
  the Mantle Transition Zone} {Constraining composition and temperature
  variations in the mantle transition zone}.{\BBCQ}
\newblock
\APACjournalVolNumPages{Nature Communications}{13}{1}{9}.
\newblock
\begin{APACrefDOI} \doi{10.1038/s41467-022-28709-7} \end{APACrefDOI}
\PrintBackRefs{\CurrentBib}

\end{thebibliography}


\begin{thebibliography}{10}
\expandafter\ifx\csname natexlab\endcsname\relax\def\natexlab#1{#1}\fi
\providecommand{\url}[1]{\texttt{#1}}
\providecommand{\href}[2]{#2}
\providecommand{\path}[1]{#1}
\providecommand{\DOIprefix}{doi:}
\providecommand{\ArXivprefix}{arXiv:}
\providecommand{\URLprefix}{URL: }
\providecommand{\Pubmedprefix}{pmid:}
\providecommand{\doi}[1]{\href{http://dx.doi.org/#1}{\path{#1}}}
\providecommand{\Pubmed}[1]{\href{pmid:#1}{\path{#1}}}
\providecommand{\bibinfo}[2]{#2}
\ifx\xfnm\relax \def\xfnm[#1]{\unskip,\space#1}\fi
\bibitem[{Hirel(2015)}]{Hirel2015a}
\bibinfo{author}{Hirel, P.}, \bibinfo{year}{2015}.
\newblock \bibinfo{title}{Atomsk: {{A}} tool for manipulating and converting
  atomic data files}.
\newblock \bibinfo{journal}{Computer Physics Communications}
  \bibinfo{volume}{197}, \bibinfo{pages}{212--219}.
\newblock \DOIprefix\doi{10.1016/j.cpc.2015.07.012}.
\bibitem[{Hoover(1985)}]{Hoover1985a}
\bibinfo{author}{Hoover, W.G.}, \bibinfo{year}{1985}.
\newblock \bibinfo{title}{Canonical dynamics - equilibrium phase-space
  distributions}.
\newblock \bibinfo{journal}{Physical Review A} \bibinfo{volume}{31},
  \bibinfo{pages}{1695--1697}.
\newblock \DOIprefix\doi{10.1103/PhysRevA.31.1695}.
\bibitem[{Jinnouchi et~al.(2019a)Jinnouchi, Karsai and Kresse}]{Jinnouchi2019a}
\bibinfo{author}{Jinnouchi, R.}, \bibinfo{author}{Karsai, F.},
  \bibinfo{author}{Kresse, G.}, \bibinfo{year}{2019}a.
\newblock \bibinfo{title}{On-the-fly machine learning force field generation:
  {Application} to melting points}.
\newblock \bibinfo{journal}{Physical Review B} \bibinfo{volume}{100},
  \bibinfo{pages}{014105}.
\newblock \DOIprefix\doi{10.1103/PhysRevB.100.014105}.
\bibitem[{Jinnouchi et~al.(2019b)Jinnouchi, Lahnsteiner, Karsai, Kresse and
  Bokdam}]{Jinnouchi2019b}
\bibinfo{author}{Jinnouchi, R.}, \bibinfo{author}{Lahnsteiner, J.},
  \bibinfo{author}{Karsai, F.}, \bibinfo{author}{Kresse, G.},
  \bibinfo{author}{Bokdam, M.}, \bibinfo{year}{2019}b.
\newblock \bibinfo{title}{Phase {Transitions} of {Hybrid} {Perovskites}
  {Simulated} by {Machine}-{Learning} {Force} {Fields} {Trained} on the {Fly}
  with {Bayesian} {Inference}}.
\newblock \bibinfo{journal}{Physical Review Letters} \bibinfo{volume}{122},
  \bibinfo{pages}{225701}.
\newblock \DOIprefix\doi{10.1103/PhysRevLett.122.225701}.
\bibitem[{Katsura et~al.(2010)Katsura, Yoneda, Yamazaki, Yoshino and
  Ito}]{Katsura2010a}
\bibinfo{author}{Katsura, T.}, \bibinfo{author}{Yoneda, A.},
  \bibinfo{author}{Yamazaki, D.}, \bibinfo{author}{Yoshino, T.},
  \bibinfo{author}{Ito, E.}, \bibinfo{year}{2010}.
\newblock \bibinfo{title}{Adiabatic temperature profile in the mantle}.
\newblock \bibinfo{journal}{Physics of the Earth and Planetary Interiors}
  \bibinfo{volume}{183}, \bibinfo{pages}{212--218}.
\newblock \DOIprefix\doi{10.1016/j.pepi.2010.07.001}.
\bibitem[{Kresse and Furthmuller(1996)}]{Kresse1996a}
\bibinfo{author}{Kresse, G.}, \bibinfo{author}{Furthmuller, J.},
  \bibinfo{year}{1996}.
\newblock \bibinfo{title}{Efficient iterative schemes for ab initio
  total-energy calculations using a plane-wave basis set}.
\newblock \bibinfo{journal}{Physical Review B} \bibinfo{volume}{54},
  \bibinfo{pages}{11169--11186}.
\newblock \DOIprefix\doi{10.1103/PhysRevB.54.11169}.
\bibitem[{Kresse and Joubert(1999)}]{Kresse1999a}
\bibinfo{author}{Kresse, G.}, \bibinfo{author}{Joubert, D.},
  \bibinfo{year}{1999}.
\newblock \bibinfo{title}{From ultrasoft pseudopotentials to the projector
  augmented-wave method}.
\newblock \bibinfo{journal}{Physical Review B} \bibinfo{volume}{59},
  \bibinfo{pages}{1758--1775}.
\newblock \DOIprefix\doi{10.1103/PhysRevB.59.1758}.
\bibitem[{Muir and Brodholt(2018)}]{Muir2018}
\bibinfo{author}{Muir, J.M.R.}, \bibinfo{author}{Brodholt, J.P.},
  \bibinfo{year}{2018}.
\newblock \bibinfo{title}{Water distribution in the lower mantle:
  {{Implications}} for hydrolytic weakening}.
\newblock \bibinfo{journal}{Earth and Planetary Science Letters}
  \bibinfo{volume}{484}, \bibinfo{pages}{363--369}.
\newblock \DOIprefix\doi{10.1016/j.epsl.2017.11.051}.
\bibitem[{Perdew et~al.(2008)Perdew, Ruzsinszky, Csonka, Vydrov, Scuseria,
  Constantin, Zhou and Burke}]{Perdew2008a}
\bibinfo{author}{Perdew, J.P.}, \bibinfo{author}{Ruzsinszky, A.},
  \bibinfo{author}{Csonka, G.I.}, \bibinfo{author}{Vydrov, O.A.},
  \bibinfo{author}{Scuseria, G.E.}, \bibinfo{author}{Constantin, L.A.},
  \bibinfo{author}{Zhou, X.L.}, \bibinfo{author}{Burke, K.},
  \bibinfo{year}{2008}.
\newblock \bibinfo{title}{Restoring the density-gradient expansion for exchange
  in solids and surfaces}.
\newblock \bibinfo{journal}{Physical Review Letters} \bibinfo{volume}{100},
  \bibinfo{pages}{4}.
\newblock \DOIprefix\doi{10.1103/PhysRevLett.100.136406}.
\bibitem[{Wang et~al.(2021)Wang, Kingsbury, McDermott, Horton, Jain, Ong,
  Dwaraknath and Persson}]{Wang2021}
\bibinfo{author}{Wang, A.}, \bibinfo{author}{Kingsbury, R.},
  \bibinfo{author}{McDermott, M.}, \bibinfo{author}{Horton, M.},
  \bibinfo{author}{Jain, A.}, \bibinfo{author}{Ong, S.P.},
  \bibinfo{author}{Dwaraknath, S.}, \bibinfo{author}{Persson, K.A.},
  \bibinfo{year}{2021}.
\newblock \bibinfo{title}{A framework for quantifying uncertainty in {{DFT}}
  energy corrections}.
\newblock \bibinfo{journal}{Scientific Reports} \bibinfo{volume}{11},
  \bibinfo{pages}{15496}.
\newblock \DOIprefix\doi{10.1038/s41598-021-94550-5}.

\end{thebibliography}

\end{document}


\maketitle

\begin{center}
    Department of Geosciences, Princeton University, Princeton, NJ 08544, USA
\end{center}

\renewcommand{\theequation}{S\arabic{equation}}
\renewcommand{\thetable}{S\arabic{table}}
\renewcommand{\thefigure}{S\arabic{figure}}
\newcommand{\brg}{\mathrm{MgSiO_3}}
\newcommand{\water}{\mathrm{H_2O}}
\newcommand{\hyd}{\mathrm{H}}
\newcommand{\hbrg}[4]{Mg$_{#1}$Si$_{#2}$O$_{#3}$H$_{#4}$}
\newcommand{\mgtoh}{(2H)\textsubscript{Mg}}
\newcommand{\sitoh}{(4H)\textsubscript{Si}}
\newcommand{\simg}{(Mg + 2H)\textsubscript{Si}}
\newcommand{\sial}{(Al + H)\textsubscript{Si}}
\setcounter{equation}{0}  
\setcounter{figure}{0}
\setcounter{table}{0}

\noindent\textbf{\\\\\\ \large Contents of this file}
\begin{enumerate}
\item Texts S1 to S2
\item Figures S1 to S10
\end{enumerate}
\noindent\textbf{\large Additional Supporting Information (Files uploaded separately)}
\begin{enumerate}
\item Captions for Table S1 and S2
\end{enumerate}

\newpage{}
\renewcommand{\thesection}{Text S1}
\section{hydrogen diffusion in Al-bearing bridgmanite}

\citet{Muir2018} reported that most water in bridgmanite may be incorporated via Al\textsuperscript{3+}-H\textsuperscript{+} pairs throughout the lower mantle. Therefore, we investigate the hydrogen diffusion in Al-bearing bridgmanite considering (Al + H)\textsubscript{Si} defects using the on-the-fly machine learning force field (MLFF) trained from \textit{ab initio} data.

We consider three supercells containing 80 ($2 \times 2 \times 1$ for {\it Pbnm} bridgmanite), 360 ($3 \times 3 \times 2$ for {\it Pbnm} bridgmanite), and 2880 ($6 \times 6 \times 4$ for {\it Pbnm} bridgmanite) atoms. The water content of the system is controlled by substituting different numbers of Si atoms with Al atoms and hydrogen atoms, generating (Al + H)\textsubscript{Si} defects that are evenly distributed in the supercell.
All \textit{ab initio} calculations and MLFF generations are performed using VASP \citep{Kresse1996a}. For \textit{ab initio} calculations, we adopt the PBEsol approximation \citep{Perdew2008a} with the projector augmented wave (PAW) method \citep{Kresse1999a}. The core radii of O (2s$^2$2p$^4$), Si (3s$^2$3p$^2$), Mg (2p$^6$3s$^2$), Al (3s$^2$3p$^1$), and H (1s$^1$) are 0.820 Å, 1.312 Å, 1.058 Å, 1.402 Å, and 0.370 Å, respectively. We employ a 500-eV energy cutoff and sample the Brillouin zone at the $\Gamma$ point. All molecular dynamics (MD) simulations are performed in the NVT ensemble at 2000 K controlled by the Nosé-Hoover thermostat \citep{Hoover1985a} with a time step of 0.5 fs. During MD simulations, MLFFs are trained using total energies, forces, and stress tensors from automatically selected DFT structures. Similar to the DeePMD method in the main text, the potential energy of the system is decomposed as the sum of local energies derived from the local environments of atoms. A Bayesian error estimate is applied to decide between DFT or MLFF forces for the MD simulation. If the predicted errors exceed the threshold, a new reference structure will be chosen and recalculated, based on which the coefficients of the MLFF will be adjusted. A detailed description of the MLFF approach can be found in \citet{Jinnouchi2019a} and \citet{Jinnouchi2019b}. 

We initiate our on-the-fly training with a small system (Mg\textsubscript{16}Si\textsubscript{14}Al\textsubscript{2}O\textsubscript{48}H\textsubscript{2}) and extend the training set by continuing the on-the-fly training using larger systems ($3 \times 3 \times 2$ for {\it Pbnm} bridgmanite) with various water contents (0.12--1.68 wt\%) at different pressures (24, 25, and 26 GPa). The Al content in the DFT training set ranges from 0.37--5.04 wt\% and is proportional to the water content. The final training set consists of 1800 DFT configurations and the corresponding MLFF is refitted for prediction-only MD runs. We compare the energies, atomic forces, and stresses from the MLFF to those from DFT calculations for 200 configurations that are not included in the training set, and the root-mean-square errors of the energies, atomic forces, and stresses are 0.55 meV atom$^{-1}$, 0.09 eV Å$^{-1}$, and 0.09 GPa, respectively. Since the training set only contains one mineral phase with one hydrogen substitution mechanism, and the MLFF is limited to $\sim$2000 K and 24--26 GPa, this error is much smaller than our MLP introduced in the main text.

Prediction-only MD simulations are performed with the MLFF at 2000 K and 25 GPa for different supercells of bridgmanite with different water contents (Mg\textsubscript{16}Si\textsubscript{14}Al\textsubscript{2}O\textsubscript{48}H\textsubscript{2}, $C_{\rm water}=1.12$ wt\%; Mg\textsubscript{72}Si\textsubscript{70}Al\textsubscript{2}O\textsubscript{216}H\textsubscript{2}, $C_{\rm water}=0.25$ wt\%; Mg\textsubscript{576}Si\textsubscript{568}Al\textsubscript{8}O\textsubscript{1728}H\textsubscript{8}, $C_{\rm water}=0.12$ wt\%). For comparison, the same supercells of bridgmanite with (Mg + 2H)\textsubscript{Si} defects are simulated using the MLP 
(Mg\textsubscript{17}Si\textsubscript{15}O\textsubscript{48}H\textsubscript{2}, $C_{\rm water}=1.12$ wt\%; Mg\textsubscript{73}Si\textsubscript{71}O\textsubscript{216}H\textsubscript{2}, $C_{\rm water}=0.25$ wt\%; Mg\textsubscript{580}Si\textsubscript{572}O\textsubscript{1728}H\textsubscript{8}, $C_{\rm water}=0.12$ wt\%). The simulation time is 100--200 ps for each run and 10 independent runs with different initial velocity distributions are performed for each system to evaluate the uncertainties. Two 100-ps trajectories of hydrogen in the largest systems with the (Al + H)\textsubscript{Si} defect and the (Mg + 2H)\textsubscript{Si} defect are shown in Fig.~\ref{fig:Al-traj}a and Fig.~\ref{fig:Al-traj}b, respectively. The hydrogen diffusion coefficients in all those systems are derived from the mean square displacements and plotted in Fig.~\ref{fig:Al-Mg}. For Mg\textsubscript{17}Si\textsubscript{15}O\textsubscript{48}H\textsubscript{2} with the (Mg + 2H)\textsubscript{Si} defect, we used both the MLP and MLFF methods to perform simulations under completely identical conditions, and obtained very close diffusion coefficients (with only a 3.4\% difference, Fig.~\ref{fig:Al-Mg}a), verifying the consistency of the results from these two methods. Consistent with the MD trajectory of hydrogen associated with (Al + H)\textsubscript{Si} reported by \citet{Muir2018}, hydrogen atoms are not strongly bound to a specific  oxygen atom in a fixed lattice site, but are able to diffuse far from the Al\textsuperscript{3+} ion. In addition, at 2000 K and 25 GPa, the hydrogen diffusion coefficients for (Al + H)\textsubscript{Si} and (Mg + 2H)\textsubscript{Si} both fall within the order of magnitude of 10$^8$ m$^2$\;s$^{-1}$, exceeding (2H)\textsubscript{Mg} and (4H)\textsubscript{Si} by at least an order of magnitude (see Fig.~5). We conclude that (Al + H)\textsubscript{Si} and (Mg + 2H)\textsubscript{Si} exhibit remarkably similar hydrogen diffusion patterns with closely matching hydrogen diffusion coefficients at the uppermost lower mantle conditions, considering all three water and aluminum contents. Owing to this intrinsic similarity between (Al + H)\textsubscript{Si} and (Mg + 2H)\textsubscript{Si}, we propose that the hydrogen diffusion in these two defects can be treated analogously, and our discussions about (Mg + 2H)\textsubscript{Si} might well apply to (Al + H)\textsubscript{Si} too.

\renewcommand{\thesection}{Text S2}
\section{Defect distribution in the supercell}

To accurately quantify the distribution of defects in the supercell, we used the average inter-defect distance of nearest neighbors ($\xi$) as a collective variable defined as
\begin{equation}
    \xi = \sum_{i=1}^N \frac{d^{\rm min}_i}{N}
\end{equation}
where $N$ is the total number of defects in the system, and $d^{\rm min}$ is the distance between a defect and its nearest neighboring defect. Taking (2H)\textsubscript{Mg} as an example, given positions of Mg atoms that will be substituted by hydrogen atoms, i.e., the positions of defects in the initial configuration for MD simulations, we compute the distance between all defects and their respective nearest neighboring defects, i.e., $d^{\rm min}$, which are then averaged to derive $\xi$. This parameter can quantify how far away the different water-containing defects are. Specifically, small $\xi$ means that defects tend to gather into clusters, and large $\xi$ means that the defects are dispersed with the largest $\xi$ corresponding to the uniform distribution of defects.

In order to determine the stability of different defect distributions and find the most energy-favored one, we conduct a simple Monte Carlo simulation. We generate 100 different hydrous bridgmanite (\hbrg{60}{64}{192}{8}) configurations with (2H)\textsubscript{Mg} defects, among which 98 configurations are randomly generated using Atomsk \citep{Hirel2015a}, that is, 4 of the 64 Mg atoms are randomly selected to be substituted by hydrogen atoms. The rest two configurations include one with the largest $\xi$ where defects are evenly distributed in the supercell, and the other with the smallest $\xi$ where 4 nearest Mg atoms within a unit cell were replaced with H atoms. We calculated the enthalpy of each configuration using both MLP and DFT methods, as shown in Fig.~\ref{fig:enthalpy}. For the MLP results, although the different defect distributions lead to some variation in the enthalpy of the system, the standard deviation is only 6.06 \si{meV.atom^{-1}} enthalpies are almost identical considering the uncertainty of our MLP, which is smaller than the RMSE of our MLP (RMSE = 6.39 \si{meV.atom^{-1}}). This is more pronounced in the DFT results, where the standard deviation of the enthalpy is as low as 0.46 \si{meV.atom^{-1}}, which is far lower than the theoretical uncertainty of calculating enthalpy using the DFT method \citep[tens of \si{meV.atom^{-1}}, ][]{Wang2021}. Therefore, we argue that the enthalpy variation in Fig.~\ref{fig:enthalpy} is negligible and all random configurations with different defect distributions may have almost the same enthalpy. Therefore, it may be true that the probability of various defect distributions occurring in nature is approximately equal, and it is difficult to find a specific defect configuration that is preferred in terms of energy. Thus, we can directly use any type of defect distribution configuration for diffusion simulation, instead of searching for the most stable one.

We investigate the impact of defect distribution on hydrogen diffusion by simulating and calculating the hydrogen diffusion coefficients using multiple different configurations with randomly distributed defects.
We then compared the results with those for which defects are uniformly distributed, in order to quantify how the defect distribution ultimately affects the hydrogen diffusion coefficient. We choose two representative bridgmanite systems at 25 GPa and 2000 K: \hbrg{16368}{16384}{49152}{32} (0.0175 wt\% water, Fig.~\ref{fig:random}a) and \hbrg{8128}{8192}{24576}{128} (0.140 wt\% water, Fig.~\ref{fig:random}b). The blue square represents the average value of hydrogen diffusion coefficients calculated from 10 independent simulations with different initial velocity distributions, considering the same supercell containing uniformly distributed defects. The orange square represents the average result obtained from 10 different supercells with randomly distributed defects. The results suggest that while using a random defect distribution may slightly increase the uncertainty of the calculated diffusivity, the impact on the mean is negligible compared to the calculation error, validating the result obtained from the uniform defect distribution in the manuscript.

\clearpage
\begin{figure*}[!h]
    \centering
    \includegraphics[width=\textwidth]{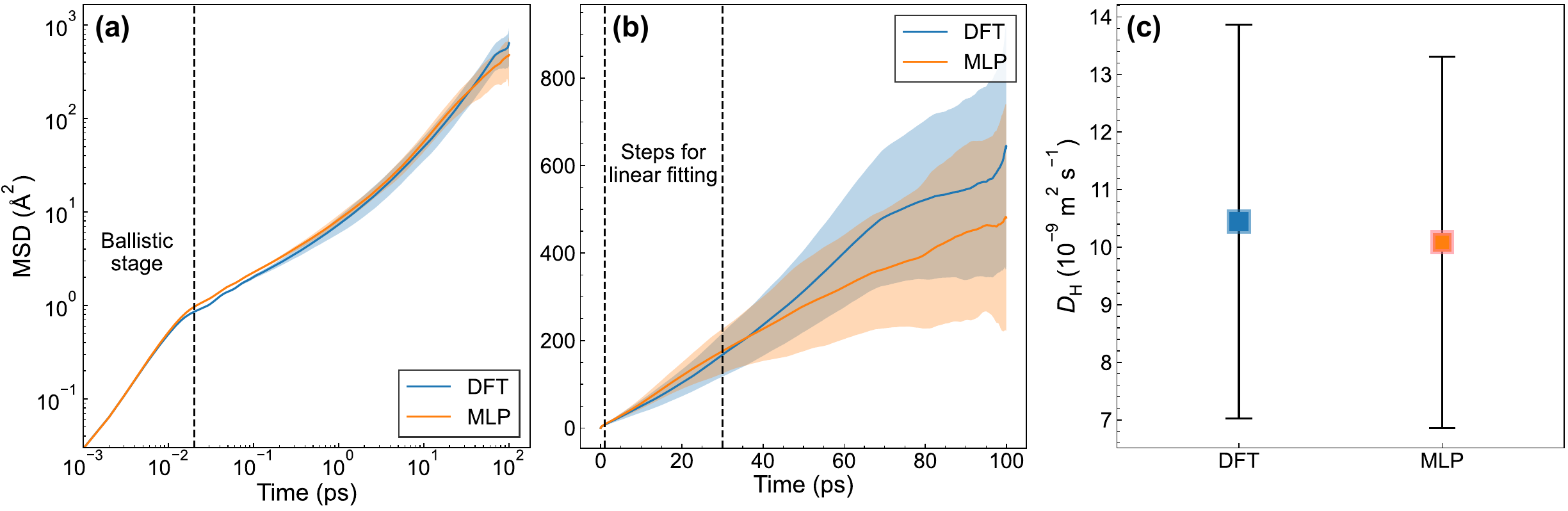}
    \caption{Comparisons of mean square displacements (MSD) and hydrogen diffusion coefficients ($D_{\rm H}$) derived with the density functional theory (DFT) and the machine learning potential (MLP) for hydrous bridgmanite (Mg$_{15}$Si$_{16}$O$_{48}$H$_{2}$) at 3000 K and 25 GPa. The shading and the error bar represent the standard deviation of 10 independent simulations. (a) MSD as a function of the simulation time using double logarithmic axes; (b) MSD as a function of the simulation time using linear axes; (c) $D_{\rm H}$ calculated using DFT and MLP.}\label{fig:DFT}
\end{figure*}

\clearpage
\begin{figure*}
    \centering
    \includegraphics[width=0.6\textwidth]{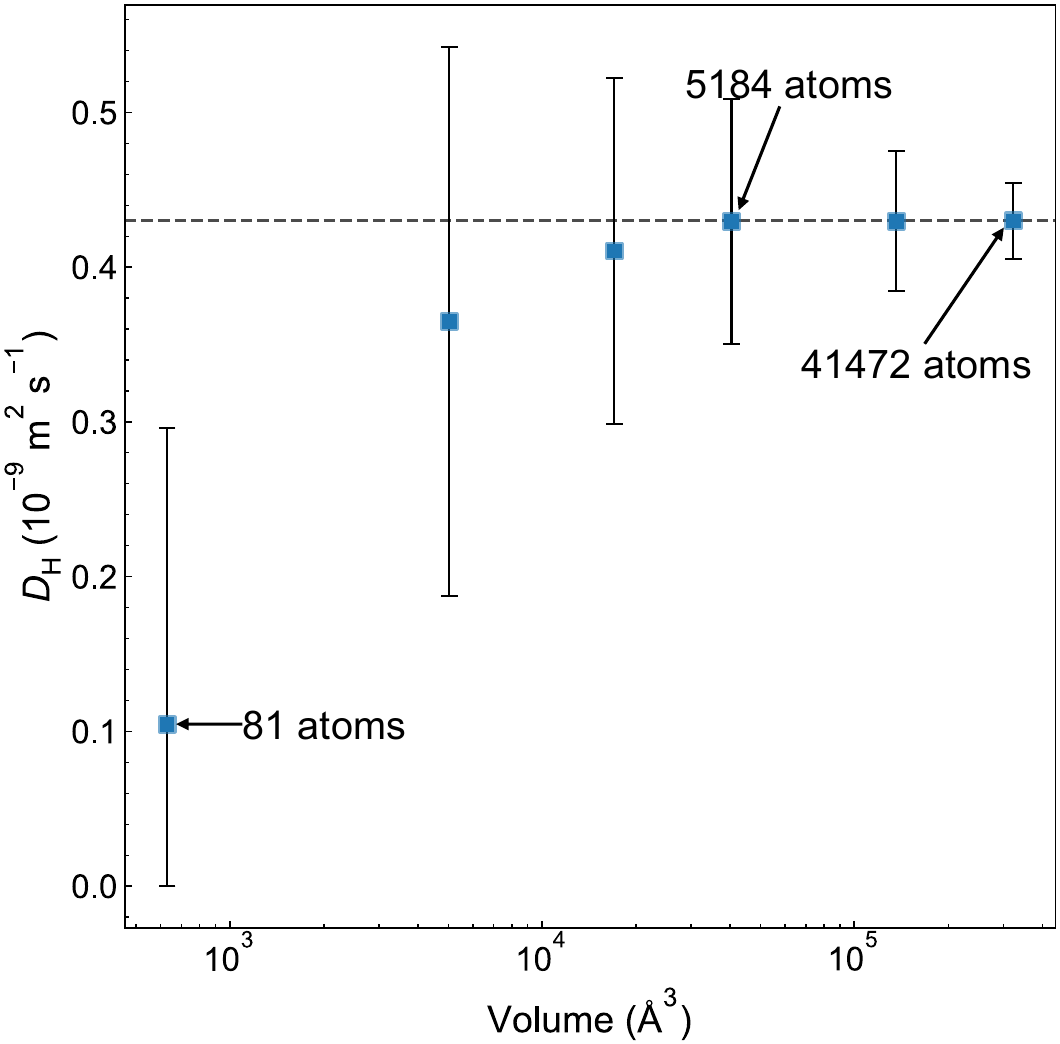}
    \caption{Hydrogen diffusion coefficients vs. supercell volumes of hydrous bridgmanite ({\hbrg{0.9375}{}{3}{0.125}}) at 2000 K and 25 GPa. The total number of atoms is indicated. The error bar represents the 2SD of 10 independent simulations. Hydrogen diffusivities increase and converge to a constant (the dashed line) with system size.}\label{fig:size}
\end{figure*}

\clearpage
\begin{figure*}
    \centering
    \includegraphics[width=0.8\textwidth]{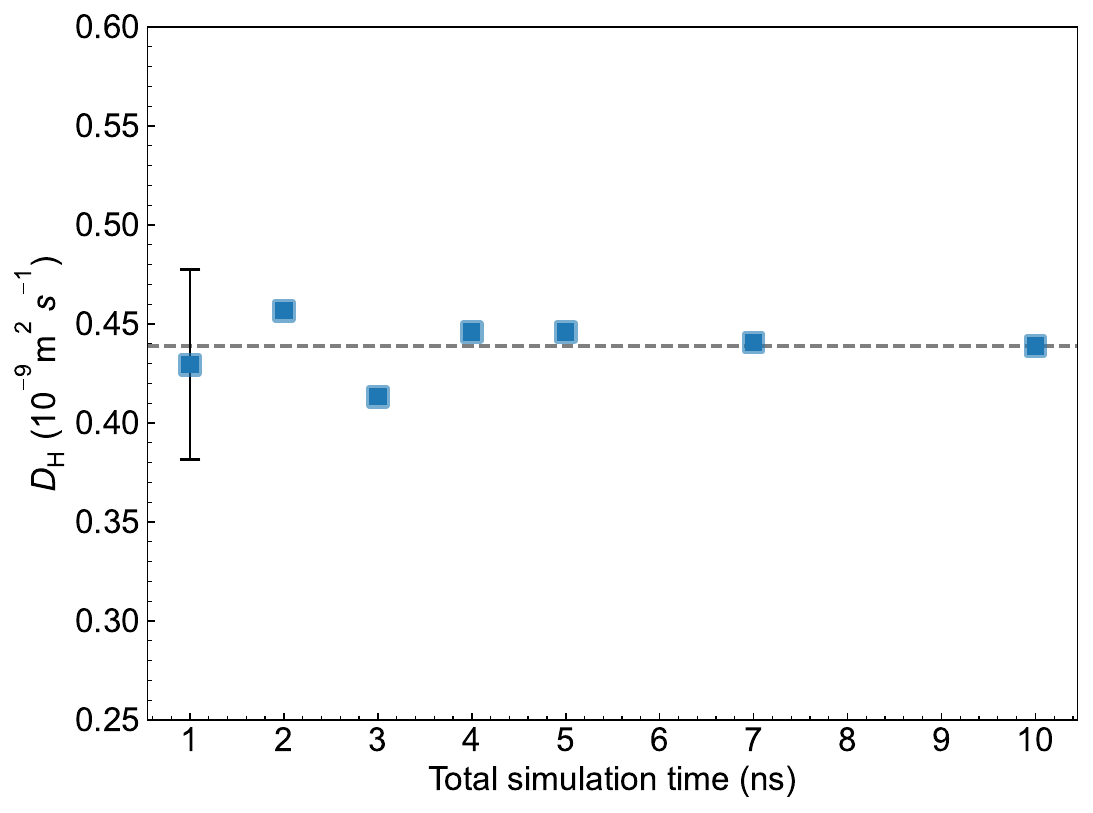}
    \caption{Hydrogen diffusion coefficients as a function of total simulation time of hydrous bridgmanite ({\hbrg{960}{1024}{3072}{128}}) at 2000 K and 25 GPa. The error bar represents the 2SD of 10 independent simulations with a simulation time of 1 ns.}\label{fig:time}
\end{figure*}

\clearpage
\begin{figure*}
    \centering
    \includegraphics[width=0.6\textwidth]{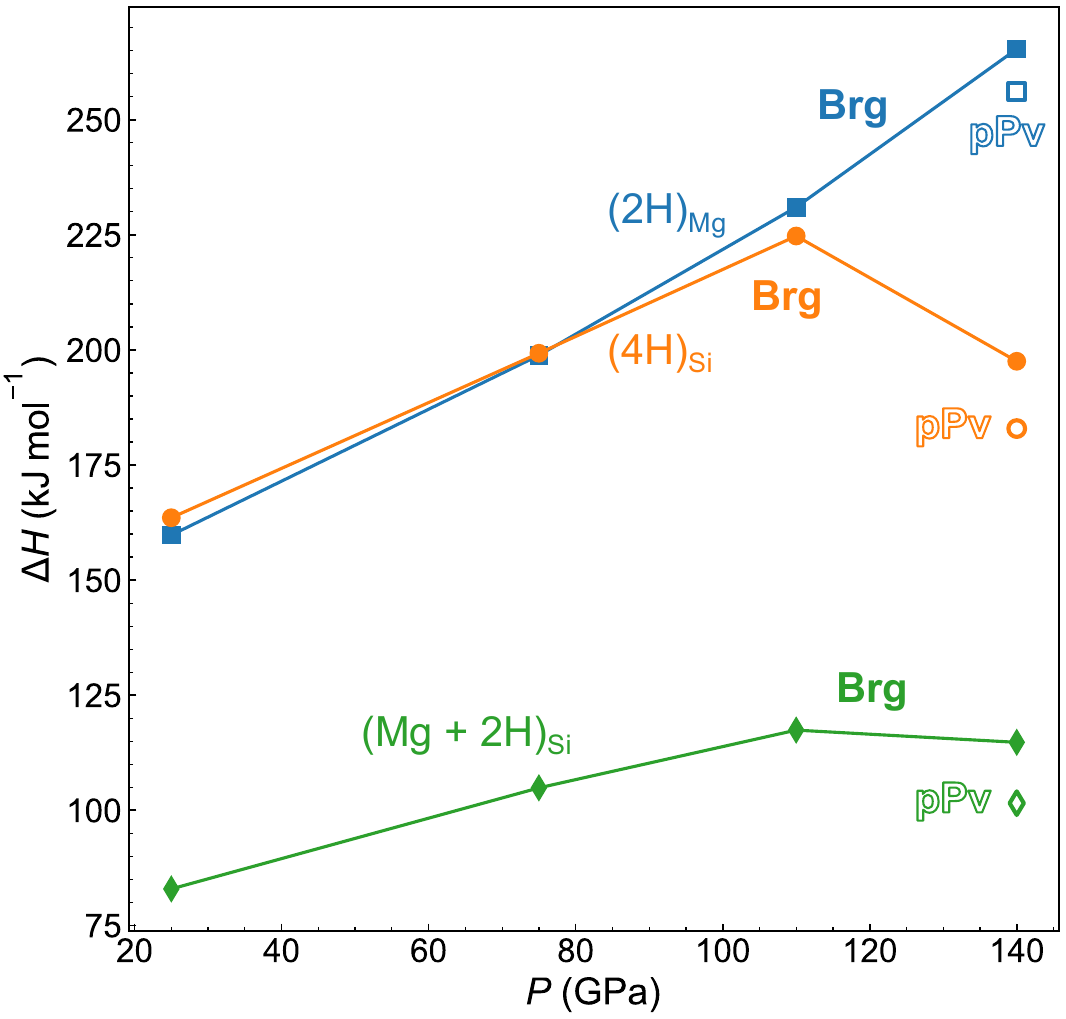}
    \caption{The activation enthalpy of hydrogen diffusion as a function of pressure for different hydrogen incorporation mechanisms. Filled and empty symbols represent bridgmanite and post-perovskite, respectively. (2H)\textsubscript{Mg} (\hbrg{7680}{8192}{24576}{1024}), (4H)\textsubscript{Si} (\hbrg{8192}{7936}{24576}{1024}), and (Mg + 2H)\textsubscript{Si} (\hbrg{8704}{7680}{24576}{1024}) defects are shown in blue, orange, and green, respectively.}\label{fig:activation}
\end{figure*}

\clearpage
\begin{figure*}
    \centering
    \includegraphics[width=\textwidth]{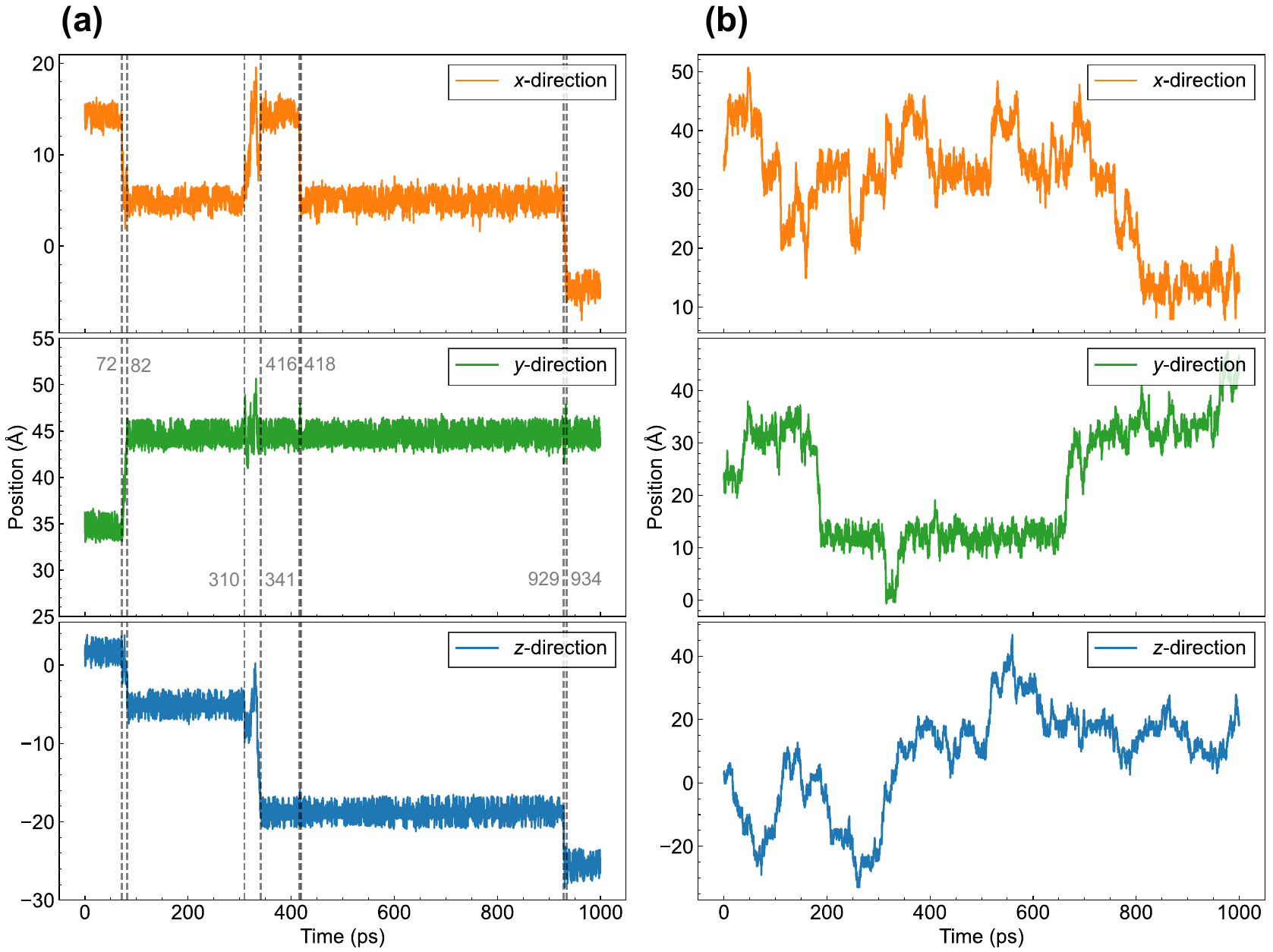}
    \caption{The atomic positions along the three crystallographic axes as a function of time in hydrous bridgmanite for (a) the (2H)\textsubscript{Mg} defect ({\hbrg{960}{1024}{3072}{128}}) and (b) the (Mg + 2H)\textsubscript{Si} defect ({\hbrg{1088}{960}{3072}{128}}) at 2000 K and 25 GPa. The trajectories of these atoms are shown in Fig. 3.}\label{fig:position}
\end{figure*}

\clearpage
\begin{figure*}
    \centering
    \includegraphics[width=\textwidth]{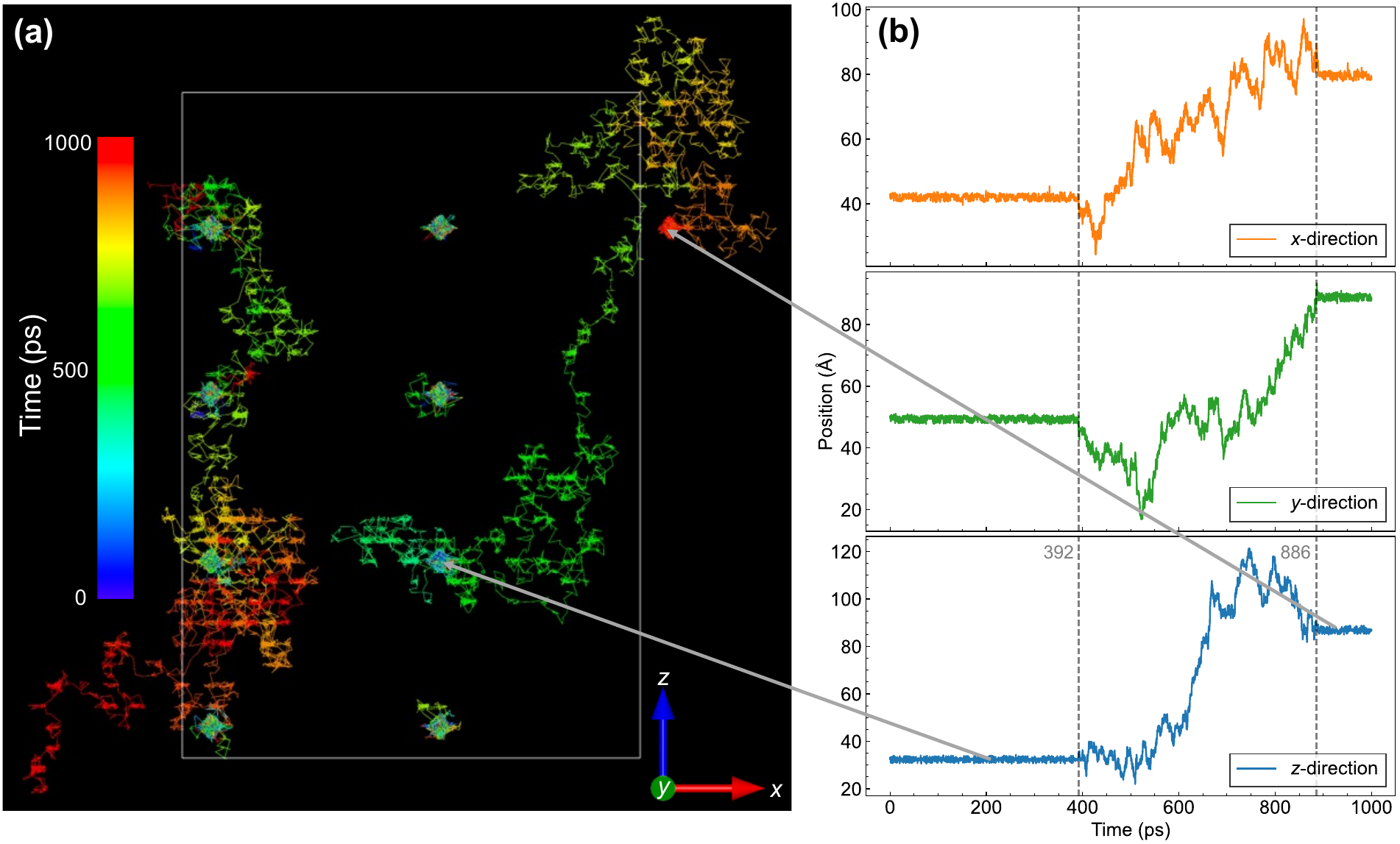}
    \caption{(a) Color mapping of the trajectory of 32 hydrogen atoms over time in hydrous bridgmanite for the (2H)\textsubscript{Mg} defect ({\hbrg{16368}{16384}{49152}{32}}, $C_{\rm water} = 0.0175$ wt\%) at 2000 K and 25 GPa. As the simulation time increases from 0 ps to 1000 ps, the color of the trajectory changes along the visible spectrum. All atoms are omitted for clarity. (b) The atomic position along the three crystallographic axes as a function of time of one hydrogen atom (pointed by gray arrows).}\label{fig:w175}
\end{figure*}

\clearpage
\begin{figure*}
    \centering
    \includegraphics[width=\textwidth]{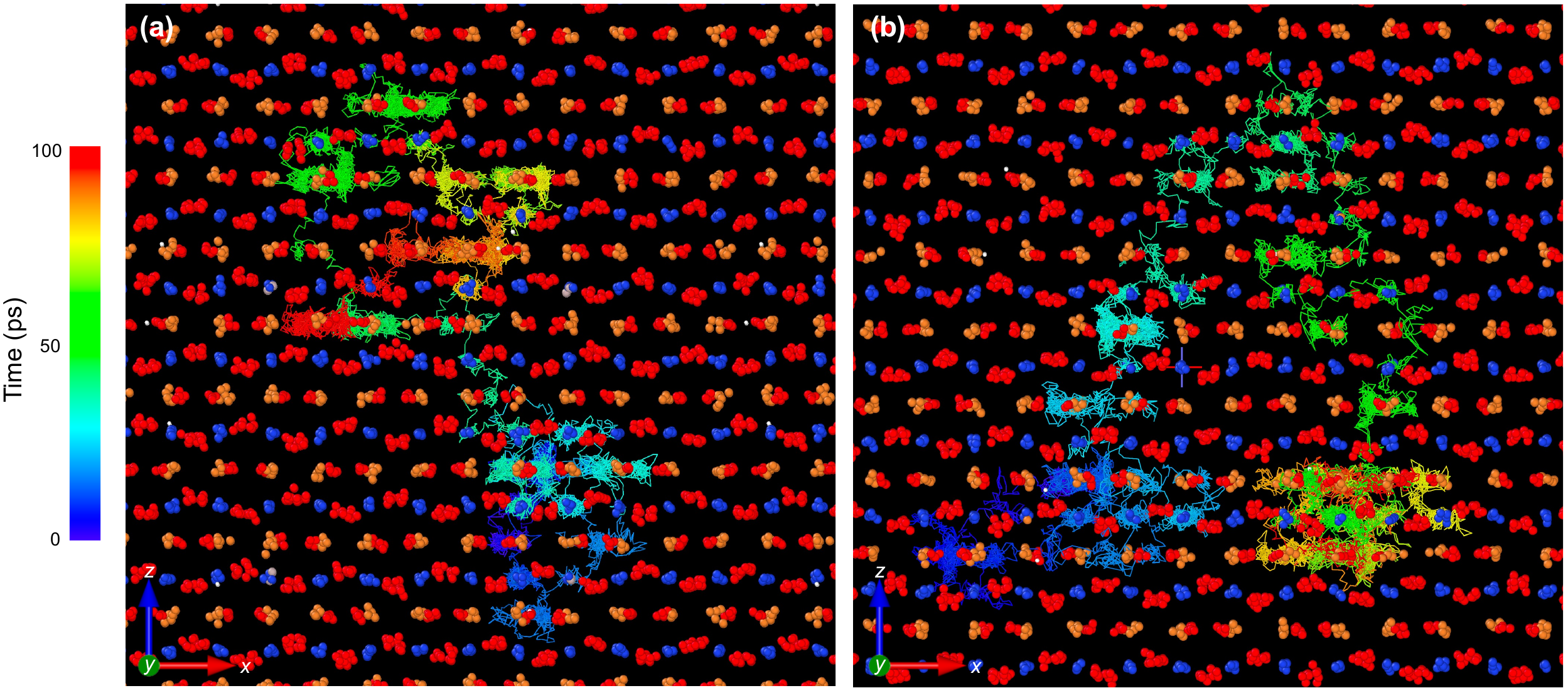}
    \caption{Color mapping of the trajectories of two hydrogen atoms over time in hydrous bridgmanite for (a) the \sial{} defect (Mg\textsubscript{576}Si\textsubscript{568}Al\textsubscript{8}O\textsubscript{1728}H\textsubscript{8}, $C_{\rm water}=0.12$ wt\%) and (b) the \simg{} defect (Mg\textsubscript{580}Si\textsubscript{572}O\textsubscript{1728}H\textsubscript{8}, $C_{\rm water}=0.12$ wt\%) at 2000 K and 25 GPa. As the simulation time increases from 0 ps to 100 ps, the color of the trajectory changes along the visible spectrum. Orange, blue, red, and white spheres represent magnesium, silicon, oxygen, and hydrogen, respectively.}\label{fig:Al-traj}
\end{figure*}

\clearpage
\begin{figure*}
    \centering
    \includegraphics[width=\textwidth]{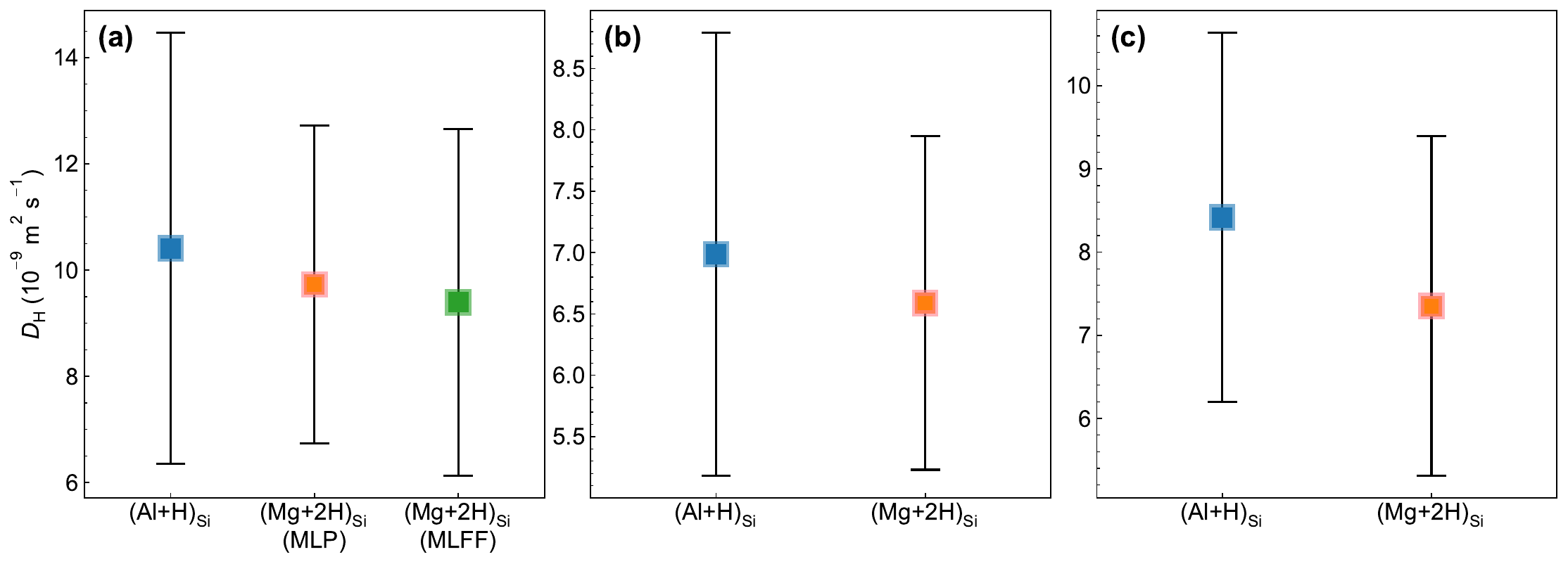}
    \caption{Comparisons of hydrogen diffusion coefficients of hydrous bridgmanite with \sial{} and \simg{} defects at 2000 K and 25 GPa. (a) \sial{}: Mg\textsubscript{16}Si\textsubscript{14}Al\textsubscript{2}O\textsubscript{48}H\textsubscript{2}, $C_{\rm water}=1.12$ wt\%; \simg{}: Mg\textsubscript{17}Si\textsubscript{15}O\textsubscript{48}H\textsubscript{2}, $C_{\rm water}=1.12$ wt\%.
    (b) \sial{}: Mg\textsubscript{72}Si\textsubscript{70}Al\textsubscript{2}O\textsubscript{216}H\textsubscript{2}, $C_{\rm water}=0.25$ wt\%; \simg{}: Mg\textsubscript{73}Si\textsubscript{71}O\textsubscript{216}H\textsubscript{2}, $C_{\rm water}=0.25$ wt\%.
    (c) \sial{}: Mg\textsubscript{576}Si\textsubscript{568}Al\textsubscript{8}O\textsubscript{1728}H\textsubscript{8}, $C_{\rm water}=0.12$ wt\%; \simg{}: Mg\textsubscript{580}Si\textsubscript{572}O\textsubscript{1728}H\textsubscript{8}, $C_{\rm water}=0.12$ wt\%. The simulation time is 200 ps for (a) and (b), and 100 ps for (c).}\label{fig:Al-Mg}
\end{figure*}

\clearpage
\begin{figure*}
    \centering
    \includegraphics[width=0.8\textwidth]{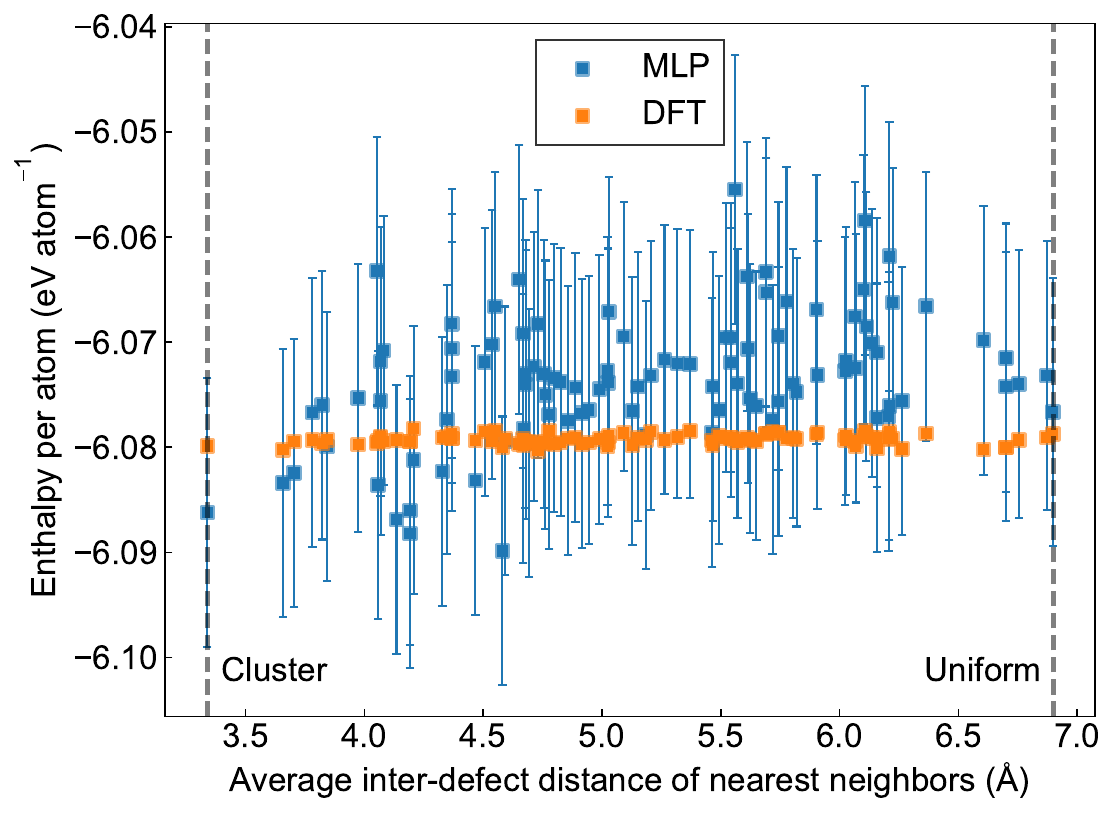}
    \caption{The static enthalpy of 100 different defect distributions of hydrous bridgmanite (\hbrg{60}{64}{192}{8}) calculated using DFT and MLP at 25 GPa. The error bars represent the 2RMSE of our MLP.}\label{fig:enthalpy}
\end{figure*}

\clearpage
\begin{figure*}
    \centering
    \includegraphics[width=\textwidth]{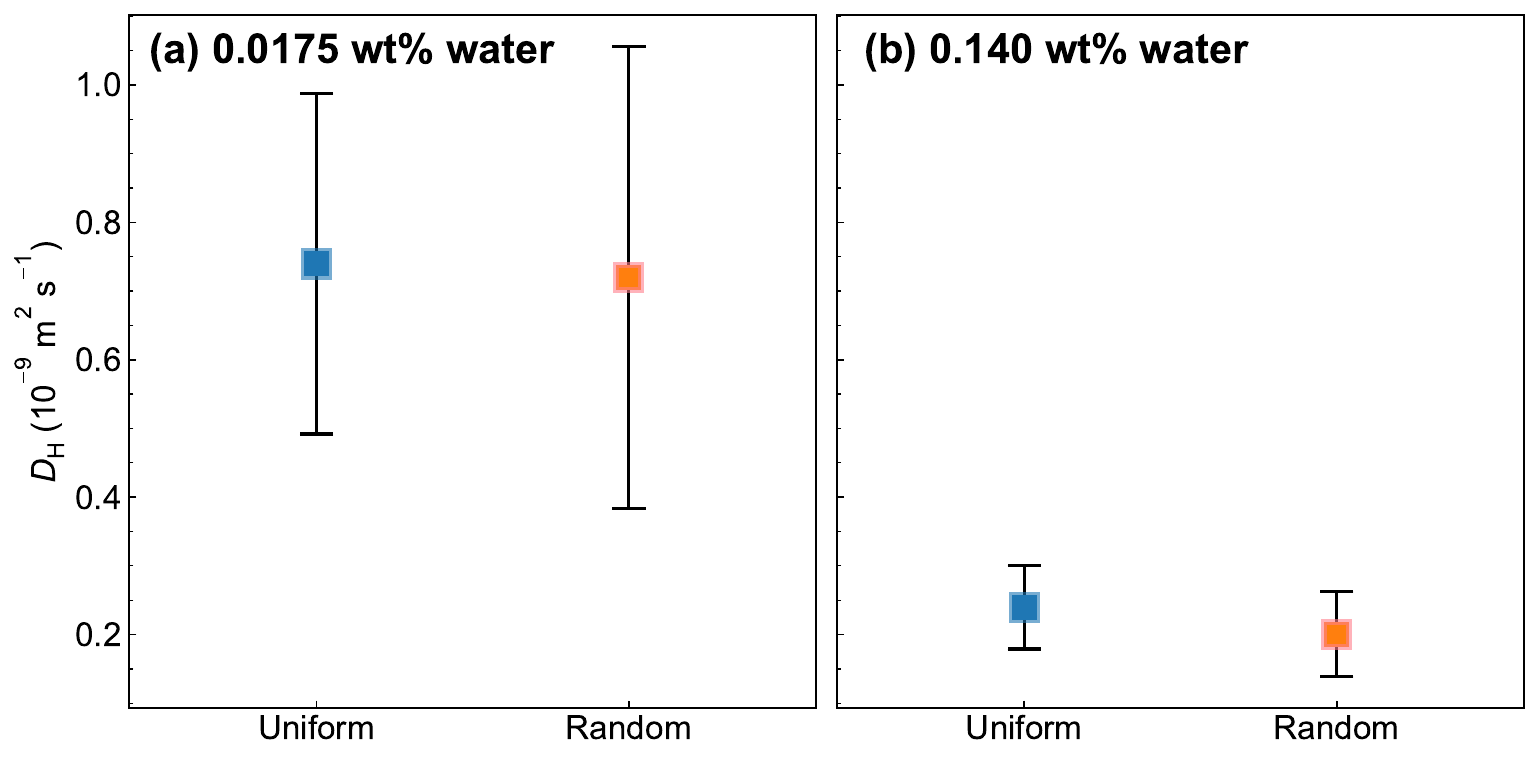}
    \caption{Comparisons of hydrogen diffusion coefficients of hydrous bridgmanite (a, \hbrg{16368}{16384}{49152}{32}; b, \hbrg{8128}{8192}{24576}{128}) calculated using uniform and random defect distributions at 2000 K and 25 GPa.}\label{fig:random}
\end{figure*}

\clearpage
\begin{table*}
\caption{Hydrogen diffusion coefficient of bridgmanite (Brg) and post-perovskite (pPv) at different pressures (25, 75, and 140 GPa) and temperatures (2000, 3000, and 4000 K). Pre-exponential factor ($D_0$) and activation enthalpy ($\Delta H$) are shown for each pressure.}
\end{table*}

\begin{table*}
\caption{Hydrogen diffusion coefficient and proton conductivity of bridgmanite (Brg) and post-perovskite (pPv) along an adiabatic geotherm \cite{Katsura2010a}. }
\end{table*}


\clearpage
\bibliographystyle{elsarticle-harv}
\bibliography{hydrogen_diffusion}